\documentclass[]{article}
\usepackage{filecontents}
\usepackage [english] {babel}
\usepackage{amsmath,amsfonts,amssymb,amsthm,epsfig,epstopdf,titling,array}
\usepackage [utf8]{inputenc}
\usepackage[T1]{fontenc}
\usepackage{csquotes}
\usepackage{mathtools,slashed}
\usepackage{ulem}
\usepackage{textcomp}
\usepackage{geometry}
\usepackage{float}
\usepackage{bm}
\usepackage[margin=10pt, font=scriptsize, labelfont=bf]{caption}
\usepackage{subcaption}
\usepackage{braket}
\usepackage{environ}
\usepackage{verbatim}
\usepackage{dsfont}
\usepackage{tikz}
\usetikzlibrary{shapes}
\usetikzlibrary{positioning}
\usetikzlibrary{patterns}
\usetikzlibrary{decorations.pathmorphing}
\usepackage[colorinlistoftodos]{todonotes}

\usepackage[colorlinks]{hyperref}

\usepackage[capitalize]{cleveref}

\usetikzlibrary{trees}
\usetikzlibrary{calc}
\usetikzlibrary{decorations.pathmorphing}
\usetikzlibrary{decorations.markings}
\usetikzlibrary{shapes}
\usetikzlibrary{fit}

\theoremstyle{plain}

\newtheorem{theorem}{Theorem}
\numberwithin{theorem}{section}

\newtheorem{Lemma}[theorem]{Lemma}

\newtheorem{Proposition}[theorem]{Proposition}

\theoremstyle{remark}
\newtheorem{remark}[theorem]{Remark}
\newtheorem{Remark}[theorem]{Remark}

\numberwithin{equation}{section}

\theoremstyle{definition}
\newtheorem{defn}[theorem]{Definition}

\definecolor{darkGreen}{rgb}{0,0.5,0}
\hypersetup{
	colorlinks,
	citecolor=darkGreen,
	filecolor=red,
	linkcolor=blue}

\tikzset{vertex/.style={circle,fill=black,inner sep=2pt},
	ctVertex/.style={diamond,fill=white,draw,inner sep=2.2pt},
	bigvertex/.style={circle,fill=black,inner sep=3pt},
	E/.append style={fill=white,draw},
	probeEP/.style={circle,fill=black,draw,inner sep=2pt,
		prefix after command= {\pgfextra{\tikzset{every pin/.style = {pin edge={decorate,decoration={snake,amplitude=2pt,segment length =4pt}}}}}}
	},
	bareProbeEP/.style={rectangle,fill=black,draw,inner sep=3pt,
		prefix after command= {\pgfextra{\tikzset{every pin/.style = {pin edge={decorate,decoration={snake,amplitude=2pt,segment length =4pt}}}}}}
	},
	nuEP/.style={circle,fill=white,draw, inner sep=2pt},
	linelabel/.style={sloped,above,very near start, inner sep=1pt,execute at begin node=$\scriptstyle,execute at end node=$},
	baseline=(current  bounding  box.center),doubled/.style={double distance= 1pt,line width=1.5pt}
}
\pgfdeclarelayer{background}
\pgfsetlayers{background,main}

\newcommand{\tn}[1]{{\left\vert\kern-0.25ex\left\vert\kern-0.25ex\left\vert #1 
		\right\vert\kern-0.25ex\right\vert\kern-0.25ex\right\vert}}

\def\blue#1{{\color{blue}#1}}

\usepackage{titling}
\thanksmarkseries{arabic}

\author{
	Alessandro Giuliani\thanks{Universit\`a degli Studi Roma Tre, Dipartimento di Matematica e Fisica, L.go S. L. Murialdo 1, 00146 Roma (Italy), \texttt{alessandro.giuliani@uniroma3.it}}, 
	Vieri Mastropietro\thanks{Sapienza Universit\`a di Roma, Dipartimento di Fisica, P.le A. Moro 5, 00185 Roma (Italy),  \texttt{vieri.mastropietro@uniroma1.it}}, 
	Slava Rychkov\thanks{Institut des Hautes \'Etudes Scientifiques, 35 Rte de Chartres, 91440 Bures-sur-Yvette (France), \texttt{slava@ihes.fr}},
	Giuseppe Scola\thanks{Scuola Internazionale Superiore di Studi Avanzati, Mathematics Area, Via Bonomea 265, 34136 Trieste (Italy), \texttt{giuseppescola89@gmail.com}}}

\date{\today}

\title{Non-trivial fixed point of a $\psi^4_d$ fermionic theory, II. Anomalous exponent and scaling operators}

\begin{document}
	
	\maketitle
	
	\begin{abstract}
		We consider the Renormalization Group (RG) fixed-point theory associated with a fermionic $\psi^4_d$ model in $d=1,2,3$ with fractional kinetic term, whose scaling dimension is fixed so that the quartic interaction is weakly relevant in the RG sense. The model is defined in terms of a Grassmann functional integral with interaction $V^*$, solving a fixed-point RG equation in the presence of external fields, and a fixed ultraviolet cutoff. 
		We define and construct the field and density scale-invariant response functions, and prove that the critical exponent of the former is the naive one, while that of the latter is anomalous and analytic. 
		We construct the corresponding (almost-)scaling operators, whose two point correlations are scale-invariant up to a remainder term, which decays like a stretched exponential at distances larger than 
		the inverse of the ultraviolet cutoff. Our proof is based on constructive RG methods and, specifically, on a convergent tree expansion for the generating function of correlations, which generalizes the approach developed by three of the authors in a previous publication \cite{GMR21}. 
	\end{abstract}
	
	\tableofcontents
	
	%%%%%%%%
	
	\section{Introduction}
	
	Since the seminal works of Kadanoff, Wilson, Fisher and others in the 1970s, renormalization group (RG) fixed points and renormalization group flows connecting them are among central objects of study of theoretical physics, with many experimental applications. {In a statistical mechanics interpretation, these fixed points physically correspond to the scaling limits of interacting systems at a second order phase transition; they determine the physical properties of the system at and near criticality, in particular the power law behavior of the correlation functions at short or long distances, and the corresponding critical exponents.} Only a small part of conjectures and intuitions accumulated about RG fixed points in theoretical physics have been put on rigorous mathematical footing. Pioneering results in this area were obtained by Gawedzki and Kupiainen, who in \cite{GK85a} showed non-perturbative stability of the Gaussian fixed point of a scalar field in four dimensions $(d=4)$, while in \cite{GK85c} they could follow, non-perturbatively, renormalization group flow of the Gross-Neveu two-dimensional model from the Gaussian fixed point of the Dirac field at short distances to some intermediate distance scale where the four-fermion coupling is still small. Later on \cite{GK85b} they constructed a non-perturbative interacting fixed point in a \emph{fractional Gross-Neveu} model, which has the fermion propagator $\slashed{k}/k^2$ replaced by $\slashed{k}/|k|^{2-\epsilon}$. This $\epsilon$ serves as a small parameter which is somewhat similar to the Wilson-Fisher deviation from the integer number of dimensions. Fractional Gross-Neveu model instead lives in the integer number of dimension, but has long-ranged interactions. Analogous fractional model can be considered also for a scalar bosonic field, where they should describe critical points of the long range Ising model \cite{FMN72, Sa73, Sa77}. Fixed points in fractional bosonic models were also rigorously constructed \cite{BMS03,Abd06,Sl18}. 
	
	The present paper is the second one in the series started by our recent work \cite{GMR21}. In \cite{GMR21}, three of us considered a fermionic model closely related to the fractional Gross-Neveu model of Gawedzki and Kupiainen. In our \emph{fractional symplectic fermion} model, the fields are spinless fermions $\psi_a$ in $d$ dimensions ($d=1,2,3$) with propagator $\Omega_{ab}/{|k|^{d/2+\epsilon}}$ where $\Omega$ is a symplectic $N\times N$ matrix, and $N$ is assumed even. The model is symmetric with respect to the symplectic group $Sp(N)$ and the bare potential includes the invariant mass and the quartic coupling terms.
	In this model, Ref.~\cite{GMR21} demonstrated the existence of an interacting fixed point for any $\epsilon$ sufficiently small, which could be positive or negative, or in fact complex. One of the main results of Ref.~\cite{GMR21} is that the fixed point potential depends analytically on $\epsilon$ in a small disk $|\epsilon|<\epsilon_0$. {See also \cite{Gr24}, where an extension of 
this result to the case of a continuous RG flow \`a la Polchinski \cite{Po84} has been recently discussed.}
This interesting analyticity property should be a general feature of fixed points of fermionic models, but it was overlooked in prior literature. For bosonic models no such property holds, as they are well defined only for a positive quartic coupling. For fermionic models there is no such contradiction, since the quartic interaction may have any sign for fermions.
	
{The fractional symplectic fermion model of \cite{GMR21} is an excellent theoretical laboratory to study various aspects of critical properties
of statistical physics models, like non mean-field critical exponents. See \cite[Sec.8]{GMR21} for a long list of open problems. Moreover, our model and closely related ones have direct physics applications. For example, similar fractional fermionic models in a large $N$ limit were considered in {\cite{Gross:2017vhb}} in relation to the famous SYK model. Furthermore, as discussed in \cite[Section 8.1.5]{GMR21}, the fractional symplectic fermion model is likely continuously connected to the local symplectic fermion model \cite{Kausch:1995py}, providing a way to compute the critical exponents of the latter in an $\epsilon$-expansions. As to the local symplectic fermion model, it is relevant e.g.~for the description of polymers and loop-erased random walks \cite{Saleur:1991hk,Wiese:2018dow}, in the
context of dS/CFT correspondence \cite{Anninos:2011ui},  and has also been studied theoretically in \cite{LeClair:2006kb,LeClair:2006mb,LeClair:2007iy,Fei:2015kta,Stergiou:2015roa}.}

{The model we investigate is also a perfect playground to test several properties of Conformal Field Theory (CFT) and of the conformal bootstrap program \cite{Poland:2018epd}, which have been extremely successful in numerically computing critical exponents of interacting critical models
at a precision comparable with, if not better than, the best Monte Carlo simulations. In brief, the goal of CFT is to construct a collection of scale invariant correlation 
functions $\langle \mathcal O_{j_1}(z_1)\cdots \mathcal O_{j_n}(z_n)\rangle$ of a complete basis of \emph{scaling operators} $\mathcal O_j$, 
starting from the requirement that these correlations are conformally covariant and satisfy the Operator Product Expansion (OPE) identities. Based on these assumptions, 
the conformal bootstrap program proposes a strategy to compute recursively the critical exponents of all the scaling operators, as well as the explicit form of their multipoint correlations functions. An outline of CFT/bootstrap axioms and a description of the results obtained in such an axiomatic approach can be found e.g.~in \cite{Poland:2018epd} (in physics language) or in \cite{Kravchuk:2021kwe,Rychkov:2020rcd} (in a more mathematical language). The relevance of this method for Statistical Mechanics (SM) comes from the conjecture 
that the correlation functions of a CFT should coincide with the scaling limit of the correlation functions of an appropriate class of SM models at a second order phase transition point: that is, CFT is expected to be an emergent description of (universality classes of) critical SM models. 
The theoretical foundation for this conjecture is Wilsonian RG. In this framework, CFT correlations are expected to be the same as those of the fixed point theory of the RG map, and scaling operators are expected to be the densities of eigenstates of the RG transformation linearized around the fixed point; 
for this reason, sometimes scaling operators are  also called \emph{eigenoperators}, see e.g.~\cite[Eq.(3.10)]{Wegner}. Unfortunately, rigorous results substantiating the validity of this conjecture 
are scarce: even the 
existence of scaling operators starting from a regularized functional integral or from a lattice Gibbs measure has been proved only in very few cases. Two examples 
are: the vertex operators of Liouville field theory, whose construction starting from a regularized functional integral has been achieved in a remarkable series of papers 
\cite{KRV19, KRV20, GKRV24}; the spin operator of the 2D Ising field theory, constructed as the limit as the lattice spacing goes to zero of an appropriately renormalized magnetization, 
whose unique limit is shown to be a conformally covariant random distribution \cite{CGN15}. In connection with the latter result, see also \cite{HKV22}, 
where a basis of lattice local fields satisfying the Virasoro algebra is constructed: these objects can be thought of as a sort of lattice counterpart of the scaling operators; however, 
the very existence of these lattice objects crucially relies on the exact solvability of the 2D Ising model at the lattice level.}

{Motivated by these problems, in this paper we consider a theory of fractional symplectic fermions defined in terms of a Grassmann functional integral, 
whose bare potential is the fixed point (FP) potential of a RG transformation in the presence of external source fields, and whose reference Gaussian part has a {\it fixed ultraviolet cutoff}.
The part of the FP potential independent of the external fields is the same as the 
one constructed in \cite{GMR21}, and it consists of a local quartic interaction, plus higher order, `irrelevant', corrections. All the large-distance critical exponents of this model are expected to be the same as the one with purely local quartic interactions, but the FP model has the advantage of being the microscopic model with the closest possible features
to those of the CFT effectively describing it, thus making it possible to 
test directly the CFT predictions.}

{In this setting, we prove two {\bf main results}:}
\begin{enumerate}
\item 
{First, we compute the critical exponents of the field-field and density-density correlations, which, contrary 
to the fixed point potential studied in \cite{GMR21}, are physically observable. We prove that, while the field-field correlation has the same critical exponent as its mean field counterpart, 
the density-density correlation acquires an anomalous (i.e., non-mean-field) exponent, which is analytic in a small disk $|\epsilon|<\epsilon_0$. This should be contrasted with the critical exponents of the fractional $\phi^4_d$ theory \cite{BMS03,Abd06,Sl18}, whose expansion in $\epsilon$ is expected to be asymptotic and conjectured to be Borel summable; note, however, that their regularity properties 
	have not been investigated yet: current results are limited to the lowest order computation of a few non-trivial, anomalous, exponents \cite{Sl18}.\footnote{Other previous results on critical exponents of interacting field theories in the mathematical physics literature include the following. In the context of models with marginally irrelevant interactions, like $\phi_4^4$ theories, the large-distance asymptotic behavior of correlation functions is characterized by logarithmic corrections with specific, fractional, exponents, computed in \cite{BBS14,BSTW17,ST16}; similar logarithmic corrections with fractional exponents are expected to characterize the short-distance asymptotic behavior of fermionic models with marginally relevant interactions, like the Gross-Neveu model in $d=2$: its exponents can be computed via a generalization of \cite{GK85c, FMRS86, DY24}, even though they have not been explicitly worked out. In the context of 2D models with `marginally marginal' interactions, i.e., marginal interactions with asymptotically vanishing beta function  \cite{BMWICMP, BFM07}, the exponents displayed by the large-distance behavior of correlation functions are analytic in the interaction strength in a small disk in the complex plane; universality for these exponents have been proved, as well, in the sense that extended scaling relations are verified \cite{BFM09, GMT20}.}}
\item {Next, we define \emph{almost-scaling operators}, whose correlation functions are scale-invariant up to remainder terms decreasing as a stretched exponential at large distances.
We believe that, in the presence of a short-distance cutoff, there is no way of defining scaling operators whose correlation functions are exactly scale invariant; in this sense, 
almost-scaling operators are the best possible approximation to scaling operators in our setting}\footnote{{An alternative to our choice of keeping the ultraviolet cutoff fixed would be to send it to infinity, after appropriate rescaling of the field or density operators, thus obtaining a set of exactly scale invariant correlation functions, as done in \cite{GKRV24,CGN15}. The corresponding 
scaling operators can in principle be obtained via a reconstruction theorem from the corresponding correlations, as done in \cite{CGN15}, and identified with random Schwarz distributions, which the renormalized operators at finite ultraviolet cutoff converge to in suitable Sobolev norms, as the cutoff is removed. However, this strategy may be difficult, if not impossible, to work out for 
composite operators: e.g, in the 2D Ising setting, there is a strong indication that the energy field cannot be constructed in the limit of removed ultraviolet cutoff
via a similar procedure, see \cite{GK25}. In this respect, the use of a fixed ultraviolet cutoff appears to be a convenient choice for constructing (almost-)scaling operators associated
to arbitrary composite fields.}}. Their behavior is to be contrasted with the behavior of correlation functions of generic (\emph{non-scaling}) local operators which are expected to decay like power laws at large distances, up to remainder terms decaying also like power laws, albeit with a larger power than the leading term. It appears plausible that generic local operators should be expressible as (infinite) linear combinations of almost-scaling operators, each of which contributes to the correlation function with a power law behavior with a different critical exponents. Such an expansion in operators with larger and larger scaling dimension is standard in physics considerations of critical phenomena, see e.g.~\cite[Eq.~(3.11)]{Wegner} or \cite[Eq.~(3.7)]{Patashinsky:1979yx}.
\end{enumerate}
	\bigskip

	{\bf Outline.} The paper is structured as follows: In Section \ref{sec2.0} we define the model and state our main results, summarized in Theorems \ref{Th:1} and \ref{Th:2}. In Section \ref{Sec2} we describe in detail 
	the RG map and derive the corresponding FP equation in the presence of the external source fields. 
	In Section \ref{Sec:3} we solve the FP equation for the potential with external source fields via a tree expansion, and prove its absolute convergence in a weighted $L^1$ norm. In Section \ref{sec5} we prove the convergence of the tree expansion in a mixed $L^1/L^\infty$ norm, required for the control of the correlation functions at fixed positions, and we conclude the proof of Theorem \ref{Th:1}. In Section \ref{sec6} we adapt the discussion of the previous section to the estimate of the remainder terms in the correlation functions of the almost-scaling operators and conclude the proof of Theorem \ref{Th:2}. The lowest order computation of the anomalous critical exponent and of the two point function are collected in two appendices. 
	
	Many aspects of the proof are generalizations of the methods introduced and discussed in detail in \cite{GMR21}: therefore, we will often refer the reader to specific sections of our previous paper and emphasize the analogies and novelties of the present work compared to \cite{GMR21}. 
	
	\section{The model and the main results}\label{sec2.0}
	
	\subsection{Generating function of the scale-invariant response functions}\label{sec2.0a}
	
	Let us recall that in \cite{GMR21} we constructed the non-trivial FP potential $H=H^*$ for a $d$-dimensional Grassmann theory with partition function $\int d\mu_{\le 0}(\psi)e^{H(\psi)}$, 
	where $\psi_a$ is an $N$-component (with $N\ge 4$ even and different from $8$) Grassmann field in $\mathcal V\subset \mathbb R^d$ with $d=1,2,3$ and $\mathcal V$ a cube whose side is eventually sent to $\infty$, and $d\mu_{\le 0}$ (denoted $d\mu_P$ in \cite{GMR21}) is a reference Gaussian integration characterized by the following two-point function, or {\it propagator}\footnote{In finite volume $\mathcal V$ the symbol $\int \frac{d^dk}{(2\pi)^d}$ should be understood a shorthand notation for the corresponding Riemann sum of discrete allowed momenta. As discussed in \cite[Appendix H]{GMR21}, the RG map leading to the definition of $H^*$ can be defined in finite volume and proved to 
		converge strongly (as an operator acting on the Banach space of interactions defined in \cite[Section 4]{GMR21}) to a well-defined infinite volume counterpart. 
		An analogous discussion can be repeated in the generalized setting considered in this paper, where external fields and correlation functions are considered. 
		In the following we will always write all the relevant equations directly in the infinite volume setting and leave 
		the details of the proof of convergence of the finite volume theory to its thermodynamic limit to the reader.}:
	\begin{equation}\begin{split}G_{ab}^{(\le 0)}(x-y)=\int d\mu_{\le 0}(\psi)\psi_a(x)\psi_b(y)&=\Omega_{ab}\int \frac{d^dk}{(2\pi)^d}\frac{\chi(k)}{|k|^{d/2+\epsilon}}e^{ik\cdot(x-y)}\\
			&\equiv \Omega_{ab} P_{\le 0}(x-y).\end{split} \end{equation}
	Here $\Omega$ is the standard $N\times N$ symplectic matrix in block-diagonal form (see \cite[Eq.(1.3)]{GMR21}), and 
	the cutoff function $\chi$ is radial, monotone decreasing in the radial direction, and such that
	\begin{equation}\label{defchi}\chi(k)=\begin{cases}
			1,\qquad \text{if}\ |k|\le 1/2, \\
			0,\qquad \text{if}\ |k|\ge1.
		\end{cases}
	\end{equation}
	As in \cite{GMR21}, we assume that $\chi$ belongs to the Gevrey class $G^s,\;s>1$ \cite[App. A]{GMR21}. The FP potential constructed in \cite{GMR21} reads:
	\begin{equation} H^*(\psi)=\nu^*\int d^dx\, \psi^2(x)+\lambda^*\int d^d x\, \psi^4(x)+H_{\text{IRR}}^*(\psi), \end{equation}
	where $\nu^*,\lambda^*$ are non-zero real-analytic functions of $\epsilon$, of order $\epsilon$, for $\epsilon$ sufficiently small, $\psi^2(x):=\sum_{a,b=1}^N\Omega_{ab}\psi_a(x)\psi_b(x)$, 
	$\psi^4(x):=(\psi^2(x))^2$, and $H^*_{\text{IRR}}$ is an infinite sum of irrelevant terms, 
	whose kernels are non-vanishing real-analytic functions of $\epsilon$ (see \cite{GMR21} for more precise claims). 
	
	\medskip
	
	In this paper, we consider the FP theory in the presence of external source fields and the corresponding generating function for correlations, formally defined as
	\begin{equation}\label{eq:1}
		W^*(\phi,J)=\log\frac{\int d\mu_{\le 0}(\psi)e^{V^*(\phi,J,\psi)}}{\int d\mu_{\le 0}(\psi)e^{H^*(\psi)}}.	
	\end{equation}	
	Here $V^*$ is a solution of the FP equation $V^*=RV^*$, with $R$ the RG map, of the form 
	\begin{equation} \label{V*form}V^*(\phi,J,\psi)=H^*(\psi)+(\phi,\psi)+(J,\psi^2)+
		\mathcal R^*(\phi,J,\psi)+\mathcal S^*(\phi,J),\end{equation}
	where $(\phi,\psi)$ and $(J,\psi^2)$ are shorthand notations for $\sum_{a=1}^N\int d^dx \phi_a(x) \psi_a(x)$ and $\int d^dx J(x) \psi^2(x)$, respectively, $\mathcal R^*(\phi,J,\psi)$ is a sum of irrelevant 
	terms depending explicitly both on $(\phi,J)$ and on $\psi$, i.e., $\mathcal R^*(\phi,J,0)=\mathcal R^*(0,0,\psi)=0$, and $\mathcal S^*(\phi,J)$ is an external potential, depending only upon the external fields $(\phi,J)$, such that $\mathcal S^*(0,0)=0$. 
The precise definition of the RG map and of the irrelevant part of $V^*$ will be given shortly, see Definition \ref{Def.1} below. Note also that, a priori, the right side of \eqref{eq:1} 
may be plagued by infrared divergences. Therefore, \eqref{eq:1} should be interpreted as\footnote{{Both $W^*$ and $W^*_{[h,0]}$ should be interpreted as 
formal infinite sums of monomials in $\phi$ and $J$, each monomial being specified by the order in $\phi$ and $J$ and by the corresponding kernel. In other words, $W^*$ and 
$W^*_{[h,0]}$ must be thought of as the infinite sequences of such kernels, whose physical interpretation is that of multipoint {\it response} functions. 
Moreover, these kernels should be thought of as the infinite volume limits of the 
corresponding finite volume kernels; the existence of the infinite volume limit is implicit in the proofs discussed in this paper, and is a corollary of the methods developed in \cite{GMR21}
and here; see \cite[Appendix H]{GMR21} for details about the non-perturbative definition of the effective potential and the existence of the infinite-volume limit.}\label{footnote:3}}:
	\begin{equation}\label{eq:1bis}\begin{split}
			& W^*(\phi,J):=\lim_{h\to-\infty}W^*_{[h,0]}(\phi,J)\\
			& W^*_{[h,0]}(\phi,J):=\log\frac{\int d\mu_{[h,0]}(\psi)e^{V^*(\phi,J,\psi)}}{\int d\mu_{[h,0]}(\psi)e^{H^*(\psi)}},\end{split}\end{equation}
	where $d\mu_{[h,0]}$ is the Grassmann Gaussian integration with propagator
\begin{equation}\label{eq:Gh0}\begin{split}G^{[h,0]}_{ab}(x-y)&=\Omega_{ab}\int \frac{d^dk}{(2\pi)^d}\frac{\chi(k)-\chi(\gamma^{-h}k)}{|k|^{d/2+\epsilon}}e^{ik\cdot (x-y)}\\
	&\equiv \Omega_{ab}P_{[h,0]}(x-y),\end{split}\end{equation}
and $\gamma>1$ is an arbitrarily chosen rescaling parameter, which is fixed once and for all.
{{More generally, for later reference, for any pair of integers $h_1<h_2$, we let 
	$$G^{[h_1,h_2]}_{ab}(x):=\Omega_{ab}\int \frac{d^dk}{(2\pi)^d}\frac{\chi(\gamma^{-h_2}k)-\chi(\gamma^{-h_1}k)}{|k|^{d/2+\epsilon}}e^{ik\cdot x}\equiv \Omega_{ab}P_{[h_1,h_2]}(x).$$
We also denote 
	\begin{equation}\begin{split}
	G^{(\le h)}_{ab}&:=\lim_{h_1\to-\infty}G^{[h_1,h]}_{ab}(x)\equiv \Omega_{ab}P_{\le h}(x)\\  G^{(\ge h)}_{ab}&:=\lim_{h_2\to+\infty}G^{[h,h_2]}_{ab}(x)\equiv \Omega_{ab}P_{\ge h}(x).
	\end{split}\label{Glehgeh}\end{equation} 
	In particular, if $h_2=h_1+1\equiv h$, we denote the {\it single-scale} propagator by 
	\begin{equation}
	g^{(h)}_{ab}(x):=G^{[h-1,h]}_{ab}(x).
	\label{gh}\end{equation}}}  
{Going back to $W^*$, our main focus will be on the two-point field-field and density-density response functions\footnote{{We distinguish between {\it response} functions, defined in terms of functional derivatives of a generating function, and {\it correlation} functions, defined as expectations of products of local or almost-local operators, see below. The two notions differ by the effect of non-linear terms in the action: e.g., when deriving the generating function twice with respect to an external field, after setting it to zero we get a two-point correlation function plus an additional term due to the terms in the action that are quadratic in the external field.
\label{footnote:responsevscorr}}}, i.e., the kernels of the 
contributions to $W^*(\phi,J)$ that are purely quadratic in $\phi$ or in $J$, respectively. These can be obtained from $W^*$ by differentiating it twice with respect to (w.r.t.) the
$\phi$ or $J$ and then setting the source fields to zero. We let:}
\begin{equation}\label{GF}
	\mathcal{G}^*_{ab}(x):=\frac{\delta^2W^*(\phi,0)}{\delta\phi_a(x)\delta\phi_b(0)}\Big|_{\phi=0}, \qquad 
	\mathcal{F}^*(x):=\frac{\delta^2W^*(0,J)}{\delta J(x)\delta J(0)}\Big|_{J=0},
\end{equation}	
which are the central objects of interest of this paper. These response functions are expected\footnote{In this paper, we won't discuss how to compute the two-point functions 
{in \eqref{ffddcorr}}
and how to prove that they have the same asymptotic behavior as the response functions: this follows from a stability analysis of the RG flow in the vicinity of the fixed point potential, which is deferred to a third paper in this series. 
	In such a third paper, we will also prove the independence of the fixed point potential from the scale parameter $\gamma$ entering the definition of the RG map, see next section, and, most importantly, the 
	universality of the critical exponents with respect to the choice of the ultraviolet cutoff function $\chi$ in \eqref{defchi}.} to have the same large distance asymptotic behavior, up to a finite 
multiplicative renormalization, as the two-point functions 
\begin{equation} \langle \psi_a(x)\psi_b(0)\rangle_{H^*}, \qquad \langle \psi^2(x);\psi^2(0)\rangle_{H^*},\label{ffddcorr}\end{equation}
respectively, where $\langle \cdots \rangle_{H^*}:=\lim_{h\to-\infty}\frac{\int d\mu_{[h,0]}(\psi)e^{H^*(\psi)}\cdots}{\int d\mu_{[h,0]}(\psi)e^{H^*(\psi)}}$ (in the second expression, the semicolon indicates truncated, or connected, expectation, i.e., $\langle A;B\rangle_{H^*}:= 
\langle A\,B\rangle_{H^*}-\langle A\rangle_{H^*}\langle B\rangle_{H^*}$); in particular, they are expected to define the same critical exponents. 
However, contrary to the two-point functions in \eqref{ffddcorr}, the response functions $\mathcal{G}^*(x)$ and $\mathcal{F}^*(x)$ are 
{\it scale-invariant}. {Our first main results can be in fact informally stated as follows (see Theorem \ref{Th:1} below for a more precise statement):}

\medskip

{{\bf Main result, I.} {\it For $\epsilon$ small enough, the response functions $\mathcal{G}^*$ and $\mathcal{F}^*$
are analytic in $\epsilon$ and such that, under rescaling by $\rho=\gamma^k$ (here $\gamma$ is the scaling parameter in \eqref{eq:Gh0} and following equations --
entering also the definition of RG map -- and $k\in\mathbb Z$ is an integer), they behave as follows: 
		$$\mathcal{G}_{a,b}^*(\rho x)=\rho^{-d/2+\epsilon}\mathcal{G}^*_{a,b}(x),\qquad\mathcal{F}^*(\rho x)=\rho^{-d+2\epsilon-2\eta_2(\epsilon)}\mathcal{F}^*(x),$$	
		where the {\bf anomalous critical exponent} $\eta_2(\epsilon)= 2\epsilon\frac{N-2}{N-8}+O(\epsilon^2)$ is analytic in $\epsilon$.}}

\medskip

{Scale invariance} may look surprising, at first sight, due to the presence of a fixed ultraviolet cutoff in our theory; but it turns out that the very definition of fixed point potential
$V^*$ fixes the irrelevant terms $\mathcal R^*$ 
and the external potential $\mathcal S^*$ in a special way, so that the resulting response functions are, in fact, scale invariant. Even in the non-interacting theory 
{{corresponding to the Gaussian fixed point}}, 
the mechanism fixing these additional terms is not completely trivial: it is in fact instructive, as a preliminary exercise, to compute $V^*$ and $\mathcal G^*,\mathcal F^*$ in the {{non-interacting case}}, see Section \ref{sec:freecase}. {As explained in footnote \ref{footnote:responsevscorr}, 
we distinguish between the notions of response functions, like $\mathcal F^*, \mathcal G^*$, defined in terms of functional derivatives of the generating function, and 
correlation functions, defined as the expectations of products of almost-local operators\footnote{{A local operator at $x$ is a (possibly infinite) linear combination of products 
of fields $\psi(z)$ supported in a finite ball centered at $x$. An almost-local operator may depend upon fields $\psi(z)$ at arbitrary large distance from $x$: however, 
the dependence upon $\psi(z)$ decays stretched-exponentially to zero as $|z-x|\to\infty$.}}.
By its very definition, the response function $\mathcal F^*$ (resp. $\mathcal G^*$) is 
equal to the two-point correlations of an almost-local operators $\mathcal O^{(1)}$ (resp. $\mathcal O^{(2)}$), plus an additional contribution, equal to the expectations of the second derivative of $\mathcal R^*(\phi,J,\psi)+\mathcal S^*(\phi,J)$ with respect to $\phi$ (resp. $J$) at $\phi=J=0$. Using a Quantum Field Theory jargon, this additional term may be called a `Schwinger term'. 
Our second main result is that the Schwinger terms associated with the response functions are stretched-exponentially small at large distances: in other words, 
the response functions $\mathcal F^*$ and $\mathcal G^*$ are equal to the two-point functions of two almost-local operators, up to stretched exponentially small corrections. 
In view of the scale invariance of the response functions, the correlation functions of these almost-local operators are then scale invariant up to stretched exponentially small corrections, 
thus allowing us to interpret them as `almost-scaling operators'. In formulae, anticipating here the statement of Theorem \ref{Th:2}, we have:}

\medskip

{{\bf Main result, II.} {\it Let $\mathcal O^{(1)}_{{a}}(x)=\frac{\delta V^*(\phi,J,\psi)}{\delta\phi_{{a}}(x)}\big|_{\phi=J=0}$, and $\mathcal O^{(2)}(x)=\frac{\delta V^*(\phi,J,\psi)}{\delta J(x)}\big|_{\phi=J=0}$. Then $\mathcal{G}^*$ and $\mathcal{F}^*$ can be written as
\begin{equation}\begin{split}&\mathcal G_{{ab}}^*(x)=\langle \mathcal O^{(1)}_{{a}}(x)\mathcal O^{(1)}_{{b}}(0)\rangle_{H^*}+{{\big[}}\mathcal E_1(x){{\big]_{ab}}},\\
&\mathcal F^*(x)=\langle \mathcal O^{(2)}(x);\mathcal O^{(2)}(0)\rangle_{H^*}+\mathcal E_2(x),\end{split}\end{equation}
where the correction terms $\mathcal E_1(x),\mathcal E_2(x)$ decay to zero at large distances like a stretched exponential, i.e., for $|x|\ge 1$ they are bounded form above by 
(const.)$e^{-c|x|^\sigma}$, for suitable $c>0$ and $0<\sigma<1$.}}

\medskip

{In the incoming subsections, before stating our main results more formally in Section \ref{sect:2.7}, we need to introduce and define a number of important concepts, such as the action of 
the RG map (see Section \ref{sec:RG}), and 
the notions of scaling dimension, relevant and irrelevant operators (see Section \ref{sec:2.3sc}); moreover, we need to specify the way in which we intend to parameterize the 
generating function or the fixed point potential: as mentioned in footnote \ref{footnote:3} potentials can be identified with infinite sequences of kernels; more specifically, we shall 
choose the sequence associated with the fixed point potential to belong to a space of sequences, called {\it trimmed}, such that the relevant terms are local (see Section \ref{sec:2.4});
a formal definition of fixed point potential is given in Section \ref{sect:2.5}, while the notion of almost-scaling operators is given in Section \ref{sect:2.6}.}

\subsection{The RG map}\label{sec:RG}

In order to clarify the notion of FP potential $V^*$ in \eqref{eq:1bis}, we need, first of all, to specify the action of the RG map $R$. Before doing this formally, let us remind the reader the logic behind the 
definition of the $R$, which maps a potential $V$ to a new potential $V'$ as follows (this discussion parallels the one in Section 2.1 of \cite{GMR21}): decomposing 
\begin{equation}
	G^{(\le 0)}(x)=G^{(\le-1)}(x)+g^{(0)}(x),
\end{equation}
see \eqref{Glehgeh} and \eqref{gh} for the definitions of $G^{(\le -1)}$ and $g^{(0)}$, we write (`addition principle for Gaussian integrals'):
\begin{equation}
	\frac1{\mathcal N_{\le 0}}\int d\mu_{\le 0}(\psi) e^{V(\phi, J,\psi)}F(\psi)=\frac1{\mathcal N_{\le 0}}\int d\mu_{\le -1}(\psi_\gamma)\int d\mu_0(\psi') e^{V(\phi, J,\psi_\gamma+\psi')}F(\psi_\gamma+\psi'),\end{equation}
where $\mathcal N_{\le 0}=\int d\mu_{\le 0}(\psi) e^{V(0,0,\psi)}$ is a normalization constant, and $d\mu_{\le -1}$ (resp. $\mu_0$) is the Grassmann Gaussian integration with propagator $G^{(\le -1)}$ (resp. 
$g^{(0)}$). 
If $F(\psi)$ is a `large scale observable', i.e., if it does not depend on the Fourier modes $\hat\psi(k)$ with $k$ in the support of $\hat g^{(0)}(k)$, then 
\begin{equation}\label{eq:below}\begin{split}
		\frac1{\mathcal N_{\le 0}}\int d\mu_{\le 0}(\psi) e^{V(\phi, J,\psi)}F(\psi)&=\frac1{\mathcal N_{\le 0}}\int d\mu_{\le -1}(\psi_\gamma)F(\psi_\gamma)\int d\mu_0(\psi')e^{V(\phi, J,\psi_\gamma+\psi')}\\
		&= \frac{\mathcal N_0}{\mathcal N_{\le 0}} \int d\mu_{\le -1}(\psi_\gamma)e^{V_{\text{eff}}(\phi,J,\psi_\gamma)} F(\psi_\gamma),\end{split}\end{equation}
where $\mathcal N_0:=\int d\mu_{0}(\psi) e^{V(0,0,\psi)}$	and $V_{\text{eff}}(\phi,J,\psi_\gamma)=\log(\mathcal N_0^{-1}\int d\mu_0(\psi')e^{V(\phi,J,\psi_\gamma+\psi')})$. 		
Thanks to the scaling property $G^{(\le -1)}(x)=\gamma^{-2[\psi]}G^{(\le 0)}(x/\gamma)$, {{with $[\psi]=\frac{d}4-\frac{\epsilon}2$ (see \cite[eqs.(2.14),(2.15)]{GMR21}),}
	the random field $\psi_\gamma$ has the same distribution as $\gamma^{-[\psi]}\psi(\cdot/\gamma)\equiv D\psi$, with $D$ the dilatation operator, and $\psi$ a random field 
	with distribution $d\mu_{\le 0}$, so that 
	\[\eqref{eq:below} =\frac1{\mathcal N_{\le 0}'}\int d\mu_{\le 0}(\psi)e^{V_{\text{eff}}(\phi,J,D\psi)}F(D\psi),\] where $\mathcal N'_{\le 0}:=\int d\mu_{\le 0}(\psi)e^{V_{\text{eff}}(0,0,D\psi)}\equiv\mathcal N_{\le 0}/\mathcal N_0$. Now, if $V$ has the same structure as \eqref{V*form}, i.e., 
	\begin{equation} \label{Vform}V(\phi,J,\psi)=H^*(\psi)+(\phi,\psi)+(J,\psi^2)+
		\mathcal R(\phi,J,\psi)+\mathcal S(\phi,J),\end{equation}
	with $\mathcal R$ and $\mathcal S$ two potentials in an $O(\epsilon)$ neighborhood of the non-interacting fixed point discussed in Section \ref{sec:freecase} below (with $\mathcal R$ a sum of irrelevant terms, in the sense discussed in Section \ref{sec:2.3sc} below, and $\mathcal S$ an external potential, depending only upon the external fields $\phi,J$), then it turns out, see Section \ref{Sec2}, that
	\begin{equation} \label{Veffform}V_{\text{eff}}(\phi,J,D\psi)=H^*(\psi)+\tilde Z_1(\phi,D\psi)+\tilde Z_2(J,(D\psi)^2)+
		\tilde{\mathcal R}(\phi,J,D\psi)+\tilde{\mathcal S}(\phi,J),\end{equation}
	with $\tilde Z_1=1+O(\epsilon)$, $\tilde Z_2=1+O(\epsilon)$ two constants produced by the integration of the fluctuation field $\psi'$, and $\tilde{\mathcal R}$, $\tilde{\mathcal S}$ two new potentials that are, again, 
	$O(\epsilon)$-close to the non-interacting fixed point, with $\tilde{\mathcal R}$ a sum of irrelevant terms. In view of \eqref{Veffform}, it is natural to rescale the external fields in a way similar to $\psi$, in such a way to recast the terms involving the external 
	fields in a form as close as possible to the one in which they originally appeared in $V$: therefore, we let $\phi\equiv D\phi'$ and $J\equiv DJ'$, with $D$ the dilatation operator, acting 
	on the external fields as: $D\phi(x)=\gamma^{-[\phi]}\phi(x/\gamma)$ and $DJ(x)=\gamma^{-[J]}J(x/\gamma)$, with the scaling dimensions $[\phi], [J]$ to be fixed appropriately, in a way 
	explained shortly. In terms of these definitions: $(\phi,D\psi)=(D\phi',D\psi)=\sum_{a=1}^N\gamma^{-[\phi]-[\psi]}\int d^dx \phi_a'(x/\gamma) \psi_a(x/\gamma)=\gamma^{d-[\phi]-[\psi]}(\phi',\psi)$, and similarly for $(J,(D\psi)^2)$. Therefore, 
	\begin{equation} \label{equindi}\begin{split}
			&V_{\text{eff}}(\phi,J,D\psi)=V_{\text{eff}}(D\phi',DJ',D\psi)\\
			&\quad =H^*(\psi)+\tilde Z_1\gamma^{d-[\phi]-[\psi]}(\phi',\psi)+
			\tilde Z_2\gamma^{d-[J]-2[\psi]}(J',\psi^2)+\tilde{\mathcal R}(D\phi',DJ',D\psi)+\tilde{\mathcal S}(D\phi',DJ')\\
			&\quad \equiv V'(\phi',J',\psi).\end{split}
	\end{equation}
	By definition, $V'$ is the image of $V$ under the RG map, i.e., $V'=RV$; more precisely, $V'$ is {\it equivalent}, in the sense of \cite[Section 5.2.1]{GMR21}, to $RV$: the missing ingredient, in the previous informal discussion, is the action of the trimming map $T$, which is an operator, equivalent to the identity, which isolates the relevant terms from the irrelevant ones, see Section \ref{Trimming} below.
	
	Note that the RG map depends, among other things, upon the choice of the scaling dimensions $[\phi]$ and $[J]$. In order to fix these dimensions, we use the requirement that, at the fixed point, 
	$Z_1\gamma^{d-[\phi]-[\psi]}=Z_2\gamma^{d-[J]-2[\psi]}=1$, with $Z_1=\tilde Z_1^*$ and $Z_2=\tilde Z_2^*$ the constants as in \eqref{Veffform}-\eqref{equindi} associated with the 
	fixed point potential $V^*$. In this way, the fixed point equation (FPE) defines an unambiguous criterium for assigning a scaling dimension to the external fields, which has a simple relation with the 
	critical exponents of the corresponding response functions (cf. \eqref{DphiJ} with \eqref{1.12}); once that $[\phi]$ and $[J]$ are fixed, the FPE becomes an equation for $\mathcal R^*, \mathcal S^*$, of the form: 
	$\tilde{\mathcal R}^*(D\phi',DJ',D\psi)=\mathcal R^*(\phi',J',\psi)$, $\tilde{\mathcal S}^*(D\phi',DJ')=\mathcal S^*(\phi',J')$.
	The uniqueness of the solution for the FPE for $[\phi], [J], \mathcal R^*, \mathcal S^*$, proved below, see Theorem \ref{Th:1}, and the fact that the critical exponent of the two-point 
	correlations \eqref{ffddcorr} are the same as those of the response functions $\mathcal G^*, \mathcal F^*$ associated with $V^*$, proved in the next paper of this series, 
	confirms the physical significance of the scaling dimensions $[\phi]$ and $[J]$, and of the fixed point potential $V^*$.
	
	\medskip
	
	Given these premises, let us define the RG map $R$ a bit more formally as follows: 
	\[R:=D^{-(h-1)}T{{S^{(h)}}}D^h\equiv DT{{S^{(0)}}}\]
	(independence from $h\in\mathbb Z$ is a consequence of the definitions of $S^{(h)},T,D$ below), where:
	\begin{itemize}
		\item {{$S^{(h)}$}} is the integration on scale $h\le0$, i.e., 
		\begin{equation}\label{eq:1.3} {{S^{(h)}}} V(\phi,J,\psi)=\log \frac{\int d\mu_{h}(\psi')e^{V(\phi,J,\psi'+\psi)}}{\int d\mu_{h}(\psi')e^{V(0,0,\psi')}},\end{equation}
		where $d\mu_h$ is the Grassmann Gaussian integration with  propagator $g^{(h)}_{ab}(x)$, see \eqref{gh};
		\item $T$ is the trimming map, defined in detail in Section \ref{Trimming} below, which is \textit{equivalent to the identity}: when acting on a potential $V$ it produces an equivalent potential $V'\sim V$, in the sense of \cite[Sect.5.2.1]{GMR21}, of the form $V'=TV=\mathcal LV+\mathcal IV$; here $\mathcal LV$ is a sum of finitely many {\it local} relevant and marginal terms (belonging to a finite dimensional subspace of the Banach space of potentials), and $\mathcal I V$ is an equivalent rewriting of $TV-\mathcal LV$ as an infinite linear combination of non-local, {\it irrelevant}, monomials in $\phi, J, \psi,\partial\psi$;
		\item $D$ is the dilatation, acting on $V(\phi,J,\psi)$ as $DV(\phi,J,\psi):=V(D\phi,DJ,D\psi)$, with $D\psi(x)=\gamma^{-[\psi]}\psi(x/\gamma)$ (recall: $[\psi]=d/4-\epsilon/2$, see \cite[eqs.(2.14),(2.15)]{GMR21}), while $D\phi(x)=\gamma^{-[\phi]}\phi(x/\gamma)$, $DJ(x)=\gamma^{-[J]}J(x/\gamma)$, where 
		$[\phi]=d-[\psi]+ O(\epsilon)$ and $[J]=d-2[\psi]+O(\epsilon)$ are two analytic functions of $\epsilon$, for $\epsilon$ sufficiently small, to be determined in such a way that $V^*$, of the form described in Definition \ref{Def.1} below, is a solution of the FP equation $V^*=RV^*$. A simple argument, spelled out in Section \ref{sec:structure} below\footnote{See in particular \eqref{eq:2.44}. Note that the discussion in Section \ref{sec:structure} is written in terms of $\Delta_1=d-[\phi]$ and $\Delta_2=d-[J]$ rather than 
			in terms of $[\phi], [J]$, but this is just a matter of notation.}  
		shows that $2(d-[\phi])$ and $2(d-[J])$ are the critical exponents of the response functions $\mathcal G^*$ and $\mathcal F^*$, respectively: therefore, denoting by $2\Delta_1$ and $2\Delta_2$ these critical exponents, we rewrite the action of the dilatation operator on the external fields in terms of $\Delta_1:=d-[\phi]$, $\Delta_2:=d-[J]$:
		\begin{equation}\label{DphiJ}
			D\phi(x)=\gamma^{-d+\Delta_1}\phi(x/\gamma),\qquad DJ(x)=\gamma^{-d+\Delta_2}J(x/\gamma);	
		\end{equation}	
		we will see that $\Delta_1=[\psi]$ and $\Delta_2=2[\psi]+\eta_2(\epsilon)$, with $\eta_2(\epsilon)=2\epsilon \frac{N-2}{N-8}+O(\epsilon^2)$.
	\end{itemize}	
	
	\subsubsection{The fixed point potential in the non-interacting case}
	\label{sec:freecase}
	
	Before introducing the notion of relevant and irrelevant operators, and giving a more formal definition of fixed point potential $V^*$, 
	it is instructive to discuss the `Gaussian' case corresponding, {in the absence of external fields, to the trivial fixed point $H^*(\psi)=0$}. 
	Even in this case, the fixed point potential {in the presence of the external fields, to be  denoted by $V_0^*$}, is not trivial, and its {structure} is crucial to guarantee that the response functions $\mathcal G^*$ and $\mathcal F^*$ are scale-invariant, notwithstanding the presence of an ultraviolet cutoff. 
	For {$H^*(\psi)=0$, we let $\Delta_1=[\psi]$ and  $\Delta_2=2[\psi]$ in \eqref{DphiJ}, so that} the action of the dilatation operator reduces to: 
	\begin{equation}\label{resc0}D\psi(x)=\gamma^{-\frac{d}4{+\frac{\epsilon}2}}\psi(x/\gamma), \quad D\phi(x)=\gamma^{-\frac34d{-\frac{\epsilon}2}}\phi(x/\gamma), \quad DJ(x)=\gamma^{-\frac{d}2{-\epsilon}}J(x/\gamma).\end{equation}
	The non-interacting fixed point potential $V_0^*$ we consider is the one obtained as the image of the `naive' potential $V_0(\phi,J,\psi):=(\phi,\psi)+(J,\psi^2)$ under the iterated action of the RG map $D{{S^{(0)}}}$, in the limit as 
	the number of iterations goes to infinity: 
	\begin{equation} \label{freeFPP}V_0^*(\phi,J,\psi)=\lim_{n\to\infty}(DS^{(0)})^nV_0(\phi,J,\psi).\end{equation}
	The action of ${{S^{(0)}}}$ on a potential $V$ returns (see \cite[Section 5.1]{GMR21}): 
	$${{S^{(0)}}} V(\phi,J,\psi)=\sum_{s\ge 1}\frac1{s!}\langle \underbrace{V(\phi,J,\psi+\cdot); \cdots ; V(\phi,J,\psi+\cdot)}_{s \text{ times}}\rangle_{{0}}^{{(0)}},$$
	where $\langle A_1; \cdots ; A_s\rangle_{{0}}^{{(0)}}$ denotes connected expectation of order $s$ with respect to (w.r.t.) the Gaussian integration $d\mu_0$ with propagator $g^{(0)}$ {(the lower label `$0$' indicates that the expectation is computed w.r.t. a Gaussian, non-interacting, measure, while the upper label `$(0)$' refers to the scale of the propagator $g^{(0)}$)}.
	{Using the fact that the naive potential $V_0(\phi,J,\psi)$ is invariant under the action of $D$, one finds that \eqref{freeFPP} can be re-expressed as 
		\begin{equation} \label{freeFPPbis}V_0^*(\phi,J,\psi)={{S^{(\ge 1)}}}V_0(\phi,J,\psi), \end{equation}
		where 
		\begin{equation} \label{freeinttris}{{S^{(\ge  1)}}} V(\phi,J,\psi)=\sum_{s\ge 1}\frac1{s!}\langle \underbrace{V(\phi,J,\psi+\cdot); \cdots ; V(\phi,J,\psi+\cdot)}_{s \text{ times}}\rangle_{0}^{(\ge 1)},\end{equation}
		and $\langle A_1; \cdots ; A_s\rangle_{0}^{(\ge 1)}$ denotes connected expectation of order $s$ w.r.t. the Gaussian integration $d\mu_{\ge 1}$ with propagator $G^{(\ge 1)}$, {see \eqref{Glehgeh}}. As well known (see also \cite[Appendix D.2]{GMR21}),
	connected expectations w.r.t. {a Gaussian Grassmann measure} can be expressed in terms of connected Feynman diagrams {associated} with the {corresponding} propagator. Using this fact, 
	{\eqref{freeinttris} implies} that the fixed point potential in \eqref{freeFPP}
	is $V_0^*(\phi,J,\psi)=(\phi,\psi)+(J,\psi^2)+\mathcal R^*_0(\phi,J,\psi)+\mathcal S^*_0(\phi,J)$, where, if the two terms $(\phi,\psi)$ and $(J,\psi^2)$ in $V_0$ are graphically represented as in Figure \ref{fig.primidue}, 
	\begin{figure}[ht]
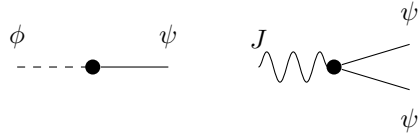

		\begin{center}
			\begin{tabular}{rcl}
				\tikz[baseline=-2pt]{\draw[dashed](0,0) node [label=above:{$\phi$}] {}--(1,0) node [vertex] (t0) {};
					\draw (t0) -- (2,0) node [label=above:{$\psi$}] {};
				}
				$\;\;\;\;\;$
				\tikz[baseline=-2pt]{\draw [decorate,decoration={snake,amplitude=2mm}](0,0) node [label=above:{$J$}]  {} -- (1,0) node[vertex] (v0) {};
					\draw (v0) node [] {} -- +(1,+0.3) node [label=above:{$\psi$}]  (v1) {};
					\draw (v0) node [] {} -- +(1,-0.3) node [label=below:{$\psi$}]  (v2) {};}
			\end{tabular}
		\end{center}\caption{{The graphical representations of the two vertices $(\phi,\psi)$ and $(J,\psi^2)$, respectively.}}
		\label{fig.primidue}
	\end{figure}
	then $\mathcal R^*_0$ and $\mathcal S^*_0$ are graphically represented as in  Figure \ref{eq:fig27}, with 
	the solid lines representing the propagator $G^{(\ge 1)}$. 
	\begin{figure}[ht]
		\begin{center}
			\begin{tabular}{rcl}
				&&$\mathcal{R}^*_{0}(\phi,J,\psi)=$\\
				&&\;\;\;$\sum_{n\ge 1}$
				\tikz[baseline=-2pt]{\draw[dashed](0,0) {}--(1.2,0) node [vertex, label=below:{$y_0$}] (t0) {};
					\draw (t0) -- (2.4,0) node [vertex,label=below:{$y_1$}] (t1){};
					\draw(1.8,0)node[label=above:{$\ge1$}]{};
					\draw[decorate,decoration={snake,amplitude=2mm}](2.4,1)--(2.4,0);
					\draw(t1)--(3.6,0)node [vertex,label=below:{$y_2$}] (t2){};
					\draw(3,0)node[label=above:{$\ge1$}]{};
					\draw[decorate,decoration={snake,amplitude=2mm}](3.6,1)--(3.6,0);
					\draw(t2)--(4.2,0);
					\draw[dotted](4.2,0)--(4.8,0);
					\draw(4.8,0)--(5.4,0)node[vertex,label=below:{$y_n$}](t3){};
					\draw[decorate,decoration={snake,amplitude=2mm}](5.4,1)--(5.4,0);
					\draw[dotted](3.9,-.3)--(5.1,-.3);
					\draw (t3)--(6.6,0);
				}\;$+\sum_{n\ge 2}$
				\tikz[baseline=-2pt]{\draw(0,0) {}--(1.1,0) node [vertex, label=below:{$y_1$}] (t0) {};
					\draw[decorate,decoration={snake,amplitude=2mm}](1.1,1)--(1.1,0);
					\draw (t0) -- (2.4,0) node [vertex,label=below:{$y_2$}] (t1){};
					\draw(1.7,0)node[label=above:{$\ge1$}]{};
					\draw[decorate,decoration={snake,amplitude=2mm}](2.4,1)--(2.4,0);
					\draw(t1)--(3,0);
					\draw[dotted](3,0)--(3.5,0);
					\draw(3.5,0)--(4.3,0)node[vertex,label=below:{$y_n$}](t3){};
					\draw[decorate,decoration={snake,amplitude=2mm}](4.3,1)--(4.3,0);
					\draw[dotted](2.7,-.3)--(3.9,-.3);
					\draw (t3)--(5.4,0);
				},
				\\ \\ \\
				&&$\mathcal{S}^*_{0}(\phi,J)=$\\
				&&$\sum_{n\ge0}$
				\tikz[baseline=-2pt]{\draw[dashed](0,0) {}--(1.2,0) node [vertex, label=below:{$y_0$}] (t0) {};
					\draw (t0) -- (2.4,0) node [vertex,label=below:{$y_1$}] (t1){};
					\draw(1.8,0)node[label=above:{$\ge1$}]{};
					\draw[decorate,decoration={snake,amplitude=2mm}](2.4,1)--(2.4,0);
					\draw(t1)--(3,0);
					\draw[dotted](3,0)--(3.6,0);
					\draw(3.6,0)--(4.2,0)node[vertex,label=below:{$y_{n}$}](t3){};
					\draw[decorate,decoration={snake,amplitude=2mm}](4.2,1)--(4.2,0);
					\draw[dotted](2.7,-.3)--(3.9,-.3);
					\draw(t3)--(5.4,0)node [vertex,label=below:{$y_{n+1}$}] (t2){};
					\draw(4.8,0)node[label=above:{$\ge1$}]{};
					\draw [dashed](t2)--(6.6,0);}\;\;\;$+\sum_{n\ge 1}$
				\tikz[baseline=-2pt]{%\draw(0,0)circle(1);
					\draw[dotted](0,0)circle(1);
					\draw (0,1)node[vertex,label=below:{$y_0$}](t1){};
					\draw (.93,.3)node[vertex,label=left:{$y_1$}](t2){};
					\draw (.6,-.8)node[vertex](t3){};
					\draw (.35,-.85)node[label={$y_2$}]{};
					\draw (-.93,.3)node[vertex,label=right:{$y_n$}](t4){};
					\draw (-.45,-.85)node[vertex](t5){};
					\draw (-.35,-.85)node[label={$y_3$}]{};
					\draw[decorate,decoration={snake,amplitude=2mm}](t1)--(0,2);
					\draw[decorate,decoration={snake,amplitude=2mm}](t2)--(2,0.8);
					\draw[decorate,decoration={snake,amplitude=2mm}](t3)--(1.1,-1.8);
					\draw[decorate,decoration={snake,amplitude=2mm}](t5)--(-1.1,-1.8);
					\draw[decorate,decoration={snake,amplitude=2mm}](t4)--(-2,0.8);
					\draw (-.45,-.52)node[](t55){};
					\draw (-.70,.22)node[](t44){};
					\draw[dotted](t55) to[out=-220,in=-95] (t44);
					\draw(.9,.6)node[label={$\ge1$}]{}; 
					\draw(-.9,.6)node[label={$\ge1$}]{}; 
					\draw(1.3,-.75)node[label={$\ge1$}]{}; 
					\draw(0.1,-1.6)node[label={$\ge1$}]{}; 
					\draw(t4) to[out=5112.5,in=-5212.5] (t1);
					\draw(t1) to[out=-2166.2,in=110] (t2);
					\draw(t2) to[out=-2233,in=44] (t3);
					\draw(t3) to[out=-9150.5,in=-9032.5] (t5);
					\draw (-.98,-.10)node[](t6){};
					\draw(t4) to[out=5110.5,in=-4219] (t6);
					\draw(-.85,-.55)node[](t7){};
					\draw(t5) to[out=4843,in=-4005] (t7);
				}
			\end{tabular}
		\end{center}
		\caption{{The graphical representations of $\mathcal R_0^*$ and $\mathcal S_0^*$, respectively.}}\label{eq:fig27}
	\end{figure}
	In formulae, recalling the definition of $P_{\ge 1}$ in \eqref{Glehgeh}:
	\begin{equation} \begin{split}\label{Free1}
			& \mathcal R^*_0(\phi,J,\psi)=\sum_{m\ge 1} (-2)^m\sum_{a=1}^N\int d^d y_0\cdots d^d y_m \phi_a(y_0)P_{\ge 1}(y_0-y_1)J(y_1)P_{\ge 1}(y_1-y_2)\cdots J(y_m)\psi_a(y_m)\\
			& \quad +\sum_{m\ge 2}(-2)^{m-1}\sum_{a,b=1}^N\Omega_{ab}\int d^d y_1\cdots d^d y_m \psi_a(y_1)J(y_1)P_{\ge 1}(y_1-y_2)\cdots J(y_m)\psi_b(y_m)\\
			&\mathcal S^*_0(\phi,J)=\sum_{m\ge 0} (-2)^{m-1}\sum_{a,b=1}^N\Omega_{ab}\int d^d y_0\cdots d^dy_{m+1} \phi_a(y_0)
			P_{\ge 1}(y_0-y_1)\cdots J(y_m)P_{\ge 1}(y_m-y_{m+1})\phi_b(y_{m+1})\\
			&\quad+\sum_{m\ge 1}(-2)^{m-1}\frac{N}{m}\int d^dy_1\cdots d^dy_m J(y_1)P_{\ge 1}(y_1-y_2)\cdots J(y_m)P_{\ge 1}(y_m-y_1)\end{split}\end{equation}
	Therefore, the non-interacting fixed point generating function is: 
	\begin{equation} \label{Free2} W^*_0(\phi,J)=\mathcal S^*_0(\phi,J)+\log\int d\mu_{\le 0}(\psi)e^{(\phi,\psi)+(J,\psi^2)+\mathcal R^*_0(\phi,J,\psi)},\end{equation}
	from which a straightforward computation shows that 
	\begin{equation}\begin{split}\label{g0starf0star}  {\mathcal{G}^*_{0;ab}(x):=}&\frac{\delta^2W^*_0(\phi,0)}{\delta\phi_a(x)\delta\phi_b(0)}\Big|_{\phi=0}=\Omega_{ab}\big[P_{\le 0}(x)+P_{\ge 1}(x)\big], \\
			{\mathcal{F}^*_{0}(x):=}&\frac{\delta^2W^*_0(0,J)}{\delta J(x)\delta J(0)}\Big|_{J=0}=-{2}N\big[P_{\le 0}^2(x)+2P_{\le 0}(x)P_{\ge 1}(x)+P_{\ge 1}^2(x)\big],\end{split}\end{equation}
	which are scale invariant, as anticipated above. 
	
	\subsection{Scaling dimension, relevant and irrelevant operators}\label{sec:2.3sc}
	
	Let us consider a monomial in $\phi,J,\psi$ of the form
	\begin{equation}\label{tot}\begin{split} & M(\phi,J,\psi)=\\ 
			&\!\!\!
			=\!\!\iiint\! d\boldsymbol{x}\, d\boldsymbol{y} \, d \boldsymbol{z}\, \phi_{a_1}(x_1)\cdots \phi_{a_n}(x_n)J(y_1)\cdots J(y_m)\partial^{p_1}_{\mu_1}\psi_{b_1}(z_1)\cdots \partial^{p_l}_{\mu_l}\psi_{b_n}(z_l) \big[H_{n,m,l,\boldsymbol{p}}(\boldsymbol{x},\boldsymbol{y},\boldsymbol{z})\big]_{\boldsymbol{a},\boldsymbol{\mu},\boldsymbol{b}}
	\end{split}	\end{equation}
	with $\boldsymbol{x}=(x_1,\ldots,x_n)$, etc., $p_i\in\{0,1\}$, $\mu_i\in\{0,1,\ldots,d\}$\footnote{If $p_i=1$, then $\mu_i\in\{1,\ldots,d\}$. If $p_i=0$, the index $\mu_i$ needs not to be specified, but, conventionally,
		in that case we set $\mu_i=0$. If $\boldsymbol{p}=\boldsymbol{0}$ we shall often drop the label $\boldsymbol{\mu}$.},
	$n+l$ even, $H_{n,m,l,\boldsymbol{p}}$ translationally invariant and of finite $L^1$ norm, i.e., 
	\begin{equation} \label{finiteL1}\|H_{n,m,l,\boldsymbol{p}}\|:=\max_{\boldsymbol{a}, \boldsymbol{\mu},\boldsymbol{b}}
		\iiint^* d\boldsymbol{x}\, d\boldsymbol{y} \, d \boldsymbol{z}\, \big|\big[H_{n,m,l,\boldsymbol{p}}(\boldsymbol{x},\boldsymbol{y},\boldsymbol{z})\big]_{\boldsymbol{a},\boldsymbol{\mu},\boldsymbol{b}}\big|<+\infty\end{equation} (here the $*$ on the integral indicates that we are \textit{not} integrating over one of the elements of $(\boldsymbol{x},\boldsymbol{y},\boldsymbol{z})$, say $x_1$). 
		
\begin{Remark}\label{remark:notation}		{{In order to make contact with the notation used in \cite[Section 4.1]{GMR21} and for later use, given the $n$-ples $\boldsymbol{a},\boldsymbol{x}$, the $m$-ple $\boldsymbol{y}$ and the $l$-ples $\boldsymbol{p},  \boldsymbol{\mu}, \boldsymbol{z}$, we shall also denote 
			$\big[H_{n,m,l,\boldsymbol{p}}(\boldsymbol{x},\boldsymbol{y},\boldsymbol{z})\big]_{\boldsymbol{a},\boldsymbol{\mu},\boldsymbol{b}}$ by $H(\boldsymbol{a},\boldsymbol{x},\boldsymbol{y},\mathbf{B},\boldsymbol{z})$, with $\mathbf{B}=((p_1,\mu_1,b_1),\ldots,(p_l,\mu_l,b_l))$. Moreover, we let
			\begin{equation}\begin{split}
					 \Phi(\boldsymbol{a},\boldsymbol{x})&:= \phi_{a_1}(x_1)\cdots \phi_{a_n}(x_n),\\ J(\boldsymbol{y})&:=J(y_1)\cdots J(y_m)\\
					  \Psi(\mathbf{B},\boldsymbol{z})&:= \partial^{p_1}_{\mu_1}\psi_{b_1}(z_1)\cdots \partial^{p_l}_{\mu_l}\psi_{b_n}(z_l).
			\end{split}\label{GMRnotation}\end{equation} 
			Without loss of generality, we can assume, and will do so from now on, that $H(\boldsymbol{a},\boldsymbol{x},\boldsymbol{y},\mathbf{B},\boldsymbol{z})$ is: anti-symmetric under simultaneous permutations of the elements of $\boldsymbol{a},\boldsymbol{x}$, symmetric under permutations of the elements of $\boldsymbol{y}$;
			anti-symmetric under simultaneous permutations of the elements of $\boldsymbol{B},\boldsymbol{z}$.}}\end{Remark}
		
	The action of $D$ on $M(\phi,J,\psi)$ is defined as $DM(\phi,J,\psi):=M(D\phi,DJ,D\psi)$. After a change of variables, this can be rewritten as: 
	\begin{equation}\label{totbis}\begin{split} &DM(\phi,J,\psi)=\\ 
			&\!\!\!
			=\!\!\iiint\! d\boldsymbol{x}\, d\boldsymbol{y} \, d \boldsymbol{z}\, \phi_{a_1}(x_1)\cdots \phi_{a_n}(x_n)J(y_1)\cdots J(y_m)\partial^{p_1}_{\mu_1}\psi_{b_1}(z_1)\cdots \partial^{p_l}_{\mu_l}\psi_{b_n}(z_l) \big[D H_{n,m,l,\boldsymbol{p}}(\boldsymbol{x},\boldsymbol{y},\boldsymbol{z})\big]_{\boldsymbol{a},\boldsymbol{\mu},\boldsymbol{b}}
	\end{split}	\end{equation}
	where
	\begin{equation} \label{eq:scal}
		D H_{n,m,l,{\boldsymbol{p}}}(\boldsymbol{x},\boldsymbol{y},\boldsymbol{z}):= \gamma^{\delta_{sc}(n,m,l,\boldsymbol{p})} H_{n,m,l,{\boldsymbol{p}}}(\gamma\boldsymbol{x},\gamma\boldsymbol{y},\gamma\boldsymbol{z}),
	\end{equation}
	with 
	\begin{equation}\label{eq:ScDim}	
		\delta_{sc}(n,m,l,\boldsymbol{p}):=	d(n+m+l)-n(d-\Delta_1)-m(d-\Delta_2)-l[\psi]-\|\boldsymbol{p}\|_1,
	\end{equation}	
	defines the action of the dilatation operator $D$ on the kernels.
	Note also that the $L^1$ norm of $H_{n,m,l,{\boldsymbol{p}}}$ rescales as follows under the action of $D$: 
	\begin{equation}
		\|D H_{n,m,l,{\boldsymbol{p}}}\|=\gamma^{D_{\text{sc}}(n,m,l,\boldsymbol{p})}\|H_{n,m,l,\boldsymbol{p}}\|,
	\end{equation}
	where 
	\begin{equation}\label{eq:ScDim1}\begin{split}
			D_{\text{sc}}(n,m,l,\boldsymbol{p})&:=\delta_{sc}(n,m,l,{\boldsymbol{p}})-d(n+m+l-1)\\ &\phantom{:}=d-n(d-\Delta_1)-m(d-\Delta_2)-l[\psi]-\|\boldsymbol{p}\|_1\end{split}
	\end{equation} 
	is the \textit{scaling dimension} of $H_{n,m,l,\boldsymbol{p}}$. Depending on whether $D_{\text{sc}}(n,m,l,\boldsymbol{p})$ is positive, vanishing or negative, we say that the corresponding monomial is	relevant, marginal or irrelevant. Note that, anticipating the fact that $\Delta_1=[\psi]$ and $\Delta_2=2[\psi]+O(\epsilon)$, the only relevant or marginal terms\footnote{To be precise, the relevance or irrelevance of the term with $(n,m,l,{\boldsymbol{p}})=(0,1,2,{\boldsymbol{0}})$ depends upon the sign of $\eta_2=\Delta_2-2[\psi]$. In any case, even if irrelevant, this term is barely so, with a scaling dimension of order $O(\epsilon)$. 
		For this reason, it is convenient to treat it differently from the irrelevant terms with scaling dimensions well separated from $0$, uniformly in $\epsilon$. Moreover, 
		the terms with $(0,0,2,\boldsymbol{p})$ with $\|\boldsymbol{p}\|_1=1$ are relevant only if $d=2,3$: in $d=1$ they are irrelevant, with scaling dimension equal to $-1/2+\epsilon$; 
		however, the treatment of these terms in $d=1$ does not present any significant simplification, as compared to the cases $d=2,3$. For this reason, and for uniformity of notation,
		we shall treat them exactly in the same way for $d=1$ and for $d=2,3$, notwithstanding this difference: in particular, the notion of 
		of trimmed sequence and of `trimming' given below, which defines the way in which we `localize' and `interpolate' the relevant terms, 
		is the same in $d=1,2,3$.} with $l>0$ are those with $(n,m,l,\boldsymbol{p})=(0,0,2,\boldsymbol{0})$, $(0,0,4,\boldsymbol{0})$, $(1,0,1,\boldsymbol{0})$, $(0,1,2,\boldsymbol{0})$, and $(0,0,2,\boldsymbol{p})$ with $\|\boldsymbol{p}\|_1=1$,  for which we have $D_{\text{sc}}(0,0,2,\boldsymbol{0})=d/2+\epsilon$, $D_{\text{sc}}(0,0,4,\boldsymbol{0})=2\epsilon$, $D_{\text{sc}}(1,0,1,\boldsymbol{0})=\Delta_1-[\psi]$, $D_{\text{sc}}(0,1,2,\boldsymbol{0})=\Delta_2-2[\psi]$, and $D_{\text{sc}}(0,0,2,(1,0))=D_{\text{sc}}(0,0,2,(0,1))=d/2+\epsilon-1$.
	
	\subsection{Trimmed sequences}\label{sec:2.4}
	
	A potential $H(\phi,J,\psi)$ is a {formal} infinite sum of monomials, of the form
	\begin{equation} H(\phi,J,\psi)=\sum_{(n,m,l,\boldsymbol{p})\in L}\ \sum_{\boldsymbol{a}, \boldsymbol{\mu}, \boldsymbol{b}}
		\iiint d\boldsymbol{x}\, d\boldsymbol{y}\, d\boldsymbol{z}\, \Phi(\boldsymbol{a}, \boldsymbol{x})J(\boldsymbol{y})\Psi(\mathbf{B},\boldsymbol{z}) [H_{n,m,l,\boldsymbol{p}}(\boldsymbol{x},\boldsymbol{y},\boldsymbol{z})]_{\boldsymbol{a},\boldsymbol{\mu},\boldsymbol{b}}\label{eq:pot}\end{equation}
	where
	\begin{equation}L:=\{(n,m,l,\boldsymbol{p}): \boldsymbol{p}=(p_1,\ldots, p_l)\ \text{with}\ p_i\in\{0,1\},\ \  n,m,l,|\boldsymbol{p}|\ge0,\ \  n+l \ \text{even}, \ \ n+m+l\ge 1\},\label{defL}\end{equation} 
	and, given $(n,m,l,\boldsymbol{p})\in L$, the sums over $\boldsymbol{a}$, $\boldsymbol{\mu}$ and $\boldsymbol{b}$ run over the sets 
	$\{1,\ldots,N\}^n$, $\{0,1\}^l$ and $\{1,\ldots,N\}^l$, respectively; moreover, we used the notations of \eqref{GMRnotation}, with $\mathbf{B}=((p_1,\mu_1,b_1),$ $\ldots,$ $(p_l,\mu_l,b_l))$. The formal sum in \eqref{eq:pot} should {just} be thought of as a way of representing the collection of kernels $H_{n,m,l,\boldsymbol{p}}$. Therefore, in the following, 
	we shall equivalently write a potential $H$ either as in \eqref{eq:pot} or as
	\begin{equation} H=\{H_{\ell}\}_{\ell\in L}.\end{equation}
		
	We shall say that a sequence $\{H_\ell\}_{\ell\in L}$, with translationally invariant kernels $H_\ell$ of finite $L^1$ norm, 
	is a {\it trimmed sequence} if the relevant or marginal kernels with $\ell=(0,0,2,\boldsymbol{0})$, $(0,0,4,\boldsymbol{0})$, 
	$(1,0,1,\boldsymbol{0})$, $(0,1,2,\boldsymbol{0})$, and $(0,0,2,\boldsymbol{p})$ with $\|\boldsymbol{p}\|_1=1$ have  the following local structure:
	\begin{equation} \label{trimm}\begin{split} 
			& \big[H_{0,0,2,\boldsymbol{0}}(\boldsymbol{z})\big]_{\boldsymbol{b}}=c_1 \delta(z_1-z_2)\Omega_{b_1b_2}, \\
			& \big[H_{0,0,2,\boldsymbol{p}}(\boldsymbol{z})\big]_{\boldsymbol{\mu},\boldsymbol{b}}=0, \quad \forall \boldsymbol{p} \ : \ \|\boldsymbol{p}\|_1=1,\\
			& \big[H_{0,0,4,\boldsymbol{0}}(\boldsymbol{z})\big]_{\boldsymbol{b}}=c_2\delta(z_1-z_2)\delta(z_1-z_3)\delta(z_1-z_4) q_{b_1b_2b_3b_4},\\
			& \big[H_{1,0,1,\boldsymbol{0}}(x_1,z_1)\big]_{a_1,b_1}=c_3\delta(x_1-z_1)\delta_{a_1,b_1}, \\
			& \big[H_{0,1,2,\boldsymbol{0}}(y_1,\boldsymbol{z})\big]_{\boldsymbol{b}}=c_4\delta(y_1-z_1)\delta(z_1-z_2)\Omega_{b_1b_2},	\end{split}
	\end{equation}
	where $c_1,c_2,c_3,c_4$ are constants, and $q_{abce}{{=\Omega_{ab}\Omega_{ce}-\Omega_{ac}\Omega_{be}+\Omega_{ae}\Omega_{bc}}}$ is the totally anti-symmetric tensor, as in \cite[Eq.(4.4)]{GMR21}. 
	
	As discussed in Section \ref{Sec2}, the RG transformation $R$ is a map from the space of trimmed sequences of kernels to itself. 
	
	\begin{remark} The non-interacting fixed point potential $V_0^*(\phi,J,\psi)=(\phi,\psi)+(J,\psi^2)+\mathcal R^*_0(\phi,J,\psi)+\mathcal S^*_0(\phi,J)$, constructed in Section \ref{sec:freecase}, can be written as $V_0^*=\{V_{0;\ell}^*\}_{\ell\in L}$, where the only non-zero kernels in the sequence are the following: 
		\begin{equation}\begin{split}
				&\big[V^*_{0;(1,m,1,\boldsymbol{0})}(x_1,{\boldsymbol{y}},z_1)\big]_{a_1,b_1}\Big|_{\epsilon=0}=(-2)^m\delta_{a_1,b_1}\delta(y_m-z_1)\prod_{i=0}^{m-1}P_{\ge 1}(y_i-y_{i+1}), \hskip2.07truecm m\ge 0 \\
				&\big[V^*_{0;(0,m,2,\boldsymbol{0})}({\boldsymbol{y}},\boldsymbol{z})\big]_{b_1,b_2}\Big|_{\epsilon=0}=(-2)^{m-1}\Omega_{b_1,b_2}\delta(z_1-y_1)\, \delta(y_m-z_2)\prod_{i=1}^{m-1}P_{\ge 1}(y_i-y_{i+1}), \qquad m\ge 1 \\
				&\big[V^*_{0;(2,m,0,\emptyset)}(\boldsymbol{x},{\boldsymbol{y}})\big]_{a_1,a_2}\Big|_{\epsilon=0}=(-2)^{m-1}\Omega_{a_1a_2}\prod_{i=0}^{m}P_{\ge 1}(y_i-y_{i+1}), 
				\hskip4.03truecm m\ge 0 \\
				&V^*_{0;(0,m,0,\emptyset)}({\boldsymbol{y}})\Big|_{\epsilon=0}=(-2)^{m-1}\frac{N}{m}\prod_{i=1}^{m}P_{\ge 1}(y_i-y_{i+1}), \hskip5.83truecm m\ge 1
		\end{split}\end{equation}
		where: $P_{\ge 1}$ was defined in \eqref{Glehgeh}; in the first two lines we dropped the $\boldsymbol{\mu}$ labels;  
		in the first line, $y_0$ should be interpreted as $x_1$, and, for $m=0$, the right side should be interpreted as that of the third line of \eqref{trimm} with $c_3=1$; for $m=1$, the right side of the second line 
		should be interpreted as that of the fourth line of \eqref{trimm} with $c_4=1$; in the third line, $y_0$ and $y_{m+1}$ should be interpreted as $x_1$ and $x_2$, respectively; in  the fourth line,
		$y_{m+1}$ should be interpreted as $y_1$.
		
		In particular, $V_0^*$ is associated with a {\it{trimmed sequence}}, such that $c_1=c_2=0$ and $c_3=c_4=1$. 
	\end{remark}
	
	\subsection{The fixed point potential}\label{sect:2.5}
	
	We are now in good position to define more precisely the fixed point potential $V^*(\phi,J,\psi)$ that we construct in this paper. 
	
	\begin{defn}\label{Def.1} $V^*=\{V^*_\ell\}_{\ell\in L}$ is the solution of the fixed point equation $V^*=RV^*$ in the space of trimmed sequences, such that: (1) 
		$V^*_{0,0,l,\boldsymbol{p}}=H^*_{l,\boldsymbol{p}}$, with $H^*_{l,\boldsymbol{p}}$ the kernels of the FP potential constructed in \cite{GMR21}; (2) 
		for $\ell=(1,0,1,\mathbf{0})$, $(0,1,2,\mathbf{0})$, $V^*_\ell$ is as in \eqref{trimm} with $c_3=c_4=1$, i.e.,  the local terms $(\phi,\psi)$ and $(J,\psi^2)$ in $V^*$ have pre-factor equal to $1$; (3)
		the scaling exponents $\Delta_1,\Delta_2$ entering the definition of $D$ and, therefore, of $R$, are analytic in $\epsilon$ for $\epsilon$ sufficiently small, and such that 
		$\Delta_1-[\psi]=O(\epsilon)$, $\Delta_2-2[\psi]=O(\epsilon)$; (4) 
		the kernels $V^*_\ell$ with $\ell\in L\setminus\{(1,0,1,\mathbf{0}),(0,1,2,\mathbf{0})\}$ are analytic in $\epsilon$ for $\epsilon$ sufficiently small and, at $\epsilon=0$, satisfy $V^*_\ell\Big|_{\epsilon=0}=V^*_{0;\ell}$, 
		with $V^*_{0;\ell}$ as in \eqref{trimm} for $\ell {{\in\{(1,m,1,\boldsymbol{0})\}_{m\ge 1}\cup \{(0,m,2,\boldsymbol{0})\}_{m\ge 2}\cup\{(2,m,0,\emptyset)\}_{m\ge 0}\cup
				\{(0,m,0,\emptyset)\}_{m\ge 1}}}$, and equal to zero otherwise.
	\end{defn}
	
	{{For the explicit form of the fixed point equation $V^*=RV^*$ as a map from the space of trimmed sequences to itself, see Section \ref{sect:FPE} below.}}
	
	Note that the existence and uniqueness of the FP potential $V^*$ specified in  Definition \ref{Def.1} is part of the results proved in this paper. As anticipated above, we will write 
	\begin{equation} \label{eqRS}V^*(\phi,J,\psi)=H^*(\psi)+(\phi,\psi)+(J,\psi^2)+\mathcal R^*(\phi,J,\psi)+\mathcal S^*(\phi,J),\end{equation}
	where, letting\footnote{{The rationale behind the labels ext, $0$, and $f$ in \eqref{eqLext} is the following: `ext' stands for `external field', it refers to the fact that $n+m>0$, i.e., there is at least one `external field' $\phi$ or $J$ field; `$f$' stands for `fluctuation field', it refers to the fact that $l>0$, i.e., there is at least one `fluctuation field' $\psi$; similarly, the label `$0$' in the first definition indicates that $l=0$, i.e., there is no fluctuation field.}}
	\begin{equation} \label{eqLext} L_{\text{ext},0}:=\{(n,m,0,\emptyset)\in L\}, \qquad L_{\text{ext},f}:= \{(n,m,l,\boldsymbol{p})\in L\ :\ n+m>0, \  l>0\}, \end{equation}
	and $L_{\text{ext},f}':=L_{\text{ext},f}\setminus \{(1,0,1,\boldsymbol{0}),
	(0,1,2,\boldsymbol{0})\}$, we let: $\mathcal R^*=\{V^*_\ell\}_{\ell\in L'_{\text{ext},f}}$ and $\mathcal S^*=\{V^*_\ell\}_{\ell\in L_{\text{ext},0}}$. 
	
	\subsection{Almost-scaling operators}\label{sect:2.6}
	
	The response functions are not correlation functions of any local operators built out the field $\psi$.
	We can however identify two quasi-local operators $\mathcal O^{(1)}$ and $\mathcal O^{(2)}$, whose two point functions will end up being the same as $\mathcal G^*$ and $\mathcal F^*$, up to corrections that decay faster than any 
	power at large distances. Because of these correction terms, we will call these operators {\it almost-scaling operators}. We do not expect that scaling operators with exactly scale invariant two point functions exist in our theory, due to the presence of the fixed ultraviolet cutoff $\chi(k)$.
	
	In order to identify $\mathcal O^{(1)}$ and $\mathcal O^{(2)}$, let us expand $V^*(\phi,J,\psi)$ as
	\begin{equation} V^*(\phi,J,\psi)=H^*(\psi)+(\phi,\mathcal O^{(1)})+(J,\mathcal O^{(2)})+\mathcal Q^*(\phi,J,\psi)+\mathcal S^*(\phi,J),\label{rewrv*}\end{equation}
	where $\mathcal Q^*$ is the part of $\mathcal R^*$ that is at least quadratic in $(\phi,J)$, and $\mathcal S^*$ is the same as in \eqref{eqRS}. In other words, $\mathcal O^{(1)}_{{a}}(x)=\frac{\delta V^*(\phi,J,\psi)}{\delta\phi_{{a}}(x)}\big|_{\phi=J=0}$, and similarly for $\mathcal O^{(2)}(x)$. From the definition of response functions, we find
	\begin{equation}\begin{split} 
			& \mathcal G_{{ab}}^*(x)=\langle \mathcal O^{(1)}_{{a}}(x)\mathcal O^{(1)}_{{b}}(0)\rangle_{H^*}+{{\big[}}\mathcal E_1(x){{\big]_{ab}}},\\
			& \mathcal F^*(x)=\langle \mathcal O^{(2)}(x);\mathcal O^{(2)}(0)\rangle_{H^*}+\mathcal E_2(x),\end{split}
		\label{eq:tadpole-def}
	\end{equation}
	where the correction terms $\mathcal E_1,\mathcal E_2$ are given by
	\begin{equation}\label{eq:E1E2}\begin{split} 
			& {{\big[}}\mathcal E_1(x){{\big]_{ab}}}:=\frac{\delta^2 \mathcal S^*(\phi,0)}{\delta\phi_{{a}}(0)\delta\phi_{{b}}(x)}\Big|_{\phi=0}+\bigg\langle \frac{\delta^2 \mathcal Q^*(\phi,0,\psi)}{\delta\phi_{{a}}(0)\delta\phi_{{b}}(x)}\Big|_{\phi=0}\bigg\rangle_{H^*},\\
			& \mathcal E_2(x):=\frac{\delta^2 \mathcal S^*(0,J)}{\delta J(0)\delta J(x)}\Big|_{J=0}+\bigg\langle \frac{\delta^2 \mathcal Q^*(0,J,\psi)}{\delta J(0)\delta J(x)}\Big|_{J=0}\bigg\rangle_{H^*}.
	\end{split}\end{equation}
	
	\subsection{Main  results}\label{sect:2.7}
	
	\begin{theorem}[Analyticity of the anomalous critical exponent and discrete scale invariance]\label{Th:1} There exists $\epsilon_0>0$ such that, for $\epsilon\in B_{\epsilon_0}(0):=\{\epsilon\in\mathbb C :  |\epsilon|<\epsilon_0\}$, there is 
		a unique potential $V^*$ as in Definition \ref{Def.1}, 
		analytic in $\epsilon$ in $B_{\epsilon_0}(0)$. Moreover, the response functions $\mathcal G^*$ and $\mathcal F^*$ defined in \eqref{GF}  are analytic in $\epsilon$ in the same domain, and behave under
		rescaling $x\to\rho x$, with $\rho=\gamma^k$, $k\in\mathbb Z$, as
		\begin{equation} \mathcal G^*_{{ab}}(\rho x)=\rho^{-2\Delta_1}\mathcal G_{{ab}}^*(x),\qquad  \mathcal F^*(\rho x)=\rho^{-2\Delta_2}\mathcal F^*(x), \label{1.12}
		\end{equation}
		with $\Delta_1=[\psi]$ and $\Delta_2=2[\psi]+\eta_2(\epsilon)$, where $\eta_2(\epsilon)$ is analytic in $\epsilon$ for $|\epsilon|<\epsilon_0$, with
		\begin{equation}\label{FO}
			\eta_2(\epsilon)=2\epsilon\frac{N-2}{N-8}+O(\epsilon^2).
		\end{equation}	 
	\end{theorem}
	
	We actually expect that $ \mathcal G^*$ and $\mathcal F^*$ are exactly scale invariant, which means that Eq.\eqref{1.12} should be true for any $\rho>0$: this will be proved in the third paper of this series, {see also the recent paper \cite{Gr24} for a proof of full scale invariance with a different choice of the cutoff function}. In this paper, we limit ourselves to show, by an  explicit lowest-order computation (see Appendix \ref{appB}), that the dominant contributions in $\epsilon$ to $\mathcal{G}^*(x)$ and $\mathcal{F}^*(x)$ are exactly scale-invariant term{{s}}, namely 
	$\mathcal G^*(x)=\mathcal G_0^*(x)+\mathcal G_{\text{h.o.}}^*(x)$, where, in the sense of distributions \cite{G08},
	\begin{equation}\label{G.1} 
		\mathcal G^*_{0{{;ab}}}(x)={\Omega_{ab}}\int \frac{d^dk}{|k|^{d/2+\epsilon}}e^{ikx}\equiv {\Omega_{ab}}C_0|x|^{-2\Delta_1}\end{equation}
	and the higher order correction $\mathcal G_{\text{h.o.}}^*$ satisfies $\big|\mathcal G_{\text{h.o.}}^*(x)\big|\le C\epsilon |x|^{-2\Delta_1}$, for some $C>0$ independent of $\epsilon$. Similarly, 
	$\mathcal F^*(x)=\mathcal F_0^*(x)+\mathcal F_{\text{h.o.}}^*(x)$, where
	\begin{equation}\label{F.1}
		\mathcal F^*_0(x)=-{2}N\int \frac{d^dk}{(2\pi)^d|k|^{d/2+\epsilon}}\int \frac{d^dp}{(2\pi)^d|p|^{d/2+\epsilon-2\eta_2}}e^{i(k+p)x}\equiv C'_0|x|^{-2\Delta_2},\end{equation}
	and $\big|\mathcal F_{\text{h.o.}}^*(x)\big|\le C'\epsilon |x|^{-2\Delta_2}$, for some $C'>0$ independent of $\epsilon$.
	
	\begin{theorem}[Almost-scaling operators]\label{Th:2} 
		There exist $C_i,c_i>0,\;i=1,2$ such that $\mathcal E_1(x)$ and $\mathcal E_2(x)$ are analytic in $\epsilon$ in the same domain $B_{\epsilon_0}(0)$ as in Theorem \ref{Th:1}, and bounded as follows:
		\begin{equation} |{{\big[}}\mathcal E_1(x){{\big]_{ab}}}|\le C_1 e^{-c_1|x|^\sigma}(\min\{1,|x|\})^{-2\Delta_1},\qquad  |\mathcal E_2(x)|\le C_2 e^{-c_2|x|^\sigma}(\min\{1,|x|\})^{-2\Delta_2},\label{eq:E}
		\end{equation}
		with $\sigma=1/s$ and $s$ the order of the Gevrey class $G^s$ which the UV cutoff function $\chi$ belongs to.
	\end{theorem}
	Putting together Eqs.~\eqref{eq:tadpole-def} with Theorems \ref{Th:1} and \ref{Th:2}, we conclude that the two point functions $\langle \mathcal{O}^{(1)}_{{a}} \mathcal{O}^{(1)}_{{b}}\rangle_{H^*}$ and $\langle \mathcal{O}^{(2)}; \mathcal{O}^{(2)}\rangle_{H^*}$ are discrete scale invariant up to stretched-exponentially small corrections at large distances, i.e.~that $\mathcal{O}^{(1)}$ and $\mathcal{O}^{(2)}$ are almost-scaling operators. As mentioned, we cannot hope for anything better than that.
	
	{\begin{remark}
			In the non-interacting case discussed in Section \ref{sec:freecase}, the stretched exponential decay of the correction terms, denoted by $\mathcal E_{0,1}, \mathcal E_{0,2}$ 
			(the `$0$' label recalling the absence of interaction), is immediate. In fact, rewriting the non-interacting fixed point potential $V_0^*$ as in \eqref{rewrv*}, 
			we see that $\mathcal O^{(1)}_a(x)=\psi_a(x)$ and $\mathcal O^{(2)}(x)=\psi^2(x)$. Therefore, for $H^*=0$, the two-point functions $\langle \mathcal O^{(1)}_a(x)\mathcal O^{(1)}_b(0)\rangle_{H^*}$ and 
			$\langle \mathcal O^{(2)}(x);\mathcal O^{(2)}(0)\rangle_{H^*}$ reduce to $\langle \psi_a(x)\psi_b(0)\rangle_{0}= \Omega_{ab}P_{\le 1}(x)$ and 
			$\langle \psi^2(x);\psi^2(0)\rangle_{0}= -2N\, P^2_{\le 0}(x)$, respectively. Therefore, from \eqref{g0starf0star} and the non-interacting analogue of \eqref{eq:tadpole-def}, 
			that is 
			\begin{equation}\begin{split} 
					& \mathcal G_{0;ab}^*(x)=\langle \psi_a(x)\psi_b(0)\rangle_{0}+\big[\mathcal E_{0,1}(x)\big]_{ab},\\
					& \mathcal F^*_0(x)=\langle \psi^2(x);\psi^2(0)\rangle_{0}+\mathcal E_{0,2}(x),\end{split}
				\label{eq:tadpole-def_non-int}
			\end{equation}
			we see that $\mathcal E_{0,1}, \mathcal E_{0,2}$ are, explicitly:
			\begin{equation}\begin{split}
					&\big[\mathcal{E}_{0,1}(x)\big]_{ab}=\Omega_{ab}P_{\ge1}(x)\\
					&\mathcal{E}_{0,2}=-2N\left[2P_{\le0}(x)P_{\ge1}(x)+P^2_{\ge1}(x)\right],
			\end{split}\end{equation}		 
			which can be easily shown to satisfy the decay bounds in \eqref{eq:E}. \end{remark}}
	
	\subsection{Structure of the proof}\label{sec:structure}
	
	The proofs of the Theorem \ref{Th:1} and \ref{Th:2} are based on the following considerations. First of all, from \eqref{eq:1bis} we have:
	\begin{equation}\label{W}
		W^*(\phi,J)=\lim_{h\to-\infty}(T{{S^{(h+1)}}}\cdots T{{S^{(0)}}}V^*)(\phi,J,0).
	\end{equation}	
	Now, recalling that the RG map $R=D^{-(h-1)}T{{S^{(h)}}} D^h\equiv DT{{S^{(0)}}}$, for all $h\le 0$, we see that 
	\eqref{W} can be written as:
	\begin{equation}\label{W*phiJ}
		W^*(\phi,J)=\lim_{h\to-\infty}D^{h} (R^{|h|}	V^*)(\phi,J,0)=\lim_{h\to-\infty}D^{h}V^*(\phi,J,0)
	\end{equation}
	where in the last equality we used that $V^*$ is a fixed point for the RG map, i.e. $RV^*=V^*$. Denote by $V^*_{n,m,l,\boldsymbol{p}}$ the kernels of $V^*$, in the sense of \eqref{tot}. 
	Using the definitions \eqref{GF}, and recalling the action \eqref{eq:scal} of $D$ on the kernels, we find, letting $\boldsymbol{x}=(x,0)$,
	\begin{equation} \label{eq:6}\begin{split}
			\mathcal G^*(x)&=2\lim_{h\to-\infty} D^hV^*_{2,0,0,\emptyset}(\boldsymbol{x})\\		
			&=2\lim_{h\to-\infty} \gamma^{2h\Delta_1}V^*_{2,0,0,\emptyset}(\gamma^{h}\boldsymbol{x}), \end{split}
	\end{equation}
	and, letting $\boldsymbol{y}=(y,0)$,  
	\begin{equation} \label{eq:7}\begin{split}
			\mathcal F^*(y)&=2\lim_{h\to-\infty} D^hV^*_{0,2,0,\emptyset}(\boldsymbol{y})\\		
			&=2\lim_{h\to-\infty} \gamma^{2h\Delta_2}V^*_{0,2,0,\emptyset}(\gamma^{h}\boldsymbol{y}). \end{split}
	\end{equation}	
	
	After having defined, in Section \ref{Sec2}, the fixed point equation for $V^*$, in Section \ref{Sec:3} we will show that the kernels $V^*_{n,m,l,\boldsymbol{p}}$, and in particular $V^*_{2,0,0,\emptyset}, V^*_{0,2,0,\emptyset}$,
	can be written in terms of a tree expansion that is absolutely convergent in a weighted $L^1$ norm. As a consequence, these kernels, as well as the scaling exponents $\Delta_1,\Delta_2$, turn out to be analytic in $
	\epsilon$ for $\epsilon$ small enough. In particular $\Delta_1=[\psi]$ and $\Delta_2=2[\psi]+\eta_2(\epsilon)$, with $\eta_2$ as in \eqref{FO} (the explicit computation of the first order contribution is deferred to Appendix \ref{appA}). 
	
	In Section \ref{sec5}, 
	we will prove the pointwise existence of the limits in \eqref{eq:6} and \eqref{eq:7}, as well as the analyticity of the limiting functions. These will require to control the absolute convergence
	of the tree expansion for the kernels of the fixed point potential in a pointwise norm. The ideas here generalize, and adapt to the present context, those used in 
	\cite{BFM07,GGM12,AGG22,AGG23} to estimate the $n$-point correlation functions. Once the pointwise existence of the limits in \eqref{eq:6}-\eqref{eq:7} is established, the scale invariant property 
	\eqref{1.12} is an immediate corollary: in fact, from \eqref{eq:6} one finds, renaming $h+k\equiv h'$,
	\begin{equation}\begin{split}\mathcal G^*(\gamma^k x)&=2\lim_{h\to-\infty} \gamma^{2h\Delta_1}V^*_{2,0,0,\emptyset}(\gamma^{h+k}\boldsymbol{x})\\ 
			&=2\gamma^{-2k\Delta_1}\lim_{h'\to-\infty} \gamma^{2h'\Delta_1}V^*_{2,0,0,\emptyset}(\gamma^{h'}\boldsymbol{x})\equiv \gamma^{-2k\Delta_1}\mathcal G^*(x),\end{split}\label{eq:2.44}
	\end{equation}
	provided the limit exists, and similarly for $\mathcal F^*$, thus concluding the proof of Theorem \ref{Th:1}. 
	
	Finally, in Section \ref{sec6}, we derive a tree expansion for the correction terms $\mathcal E_1, \mathcal E_2$, and show that its convergence in a pointwise norm
	introduced in Section \ref{sec5} implies Theorem \ref{Th:2}. 
	
	\section{Renormalization map and fixed point equation}\label{Sec2}
	
	In this section we describe more in detail the RG transformation introduced in Section \ref{sec:RG}. Moreover, we write down the fixed point equation for $V^*$, including the equations for the 
	scaling exponents $\Delta_1,\Delta_2$. The discussion follows the analogous one in \cite[Section 5]{GMR21}, which we refer to for additional details. {Note that in this 
	section we shall write down the components of the RG fixed point equation in the form of series, see in particular \eqref{FPE}-\eqref{FPEnonloc}, 
	to be considered as formal as long as convergence in an appropriate norm is 
	proved. Absolute convergence of these sums in a suitable weighted $L_1$ norm, see \eqref{norm} below, is in fact addressed and proved in Section \ref{Sec:3}.}
		
	\subsection{Integrating out}
	
	Consider a potential $H(\phi,J,\psi)=\{H_\ell\}_{\ell\in L}$, whose sequence of kernels is trimmed, in  the sense of Section \ref{sec:2.4}. 
	Let us discuss the effect of integrating out the fluctuation field on scale $0$ from $H(\phi,J,\psi)$: 
	\begin{equation}\label{HeffS0}
		H_{\text{eff}}(\phi,J,\psi)= {{S^{(0)}}} H(\phi,J,\psi), 
	\end{equation}	
	with ${{S^{(0)}}}$ the integrating-out map on scale $0$ defined in \eqref{eq:1.3}. 
	In analogy with \cite[Eq.(5.3)]{GMR21}, using the notation introduced in {Remark \ref{remark:notation}}, the kernels of the effective interaction can be written as: 
	\begin{equation}\label{HBxy}
		H_{\text{eff}}(\boldsymbol{a},\boldsymbol{x},\boldsymbol{y},\mathbf{B},\boldsymbol{z}_{\mathbf{B}})=\mathcal{P}\sum_{s\ge1}\frac{1}{s!}\sum_{\substack{\mathbf{B}_1,...,\mathbf{B}_s\\\sum_{i}\mathbf{B}_i=\mathbf{B}}}\sum_{\substack{\mathbf{A}_1,...,\mathbf{A}_s\\ \mathbf{A}_i\supset\mathbf{B}_i}}
		\sum_{\substack{\boldsymbol{a}_1,\ldots,\boldsymbol{a}_s\\ \boldsymbol{x}_1,\ldots,\boldsymbol{x}_s\\ \boldsymbol{y}_1,\ldots,\boldsymbol{y}_s}}
		(-)^{\sharp}\int d\boldsymbol{z}_{\bar{\mathbf{B}}}\mathcal{C}(\boldsymbol{z}_{\bar{\mathbf{B}}})\prod_{i=1}^sH(\boldsymbol{a}_i,\boldsymbol{x}_i,\boldsymbol{y}_i,\mathbf{A}_i,\boldsymbol{z}_{\mathbf{A}_i})	
	\end{equation}	
	where: $\mathcal P$ is the operator that anti-symmetrizes under simultaneous permutations of the elements of $\boldsymbol{a}$ and $\boldsymbol{x}$, symmetrizes over permutations 
	of the elements of $\boldsymbol{y}$, and anti-symmetrizes under simultaneous permutations of the elements of $\mathbf{B}$ and $\boldsymbol{z}_{\mathbf{B}}$ (which generalizes  
	the antisymmetrization operator $\mathcal A$ of \cite[Eq.(5.3)]{GMR21}); 
	the sums over $\mathbf{B}_i$ and $\mathbf{A}_i$ must be interpreted as described after \cite[Eq.(5.3)]{GMR21}; the sum over $\boldsymbol{a}_1,\ldots, \boldsymbol{a}_s$ is over 
	all ways to represent $\boldsymbol{a}$ as a concatenation $\boldsymbol{a}_1+\cdots+\boldsymbol{a}_s$, and similarly for the sums over $\boldsymbol{x}_1,\ldots,\boldsymbol{x}_s$ and 
	over $\boldsymbol{y}_1,\ldots,\boldsymbol{y}_s$. Finally, letting, as in \cite[Eq.(5.3)]{GMR21}, $\bar{\mathbf{B}}_i=\mathbf{A}_i\setminus\mathbf{B}_i$ and $\bar{\mathbf{B}}=\bar{\mathbf{B}}_1+\cdots{\bar{\mathbf{B}}}_s$, 
	we denote $\mathcal{C}(\boldsymbol{z}_{\bar{\mathbf{B}}}):=\langle\Psi'(\bar{\mathbf{B}}_1,\boldsymbol{z}_{\bar{\mathbf{B}}_1});\cdots;\Psi'(\bar{\mathbf{B}}_n,\boldsymbol{z}_{\bar{\mathbf{B}}_n})\rangle_0^{{(0)}}$
	where $\Psi'(\bar{\mathbf{B}}_i,\boldsymbol{z}_{\bar{\mathbf{B}}_i})$ is interpreted as in \eqref{GMRnotation}, and $\langle\cdots\rangle_0^{{(0)}}$ denotes the expectation w.r.t. $d\mu_0(\psi')$. 
	
	We compactly rewrite \eqref{HBxy} as
	\begin{equation}\label{Heffell}
		(H_{\text{eff}})_{\ell}=\sum_{s\ge 1}\sum_{(\ell_i)_{i=1}^s}S^{\ell_1,...,\ell_s}_{\ell}(H_{\ell_1},\ldots, H_{\ell_s}),
	\end{equation}	 
	where $\ell$ belongs to the label set $L$ in \eqref{defL}, while 
	$\ell_1,\ldots, \ell_s$ belong to the subset of $L$ `with at least one fluctuation field label', i.e., to 
	\begin{equation}L_f=\{(n,m,l,\mathbf{p})\in L : l>0\},\label{defL0}\end{equation}
	the label $f$ standing for `fluctuation field'.
	
	\subsection{Trimming}\label{Trimming}
	Even if the input potential $H=\{H_\ell\}_{\ell\in L}$ is trimmed, the output $H_{\text{eff}}$ will not in general  be so. 
	However, we can act on $H_{\text{eff}}$ with a linear operator, equivalent to 
	the identity in the sense of \cite[Section 5.2.1]{GMR21}, called the trimming operator $T$, which returns an equivalent trimmed potential. 
	Denoting equivalence between potentials by the symbol $\sim$, the action of the trimming operator will allow us to rewrite 
	\begin{equation}H_{\text{eff}}\sim \mathcal LH_{\text{eff}}+\mathcal I H_{\text{eff}},\end{equation}
	with $\mathcal LH_{\text{eff}}$ the local, relevant, part of $H_{\text{eff}}$, and $\mathcal IH_{\text{eff}}$ the non-local, irrelevant, part of $H_{\text{eff}}$. 
	The action of $T$ on the kernels with $n=m=0$ has already been described in \cite[Sect.5.2 and App.C]{GMR21} and won't be repeated here. The action of $T$ on the kernels with $n+m>0$ is non-trivial
	iff $(n,m,l)$ equals $(1,0,1)$ or $(0,1,2)$; in the complementary case, we let $(\mathcal IH_{\text{eff}})_{n,m,l,\boldsymbol{p}}= (H_{\text{eff}})_{n,m,l,\boldsymbol{p}}$ and $(\mathcal LH_{\text{eff}})_{n,m,l,\boldsymbol{p}}=0$. 
	
	Let us now consider the cases $(n,m,l)=(1,0,1)$, $(0,1,2)$. As in the case without source fields, trimming involves localization and interpolation: localization extracts the local parts of $(H_{\text{eff}})_{1,0,1,\boldsymbol{0}}$ and $(H_{\text{eff}})_{0,1,2,\boldsymbol{0}}$ (cf. with \cite[Eq.(5.12)]{GMR21}), as follows: 
	\begin{equation}\label{eq:loc}
		(\mathcal LH_{\text{eff}})_{1,0,1,\boldsymbol{0}}=T^{1,0,1,\boldsymbol{0}}_{1,0,1,\boldsymbol{0}}(H_{\text{eff}})_{1,0,1,\boldsymbol{0}},\quad (\mathcal LH_{\text{eff}})_{0,1,2,\boldsymbol{0}}=T^{0,1,2,\boldsymbol{0}}_{0,1,2,\boldsymbol{0}}(H_{\it{eff}})_{0,1,2,\boldsymbol{0}},
	\end{equation}	
	(the operators $T^{1,0,1,\boldsymbol{0}}_{0,1,2,\boldsymbol{0}}$ and $T^{0,1,2,\boldsymbol{0}}_{0,1,2,\boldsymbol{0}}$ are defined in the following subsections), 
	and we let $$(\mathcal LH_{\text{eff}})_{1,0,1,\boldsymbol{p}}= (\mathcal LH_{\text{eff}})_{0,1,2,\boldsymbol{p}}=0 \quad \text{for}\quad \boldsymbol{p}\neq\boldsymbol{0}.$$ On the other hand, 
	letting $P_1=\{(0),(1)\}$ and $P_2=\{(0,0),(1,0),(0,1),(1,1)\}$, interpolation rearranges the difference between $\{(H_{\text{eff}})_{1,0,1,\boldsymbol{p}}\}_{\boldsymbol{p}\in P_1}$ (resp. $\{(H_{\text{eff}})_{0,1,2,\boldsymbol{p}}\}_{\boldsymbol{p}\in P_2}$) and its local part
	$\{(\mathcal LH_{\text{eff}})_{1,0,1,\boldsymbol{p}}\}_{\boldsymbol{p}\in P_1}$ (resp. $\{(\mathcal LH_{\text{eff}})_{0,1,2,\boldsymbol{p}}\}_{\boldsymbol{p}\in P_2}$)
	in such a way to equivalently rewrite it as $\{(\mathcal I H_{\text{eff}})_{1,0,1,\boldsymbol{p}}\}_{\boldsymbol{p}\in P_1}$ (resp. $\{(\mathcal I H_{\text{eff}})_{0,1,2,\boldsymbol{p}}\}_{\boldsymbol{p}\in  P_2}$) 
	with $(\mathcal I H_{\text{eff}})_{1,0,1,\boldsymbol{0}}=(\mathcal I H_{\text{eff}})_{0,1,2,\boldsymbol{0}}=0$. More precisely, we let
	\begin{equation}\label{RH}
		\begin{split} 
			& (\mathcal I H_{\text{eff}})_{1,0,1,\boldsymbol{p}}=\begin{cases} 0 & \text{if}\ \boldsymbol{p}=\boldsymbol{0}\\ (H_{\text{eff}})_{1,0,1,\boldsymbol{p}} +T^{1,0,1,\boldsymbol{0}}_{1,0,1,\boldsymbol{p}}(H_{\text{eff}})_{1,0,1,\boldsymbol{0}}& \text{if}\ \|\boldsymbol{p}\|_1=1,  \end{cases}\\
			& (\mathcal I H_{\text{eff}})_{0,1,2,\boldsymbol{p}}=\begin{cases} 0 & \text{if}\ \boldsymbol{p}=\boldsymbol{0}\\(H_{\text{eff}})_{0,1,2,\boldsymbol{p}}+  T^{0,1,2,\boldsymbol{0}}_{0,1,2,\boldsymbol{p}}(H_{\text{eff}})_{0,1,2,\boldsymbol{0}}& \text{if}\ \|\boldsymbol{p}\|_1=1 \\
				(H_{\text{eff}})_{0,1,2,\boldsymbol{p}} 
				& \text{if}\ \|\boldsymbol{p}\|_1=2. \end{cases}\end{split}
	\end{equation}
	Identifying the kernels with the corresponding Grassmann monomials as in \cite[Eq.(C.1)]{GMR21}, the manipulations equivalent to the identity lead to the definitions of $T^{1,0,1,\boldsymbol{0}}_{1,0,1,\boldsymbol{p}}$ and $T^{0,1,2,\boldsymbol{0}}_{0,1,2,\boldsymbol{p}}$ are described in the following subsections.
	
	\subsubsection{Case \texorpdfstring{$(n,m,l)=(1,0,1)$}{(n,m,l=(1,0,1))}}\label{sec.2.2.1}
	We will use the following interpolation identity (in most cases, we drop and leave implicit the component indices of the Grassmann fields; moreover, summation over repeated indices is understood), 
	which is the same as \cite[Eq.(5.8)]{GMR21}:
	\begin{equation}
		\psi(y)=\psi(x)	+(y-x)_\mu\int_{0}^1 ds\,\partial_\mu\psi (x+s(y-x)).
	\end{equation}	
	This is used to split a non-local but relevant term into a local relevant term plus an irrelevant one (for te definition of relevant and irrelevant, see after \eqref{eq:ScDim1}). 
	Hence considering $\int d^dx d^dz\, \phi(x) G(x,z)\psi(z)$ with $G$ translationally invariant, playing the role of $(H_{\text{eff}})_{1,0,1,\mathbf{0}}$, we get:
	\begin{eqnarray}
		\int d^dx\, d^dz\, \phi(x) G(x,z)\psi(z)	&=&\int d^dx\, \phi(x)\hat G(0)\psi(x)\\
		&+&\int d^dx\,d^dz\, \phi(x)G(x,z)(z-x)_\mu\int_0^1ds\,\partial_\mu\psi(x+s(z-x))\nonumber\\
		&\equiv&\int d^dx\, \phi(x)\hat G(0)\psi(x)+\int d^dx\,d^dz'\, \phi(x) G^{\mu}(x,z')\partial_{\mu}\psi(z'),\nonumber
	\end{eqnarray}	
	with  $\hat G(0):=\int d^dz\, G(x,z)$ and 
	\begin{equation}\label{T}
		G^\mu(x,z)=\int_0^1\frac{ds}{s^{d+1}}G(x,x+(z-x)/s)(z-x)_\mu. 
	\end{equation}	
	In light of this, 
	\begin{equation}
		(T^{1,0,1,\boldsymbol{0}}_{1,0,1,\mathbf{0}}G)(x,z)=\hat G(0)\delta(x-z)
	\end{equation}	
	and, for $\boldsymbol{p}=(1)$, 
	\begin{equation}
		\big[(T^{1,0,1,\boldsymbol{0}}_{1,0,1,(1)}G)(x,z)\big]_\mu=G^{\mu}(x,z)
	\end{equation}	
	By the same considerations as in \cite[Appendix C]{GMR21}, we have that $T^{1,0,1,\boldsymbol{0}}_{1,0,1,\boldsymbol{p}}$ satisfies the following norm bounds:
	\begin{equation}\label{norm1,0,1}\|T^{1,0,1,\boldsymbol{0}}_{1,0,1,\boldsymbol{p}}G\|\le \|G\|, \qquad \|T^{1,0,1,\boldsymbol{0}}_{1,0,1,(1)}G\|\le \max_{a,b}\int d^dz\, |[G(0,z)]_{a,b}|\, |z|\end{equation}
	and similarly for the weighted $L^1$ norms to be used below. 
	
	\subsubsection{Case \texorpdfstring{$(n,m,l)=(0,1,2)$}{(n,m,l)=(0,1,2)}}	\label{sec.2.2.2}
	Consider now $\int d^dy\, d^dz_1\,d^dz_2\, J(y) \psi(z_1) F(y,\boldsymbol{z})\psi(z_2)$ with $F$ translationally invariant, playing the role of $(H_{\text{eff}})_{(0,1,2,\mathbf{0})}$. Proceeding analogously we find: 
	\begin{eqnarray}\label{TJ}
		&&\int d^dy\,d\boldsymbol{z}\, J(y) \psi(z_1)F(y,\boldsymbol{z})\psi(z_2)\\
		&&=\int d^dy\,d\boldsymbol{z}\, J(y) \Big[\psi(y)F(y,\boldsymbol{z})\psi(y)+(z_1-y)_\mu\int_0^1 ds\, \partial_\mu\psi(y+s(z_1-y))F(y,\boldsymbol{z})\psi(y+s(z_2-y))\nonumber\\
		&&+(z_2-y)_\mu\int_0^1ds\,\psi(y+s(z_1-y))F(y,\boldsymbol{z})\partial_\mu\psi(y+s(z_2-y))\Big]\nonumber\\
		&&\equiv 
		\sum_{\boldsymbol{p}:\,\|\boldsymbol{p}\|_1\le 1}
		\int d^dy\,d\boldsymbol{z}'\, J(y) \partial^{p_1}_{\mu_1}\psi(z_1')\big[F^{\boldsymbol{p}}(y,\boldsymbol{z}')\big]_{\boldsymbol{\mu}}\partial^{p_2}_{\mu_2}\psi(z_2'),
		\nonumber\end{eqnarray}	
	with  $F^{\boldsymbol{0}}(y,\boldsymbol{z})=\delta(z_1-y)\delta(z_2-y)\int d^{2d}\boldsymbol{z}' F(x,\boldsymbol{z}')$, 
	\begin{equation}
		F^{(1,0)}(y,\boldsymbol{z})=(z_1-y)_{\mu}\int_0^1\frac{ds}{s^{2d+1}}F(y,y+(z_1-y)/s,y+(z_2-y)/s),	
	\end{equation}	
	and similarly for $F^{(0,1)}$.
	Then, identifying $\boldsymbol{p}$ with the pair $(p_1,p_2)$:
	\begin{equation}\begin{split}
			&	(T^{0,1,2,\mathbf{0}}_{0,1,2,\mathbf{0}}F)(y,\boldsymbol{z})=\delta(z_1-y)\delta(z_2-y)\int d^{2d}\boldsymbol{z}'\,F(0,\boldsymbol{z}'),\\
			& 	(T^{0,1,2,\mathbf{0}}_{0,1,2,\mathbf{p}}F)(y,\boldsymbol{z})=F^{\boldsymbol{p}}(y,\boldsymbol{z}),\qquad \text{if}\qquad \|\boldsymbol{p}\|_1=1.
		\end{split}
	\end{equation}	 
	In analogy with \eqref{norm1,0,1}, we have: $\|T^{0,1,2,\boldsymbol{0}}_{0,1,2,\boldsymbol{0}}F\|\le \|F\|$, while, if $\boldsymbol{p}=(1,0)$,
	\begin{equation}\label{norm0,1,2} \|T^{0,1,2,\boldsymbol{0}}_{0,1,2,(1,0)}F\|\le \max_{a,b}\int d\boldsymbol{z}\, |[F(0,\boldsymbol{z})]_{a,b}|\, |z_1|,\end{equation}
	and similarly for $\boldsymbol{p}=(0,1)$.
	
	\subsection{Fixed point equation}\label{sect:FPE}
	
	We are now ready to write the definition of the renormalization map $H\to RH=DTS^{(0)}H$ component-wise. Recalling \eqref{HeffS0}-\eqref{Heffell}, the definition of the trimming  map in the previous subsection, and 
	the definition of the dilatation $D$ in  \eqref{eq:scal}, for any $\ell\in L$, we can write 
	\begin{equation}\label{RGmapincomponents}(RH)_{\ell}=\sum_{s\ge 1}\sum_{(\ell_i)_{i=1}^s}R^{\ell_1,\ldots,\ell_s}_{\ell}(H_{\ell_1},\ldots, H_{\ell_s}),\end{equation}
	where the sum runs over $s$-ples of labels in the set $L_f$ defined in \eqref{defL0} and, letting $\mathfrak L=\{(0,0,2),(0,0,4)$, $(1,0,1),(0,1,2)\}$,
	\begin{equation}\label{R}
		R_{\ell}^{\ell_1,\ldots,\ell_s}=D \begin{cases}
			S_{\ell}^{\ell_1,\ldots,\ell_s},&\text{if}\ \ell=(n,m,l,\mathbf{p})\ \text{with}\ (n,m,l)\not\in\mathfrak L,\\
			T_{n,m,l,\mathbf{0}}^{n,m,l,\mathbf{0}}S_{(n,m,l,\boldsymbol{0})}^{\ell_1,\ldots,\ell_s},&\text{if}\ \ell=(n,m,l,\mathbf{0}) \ \text{with}\ (n,m,l)\in\mathfrak L,\\
			\sum_{\boldsymbol{p}'}T_{n,m,l,\boldsymbol{p}}^{n,m,l,\boldsymbol{p}'}S_{(n,m,l,\boldsymbol{p}')}^{\ell_1,\ldots,\ell_s},& \text{if} \ \ell=(n,m,l,\mathbf{p}) \ \text{with}\ (n,m,l)\in\mathfrak L\ \text{and}\ \mathbf{p}\neq\mathbf{0},
		\end{cases}		
	\end{equation}	
	and in the second and third lines the operator $T_{n,m,l,\boldsymbol{p}}^{n,m,l,\boldsymbol{p}'}$ is defined as follows: it is the identity, if $\boldsymbol{p}=\boldsymbol{p}'\neq {\boldsymbol{0}}$; 
	it is the one defined in subsections \ref{sec.2.2.1}-\ref{sec.2.2.2} above, if $(n,m,l)=(1,0,1),(0,1,2)$ and either $\boldsymbol{p}=\boldsymbol{p}'=\boldsymbol{0}$ or 
	$\boldsymbol{p}'=\boldsymbol{0}$, $\|\boldsymbol{p}\|_1=1$; it is given by \cite[eq.s (5.12),(5.13),(5.14)]{GMR21} if $(n,m,l)=(0,0,2)$ and either
	$\boldsymbol{p}=\boldsymbol{p}'=\boldsymbol{0}$ or $\|\boldsymbol{p}'\|_1<\|\boldsymbol{p}\|_1=2$, or if $(n,m,l)=(0,0,4)$ and either
	$\boldsymbol{p}=\boldsymbol{p}'=\boldsymbol{0}$ or $\boldsymbol{p}'=\boldsymbol{0}$, $\|\mathbf{p}\|_1=1$; it vanishes, otherwise. 
	
	In view of these definitions, {and recalling the fact that we look for a fixed point potential with pre-factor in front of the local terms $(\phi,\psi)$ and $(J,\psi^2)$ in $V^*$ equal to $1$
		(see item (2) of Definition \ref{Def.1})}, the fixed point equation (FPE) for the `local' components $\ell\in\{(n,m,l,\boldsymbol{0})\}_{(n,m,l)\in\mathfrak L}$ reads:
	\begin{equation}\label{FPE}\begin{split}
			&\nu=\gamma^{d/2+\varepsilon}\nu+\sum_{s\ge 1}\sum_{(\ell_i)_{i=1}^s}^*R_{(0,0,2,\boldsymbol{0})}^{\ell_1,\ldots,\ell_s}(H_{\ell_1},\ldots,H_{\ell_s})\\
			&\lambda=\gamma^{2\varepsilon}\lambda+\sum_{s\ge 1}\sum_{(\ell_i)_{i=1}^s}^*R_{(0,0,4,\boldsymbol{0})}^{\ell_1,\ldots,\ell_s}(H_{\ell_1},\ldots,H_{\ell_s})\\
			& 1= \gamma^{\Delta_1-[\psi]}+\sum_{s\ge 1}\sum_{(\ell_i)_{i=1}^s}^*R_{(1,0,1,\boldsymbol{0})}^{\ell_1,\ldots,\ell_s}(H_{\ell_1},\ldots,H_{\ell_s})\equiv \gamma^{\Delta_1-[\psi]}(1+\zeta_1) \\
			& 1= \gamma^{\Delta_2-2[\psi]}+\sum_{s\ge 1}\sum_{(\ell_i)_{i=1}^s}^*R_{(0,1,2,\boldsymbol{0})}^{\ell_1,\ldots,\ell_s}(H_{\ell_1},\ldots,H_{\ell_s})\equiv \gamma^{\Delta_2-2[\psi]}(1+\zeta_2)\end{split}
	\end{equation}	
	where, again, the labels $\ell_i$ in the sums in the right hand sides are summed over $L_f$, and the $*$ indicates the constraint that the term with $s=1$ and $\ell_1=\ell$ should be excluded from the corresponding sums. 
	The FPE for the components $(0,0,2,\boldsymbol{p})$ with $\|\boldsymbol{p}\|_1=1$ is by construction trivial, $0=0$, while the FPE for all the other components reads: 
	\begin{equation}
		H_{\ell}=\sum_{s\ge 1}\sum_{(\ell_{i})_{i=1}^s}R_{\ell}^{\ell_1,...,\ell_s}(H_{\ell_1},\ldots,H_{\ell_s}).
		\label{FPEnonloc}
	\end{equation}
	Note that the FPE for the components $\ell=(n,m,l,\boldsymbol{p})$ with $n=m=0$ is the same as the one studied and solved in \cite{GMR21}, the solution being the sequence of kernels $H^*_{l,\boldsymbol{p}}$ 
	constructed there. In particular, the first two components of \eqref{FPE} are solved by the fixed point values $\lambda^*,\nu^*$ of $\lambda,\nu$ computed in \cite{GMR21}, and proved there to be analytic in 
	$\epsilon$ for $\epsilon$ sufficiently small. So, from now on, we will set $\lambda=\lambda^*$ and $\nu=\nu^*$. Moreover, we will prove below that $\zeta_1$ and $\zeta_2$ are sums of convergent series in $\lambda,\nu$, for $\epsilon$ sufficiently small and this implies that they are themselves analytic in $\epsilon$ for $\epsilon$ small. Letting $Z_1:=1+\zeta_1$ and $Z_2:=1+\zeta_2$, the FPE requires 
	\begin{equation}\label{Delta}
		\Delta_1=[\psi]-\log_{\gamma}Z_1, \qquad \Delta_2=2[\psi]-\log_{\gamma}Z_2\equiv 2[\psi]+\eta_2. 
	\end{equation}	 
	We will see in Section \ref{subsecDelta_1} that $\zeta_1=0$, so that $\Delta_1=[\psi]$. Moreover, in Section \ref{subsecDelta_2} and Appendix \ref{appA}, we will show that 
	\begin{equation}Z_2=1-\frac{2(N-2)}{N-8}\epsilon\log\gamma+O(\epsilon^2), \end{equation}
	which gives $\eta_2=2\epsilon\frac{N-2}{N-8}+O(\epsilon^2)$.	
	
	\section{Solution to the FPE via the tree expansion}\label{Sec:3}
	Let us now discuss how to determine the solution to the FPE introduced in the previous section, satisfying the properties spelled out in Definition \ref{Def.1}. 
	We use an analogue of the tree expansion discussed in \cite[App.J]{GMR21}, which provides an explicit analytic solution. 
	Our construction automatically proves the uniqueness of the solution within the class of analytic potentials with a prescribed form at $\epsilon=0$. 	
{We stress that our proof of uniqueness of the fixed point potential within the class of analytic functions via the use of tree expansions does not require that the scaling parameter $\gamma$ is large, as most proofs based on contraction arguments usually do (including the one in \cite[Section 6]{GMR21}); the proof in this and following sections just requires that $\gamma>1$. 
A priori, the radius of convergence may depend upon the choice of $\gamma$; however, the methods in \cite{Gr24}, also based on the tree expansion, should even allow one to 
prove uniformity of the analyticity region as $\gamma\to 1^+$.
These are remarkable advantages of the  tree expansion method as compared to the use of Banach fixed point theorem. The drawback is that uniqueness is guaranteed in 
a much smaller space.}
Uniqueness of the solution in the larger Banach space of potentials 
	that are close enough to the prescribed unperturbed potential in the appropriate weighted $L^1$ norm, could be proved via a contraction argument similar to the one underlying the proof of \cite[Key Lemma]{GMR21}, {under more restrictive assumptions on the scaling parameter $\gamma$}, but we will not belabor the details of this proof here. {Of course, 
	we do not expect that the fixed point depends upon $\gamma$ at all, but this remains to be proved, and we plan to do this in a third paper of this series; see also \cite[Section 6.4]{GMR21}.}

\medskip

	As already mentioned above, we fix $\lambda=\lambda^*$, $\nu=\nu^*$, the fixed point values of the quartic and quadratic interactions computed in \cite{GMR21}, once and for all. We also recall that our ansatz for 
	the fixed point potential $V^*$ requires that the local terms $(\phi,\psi)$ and $(J,\psi^2)$ have pre-factor equal to $1$, see condition (2) in Definition \ref{Def.1}, and that the 
	components $(0,0,2,\boldsymbol{p})$ with $\|\boldsymbol{p}\|_1=1$ are zero, by the requirement that $\{V^*_\ell\}_{\ell\in L}$ is trimmed, see the second line of \eqref{trimm}. 
	We first describe how to solve the FPE for the 
	remaining components, i.e., those in 
	\begin{equation} L':=L\setminus \Big(\{(n,m,l,\boldsymbol{0})\}_{(n,m,l)\in\mathfrak L}\cup \{(0,0,2,\boldsymbol{p})\}_{\|\boldsymbol{p}\|_1=1}\Big), \end{equation}
	and then we will discuss the local components $\ell=(1,0,1,\boldsymbol{0}),(0,1,2,\boldsymbol{0})$. 
	
	\subsection{The FPE for \texorpdfstring{$\ell\in L'$}{ell in L'}}\label{Caso1}
	
	The FPEs for the components $\ell\in L'$ are those in \eqref{FPEnonloc}. We proceed in a way similar to \cite[Appendix J]{GMR21}. 
	For any $\ell\in L'$, we isolate from the right side of the FPE the term with $s=1$ and $\ell_1=\ell$, i.e. the term $DH_{\ell}$, move it to the left side, and multiply both sides by $(1-D)^{-1}$. 
	The resulting equation takes the form
	\begin{equation}\label{1}
		H_{\ell}=\sum_{s\ge1}\sum_{(\ell_i)_{i=1}^s}^*(1-D)^{-1}R_{\ell}^{\ell_1,\ldots,\ell_s}(H_{\ell_1},\ldots,H_{\ell_s})	
	\end{equation}	
	where we recall that the second sum runs over $s$-ples of labels in $L_f$, see \eqref{defL0}, and $*$ denotes the constraint that, if $s=1$, then $\ell_1\ne\ell$, while $R_{\ell}^{\ell_1,\ldots,\ell_s}$ is defined in \eqref{R}. As shown below, the (diagonal in $\ell$) operator $(1-D)$ is invertible in $L^1$
	on all the components $\ell\in L'$: this is immediate for the components such that the scaling dimension at $\epsilon=0$ is different from zero, $D_{\text{sc}}(\ell)\big|_{\epsilon=0}\neq0$, see \eqref{eq:ScDim1}, that is, for all the indices in $L'$ but $(0,2,0,\emptyset)$. In order for $(1-D)$ to be invertible on this component as well, we need extra information about $\Delta_2$, besides knowing that $\Delta_2=2[\psi]+O(\epsilon)$, as we are assuming (see condition (3) in Definition \ref{Def.1}). Anticipating the fact that 
	$\Delta_2=2[\psi]+2\epsilon\frac{N-2}{N-8}+O(\epsilon^2)$, the desired invertibility for $\epsilon\neq 0$ follows, because this explicit expression for $\Delta_2$ implies that $D_{\text{sc}}(0,2,0,\emptyset)=2\epsilon\frac{N+4}{N-8}+O(\epsilon^2)$, which is different from zero for $\epsilon\neq0$.
	
	We look for a solution in the form of a sum over rooted trees, 
	\begin{equation}\label{T1}
		V^*_{\ell}=\sum_{\tau}H_{\ell}[\tau],
	\end{equation}	 
	The value $H_\ell[\tau]$ is fixed so that the following recursive equation is satisfied: 
	\begin{equation}\label{T2}
		H_{\ell}[\tau]=\sum_{(\ell_i)_{i=1}^{s_{v_0}}}^*(1-D)^{-1}R_{\ell}^{\ell_1,...,\ell_{s_{v_0}}}(H_{\ell_1}[\tau_1],...,H_{\ell_n}[\tau_{s_{v_0}}]).
	\end{equation}
	where $\tau_1,\ldots,\tau_{s_{v_0}}$ are the subtrees of $\tau$ rooted in the vertices $v_1,\ldots, v_{s_{v_0}}$ that are `children' of the root vertex $v_0$ of $\tau$. 
	The rooted trees (with root $v_0$) involved in the sum \eqref{T1} have the structure exemplified in Fig.\ref{eq:figtree3.0}. 
	\begin{figure}[ht]
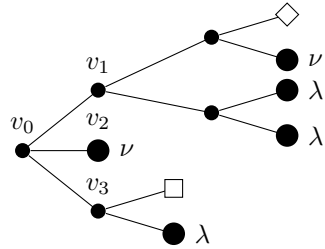

		\begin{center}
			\begin{tabular}{rcl}
				\tikz[baseline=-2pt]{ \draw (0,0) node [vertex, label=$v_0$] (v0) {} -- +(1,0.8) node [vertex, label=$v_1$] (v1) {} ;
					\draw (v0) --+(1,0) node[bigvertex,label=right:$\nu$, label=$v_2$] (v7){}; 
					\draw (v1) -- +(1.5,0.7) node [vertex] (v2) {};
					\draw (v1) -- +(1.5,-0.3) node [vertex] (v4) {};
					\draw (v2) -- +(1,0.3) node [ctVertex] (v3) {};
					\draw (v2) -- +(1,-0.3) node[bigvertex,label=right:$\nu$] (v5) {};
					\draw (v4) --+(1,-0.3) node[bigvertex, label=right:$\lambda$] {};
					\draw (v4) --+(1,+0.3) node[bigvertex,, label=right:$\lambda$] {};
					\draw (v0) --+(1.,-0.8)node[vertex, label=$v_3$](v6){};		
					\draw (v6)--+(1,0.3)node[E]{}(v7);   
					\draw (v6)--+(1,-0.3)node[bigvertex,label=right:$\lambda$]{};
				}
			\end{tabular}
		\end{center}\caption{{An example of a tree rooted in $v_0$, contributing to the sum in \eqref{T1}. Here $s_{v_0}=3$ and the 3 children of $v_0$ are $v_1,v_2,v_3$. }}
		\label{eq:figtree3.0}
	\end{figure}
	
	The iterative application of \eqref{T2} leads to a representation of the tree value $H_\ell[\tau]$ in terms of an iterated action of the operator $(1-D)^{-1}R_{\ell_v}^{\ell_{v_1},\ldots,\ell_{s_v}}$, one per vertex $v$, summed over the labels $\ell_v$, $\ell_{v_i}$ (here the labels $\ell_{v_i}$ at exponent refer to the vertices $v_1,\ldots, v_{s_v}$, which are the children of $v$ on $\tau$; the label $\ell_{v_0}$ associated with the root $v_0$ is the only one that is kept fixed, all the others are summed over). The labels $\ell_v$ associated with vertices $v\neq v_0$ that are not endpoints are summed over $L_{f}':=L_f\cap L'$, 
	while, if $v$ is an endpoint, then $\ell_v$ takes one of the values in $\{(n,m,l,\boldsymbol{0})\}_{(n,m,l)\in\mathfrak L}$, depending on the nature of the endpoint, as graphically described in Fig.\ref{eq:figtree3.1}. 
	\begin{figure}[ht]
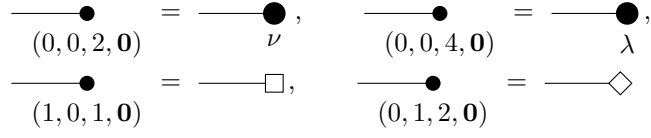

		\begin{center}
			\begin{tabular}{rcl}
				&&	\tikz[baseline=-2pt]{ \draw (0,0) node {} -- (1,0) node[vertex, label=below:{$(0,0,2,\boldsymbol{0})$}] {} ; }
				$=$
				\tikz[baseline=-2pt]{ \draw (0,0) node {} -- (1,0) node[bigvertex, label=below:{$\nu$}] {} ; }\,,\qquad
				
				\tikz[baseline=-2pt]{ \draw (0,0) node {} -- (1,0) node[vertex, label=below:{$(0,0,4,\boldsymbol{0})$}] {} ; }
				$=$
				\tikz[baseline=-2pt]{ \draw (0,0) node {} -- (1,0) node[bigvertex, label=below:{$\lambda$}] {} ; },\\
				
				&&	\tikz[baseline=-2pt]{ \draw (0,0) node {} -- (1,0) node[vertex, label=below:{$(1,0,1,\boldsymbol{0})$}] {} ; }
				$=$
				\tikz[baseline=-2pt]{ \draw (0,0) node {} -- (1,0) node[E] (v0) {} ; }\,,\qquad
				
				\tikz[baseline=-2pt]{ \draw (0,0) node {} -- (1,0) node[vertex, label=below:{$(0,1,2,\boldsymbol{0})$}] {} ; }
				$=$
				\tikz[baseline=-2pt]{ \draw (0,0) node {} -- (1,0) node[ctVertex] {} ; }		
			\end{tabular}
		\end{center}\caption{{The four types of endpoints and the corresponding $\ell$ labels.}}
		\label{eq:figtree3.1}
	\end{figure}
	
	In the evaluation of the tree value $H_\ell[\tau]$, each of these endpoints is associated with the kernel of the corresponding `interaction vertex', see Fig.\ref{eq:fig3.2}. 
	
	\begin{figure}[ht]
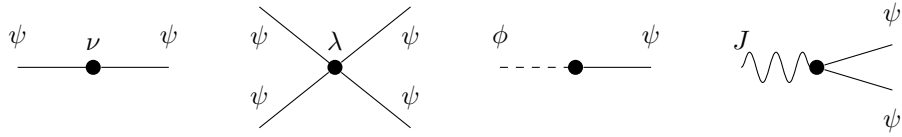

		\begin{center}
			\begin{tabular}{rcl}
				\tikz[baseline=-2pt]{\draw (0,0) node [label=above:{$\psi$}] {} -- (1,0) node [vertex, label={$\nu$}] (u0) {};
					\draw (u0) -- (2,0) node [label=above:{$\psi$}] {} ;}
				$\;\;\;\;\;$
				\tikz[baseline=-2pt]{\draw (0,0.8) node [label=below:{$\psi$}] {}  -- (1,0)node [vertex, label={$\lambda$}] (w0) {};
					\draw (w0) -- (2,0.8) node [label=below:{$\psi$}] {};
					\draw (w0) -- (0,-0.8) node [label=above:{$\psi$}]  {};
					\draw (w0) -- (2,-0.8) node [label=above:{$\psi$}]  {}; }
				$\;\;\;\;\;$
				\tikz[baseline=-2pt]{\draw[dashed](0,0) node [label=above:{$\phi$}] {}--(1,0) node [vertex] (t0) {};
					\draw (t0) -- (2,0) node [label=above:{$\psi$}] {};
				}
				$\;\;\;\;\;$
				\tikz[baseline=-2pt]{\draw [decorate,decoration={snake,amplitude=2mm}](0,0) node [label=above:{$J$}]  {} -- (1,0) node[vertex] (v0) {};
					\draw (v0) node [] {} -- +(1,+0.3) node [label=above:{$\psi$}]  (v1) {};
					\draw (v0) node [] {} -- +(1,-0.3) node [label=below:{$\psi$}]  (v2) {};}
				
			\end{tabular}
		\end{center}\caption{{The four `interaction vertices', graphically representing the contributions associated with the four types of endpoints depicted in Fig.\ref{eq:figtree3.1}.}}
		\label{eq:fig3.2}
	\end{figure}
	
	For example, for the tree $\tau$ represented in Fig.\ref{eq:figtree3.0}, $s_{v_0}=3$ and the three subtrees $\tau_1,\tau_2,\tau_3$ `exiting' from the root $v_0$ are those represented in Fig.\ref{eq:figtree3.3} (note that $\tau_2$ is `trivial', in that it consists of a single vertex, which is both the root and the endpoint of $\tau_2$).
	
	\begin{figure}[ht]
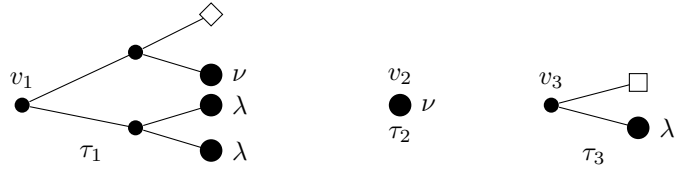

		\begin{center}
			\begin{tabular}{rcl}			
				\tikz[baseline=-2pt]{  \draw (0,0) node [vertex,label=$v_1$] (v1) {} ;
					\draw (v1) -- +(1.5,0.7) node [vertex] (v2) {};
					\draw (v1) -- +(1.5,-0.3) node [vertex,label=below:$\tau_1\;\;\;\;\;\;\;\;\;\;\;\;$] (v4) {};
					\draw (v2) -- +(1,0.5) node [ctVertex ] (v3) {};
					\draw (v2) -- +(1,-0.3) node[bigvertex,label=right:$\nu$] (v5) {};
					\draw (v4) --+(1,-0.3) node[bigvertex,label=right:$\lambda$] {};
					\draw (v4) --+(1,+0.3) node[bigvertex,label=right:$\lambda$] {};
					\draw (5,0) node[bigvertex,label=$v_2$,label=below:$\tau_2$,label=right:$\nu$ ]{};
					\draw (7,0)node[vertex,label=$v_3$](v6){};		
					\draw (v6)--+(1.15,0.3)node[E]{}(v7);   
					\draw (v6)--+(1.15,-0.3)node[bigvertex, label=below:$\tau_3\;\;\;\;\;\;\;\;\;\;\;\;$,label=right:$\lambda$]{};}
			\end{tabular}
		\end{center}\caption{The three subtrees $\tau_1, \tau_2, \tau_3$ exiting from the root $v_0$ of the tree $\tau$ represented in Fig.\ref{eq:figtree3.0}.}
		\label{eq:figtree3.3}
	\end{figure}
	
	We now intend to use the formula for the tree values described above in order to derive norm bounds on $H_\ell[\tau]$ and to prove that the tree expansion for $H_\ell$ is absolutely convergent in the appropriate norms.
	We start by discussing bounds for the following \textbf{weighted $L_1$ norm} of $H_\ell[\tau]$, generalizing the definition in \eqref{finiteL1}. 
	Recall that $H_{\ell}[\tau](\boldsymbol{x},\boldsymbol{y},\boldsymbol{z})$ with $\ell=(n,m,l,\boldsymbol{p})$ must be understood as a tensor-valued function of components 
	$\big[H_{\ell}[\tau](\boldsymbol{x},\boldsymbol{y},\boldsymbol{z})\big]_{\boldsymbol{a},\boldsymbol{\mu},\boldsymbol{b}}$, as in \eqref{tot}. For notational convenience, in some of the equations below, we will 
	equivalently rewrite these components as $H_{\ell}[\tau](\boldsymbol{a},\boldsymbol{x},\boldsymbol{y},\mathbf{B},\boldsymbol{z})$, with  $\mathbf{B}=((p_1,\mu_1,b_1),\ldots,(p_l,\mu_l,b_l))$, as in Remark \ref{remark:notation} (the label $\ell$ attached to $H_{\ell}[\tau](\boldsymbol{a},\boldsymbol{x},\boldsymbol{y},\mathbf{B},\boldsymbol{z})$ is actually redundant, but we will keep it for clarity); given such a $\mathbf{B}$, 
	we shall write $\boldsymbol{p}(\mathbf{B})=\boldsymbol{p}$. In analogy with \eqref{finiteL1}, we let 
	\begin{equation}\label{norm}
		\|H_{\ell}[\tau]\|_{w}:=\max_{\substack{\boldsymbol{a}, \mathbf{B}\, :\\ \boldsymbol{p}(\boldsymbol{B})=\boldsymbol{p}}}\iiint^* d{\boldsymbol{x}}\, d\boldsymbol{y}\, d\boldsymbol{z}|H_{\ell}[\tau](\boldsymbol{a},\boldsymbol{x},\boldsymbol{y},\mathbf{B},\boldsymbol{z})|w(\boldsymbol{x},\boldsymbol{y},\boldsymbol{z})
	\end{equation}	 
	where
	\begin{equation}\label{w}
		w(\boldsymbol{x},\boldsymbol{y},\boldsymbol{z}):=e^{\bar{C}(St(\boldsymbol{x},\boldsymbol{y},\boldsymbol{z})/\gamma)^{\sigma}},
	\end{equation}	
	with: $St(\boldsymbol{x},\boldsymbol{y},\boldsymbol{z})$ the Steiner diameter, or `tree distance', of $(\boldsymbol{x},\boldsymbol{y},\boldsymbol{z})$, see \cite[footnote 19]{GMR21}; $\bar C\equiv\frac12 C_{\chi^2}$ with 
	$C_{\chi^2}$ the positive constant in \eqref{M} below; and $\sigma=1/s$ with $s$ the Gevrey regularity of $\chi$, see the line after \eqref{defchi}. 
	
	In order to recursively estimate the norm \eqref{norm} of $H_\ell[\tau]$ via \eqref{T2}, note that, from \eqref{HBxy} and \eqref{R}, if $\ell=(n,m,l,\mathbf{p})\in L'$ with $(n,m,l)\not\in\mathfrak L$, we can write 
	\begin{equation}\begin{split}
			& H_{\ell}[\tau](\boldsymbol{a},\boldsymbol{x},\boldsymbol{y},\mathbf{B},\boldsymbol{z}_{\mathbf{B}})\\
			&\ =\frac{D}{1-D}\mathcal{P}\frac{1}{s_{v_0}!}\sum_{\substack{\mathbf{B}_1,\ldots,\mathbf{B}_{s_{v_0}}\\\sum\mathbf{B}_i=\mathbf{B}}}\sum_{\substack{\mathbf{A}_1,\ldots,\mathbf{A}_{s_{v_0}}\\\mathbf{A}_i\supset\mathbf{B}_i}}\sum_{\substack{\boldsymbol{a}_1,\ldots,\boldsymbol{a}_{s_{v_0}}\\ \boldsymbol{x}_1,\ldots,\boldsymbol{x}_{s_{v_0}}\\ \boldsymbol{y}_1,\ldots,\boldsymbol{y}_{s_{v_0}}}}
			(-)^{\sharp}\int d\boldsymbol{z}_{\bar{\mathbf{B}}}\mathcal{C}(\boldsymbol{z}_{\bar{\mathbf{B}}})\prod_{i=1}^{s_{v_0}}H_{\ell_i}[\tau_i](\boldsymbol{a}_i,\boldsymbol{x}_i,\boldsymbol{y}_i,\mathbf{A}_i,\boldsymbol{z}_{\mathbf{A}_i}),\label{eh.1}	\end{split}
	\end{equation}		
	where $\ell_i\equiv\ell_{v_i}$, and we used that for these components the trimming map is the identity. If, instead, $\ell=(n,m,l,\boldsymbol{p})\in L'$ with $(n,m,l)\in \mathfrak L$ and $\boldsymbol{p}\neq 0$, 
	then trimming acts non-trivially, and the FPE reads: 
	\begin{equation}\begin{split}
			H_{n,m,l,\boldsymbol{p}}[\tau](\boldsymbol{a},\boldsymbol{x},\boldsymbol{y},\mathbf{B},\boldsymbol{z}_{\mathbf{B}})
			&=\frac{D}{1-D}\sum_{\boldsymbol{p}'}T_{n,m,l,\boldsymbol{p}}^{n,m,l,\boldsymbol{p}'}\mathcal{P}\frac{1}{s_{v_0}!}\sum_{\substack{\mathbf{B}_1,\ldots,\mathbf{B}_{s_{v_0}}\\\sum\mathbf{B}_i=\mathbf{B}'}}\sum_{\substack{\mathbf{A}_1,\ldots,\mathbf{A}_{s_{v_0}}\\\mathbf{A}_i\supset\mathbf{B}_i}}\sum_{\substack{\boldsymbol{a}_1,\ldots,\boldsymbol{a}_{s_{v_0}}\\ \boldsymbol{x}_1,\ldots,\boldsymbol{x}_{s_{v_0}}\\ \boldsymbol{y}_1,\ldots,\boldsymbol{y}_{s_{v_0}}}}\times\\
			&\times
			(-)^{\sharp}\int d\boldsymbol{z}_{\bar{\mathbf{B}}}\mathcal{C}(\boldsymbol{z}_{\bar{\mathbf{B}}})\prod_{i=1}^{s_{v_0}}H_{\ell_i}[\tau_i](\boldsymbol{a}_i,\boldsymbol{x}_i,\boldsymbol{y}_i,\mathbf{A}_i,\boldsymbol{z}_{\mathbf{A}_i}),\end{split}\label{eh.2}
	\end{equation}		
	where, given $\mathbf{B}=((p_1,\mu_1,b_1),\ldots,(p_l,\mu_l,b_l))$ and an $l$-ple $\boldsymbol{p}'$ such that 
	$T_{n,m,l,\boldsymbol{p}}^{n,m,l,\boldsymbol{p}'}\neq0$, the set $\mathbf{B}'$ appearing in the right side in the condition $\sum\mathbf{B}_i=\mathbf{B}'$ is equal to: 
	$\mathbf{B}$ if $\boldsymbol{p}'=\boldsymbol{p}$; 
	$\mathbf{B}_0:=((0,0,b_1),\ldots,(0,0,b_l))$ if $\boldsymbol{p}'=\boldsymbol{0}$; $\mathbf{B}_{(1,0)}=((1,\mu_1,b_1),(0,0,b_2))$ if $l=2$, $\boldsymbol{p}=(1,1)$ and $\boldsymbol{p}'=(1,0)$; and
	$\mathbf{B}_{(0,1)}=((0,0,b_1),(1,\mu_2,b_2))$ if $l=2$, $\boldsymbol{p}=(1,1)$ and $\boldsymbol{p}'=(0,1)$ (these two last cases are the only ones where $\boldsymbol{p}'\neq\boldsymbol{0},\boldsymbol{p}$).
	
	\medskip
	
	Now, in order to bound the right sides of \eqref{eh.1} and \eqref{eh.2} we recall a few basic bounds from \cite{GMR21} (or their analogues adapted to the present more general context). First of all, as in \cite[(5.37)]{GMR21}, for $\ell\in L'$, 
	\begin{equation}\label{D}
		\| D(H_{\ell}[\tau])\|_{w}\le \gamma^{D_{\text{sc}}(\ell)} \| H_{\ell}[\tau]\|_{w(\cdot/\gamma)}\le \gamma^{D_{\text{sc}}(\ell)} \| H_{\ell}[\tau]\|_{w}, \end{equation}
	with $D_{\text{sc}}(\ell)$ as in \eqref{eq:ScDim1}. Note that, assuming $\Delta_1$ and $\Delta_2$ to be $\epsilon$-close to $[\psi]$ and $2[\psi]$, respectively, with $\epsilon$ small enough, 
	then $\min_{\ell\in L'\setminus \{(0,2,0,\mathbf{0})\}}
	|D_{\text{sc}}(\ell)|=\min\{1,d/2\}+O(\epsilon)$. Therefore, we also have that, for $\ell\in L'\setminus\{(0,2,0,\emptyset)\}$,
	\begin{equation}\label{1-D}
		\| (1-D)^{-1}(H_{\ell}[\tau])\|_{w}\le d_\gamma  \| H_{\ell}[\tau]\|_{w}, \end{equation}
	with $d_\gamma =\max_{\ell\in L'\setminus\{(0,2,0,\emptyset)\}}\big|1-\gamma^{D_{\text{sc}}(\ell)}\big|^{-1}$, which is finite and bounded from above, uniformly in $\epsilon$. On the other hand, recalling that $\Delta_2=2[\psi]+\eta_2$ with $[\psi]=d/4-\epsilon/2$, as in \eqref{Delta},
	\begin{equation}\label{1-D0200}
		\| (1-D)^{-1}(H_{(0,2,0,\mathbf{0})}[\tau])\|_{w}\le \alpha_\gamma \| H_{(0,2,0,\emptyset)}[\tau]\|_{w}, \end{equation}
	where $\alpha_\gamma:=\big|1-\gamma^{-2\epsilon+2\eta_2}\big|^{-1}$, which is finite and of order $\epsilon^{-1}$ iff $\epsilon\neq0$ and $\eta_2\neq \epsilon$. From the first order computation of $\eta_2$ in Appendix \ref{appA}, which implies $\eta_2=2\epsilon\frac{N-2}{N-8}+O(\epsilon^2)$, see \eqref{FO}, we see that this condition is always verified 
	for $\epsilon\neq0$ small enough.
	
	Concerning the action of the trimming operator, for any $\ell=(n,m,l,\boldsymbol{p})\in L'$ and $\boldsymbol{p}'\neq\boldsymbol{p}$ such that $T_{n,m,l,\boldsymbol{p}}^{n,m,l,\boldsymbol{p}'}$ does not vanish,
	we have the analogue of \cite[(5.43)]{GMR21}: 
	\begin{equation}\label{Tr}
		\|T_{n,m,l,\boldsymbol{p}}^{n,m,l,\boldsymbol{p}'}H_{(n,m,l,\boldsymbol{p}')}\|_{w(\cdot/\gamma)}\le C_R\gamma^{\|\boldsymbol{p}-\boldsymbol{p}'\|_1}\|H_{(n,m,l,\boldsymbol{p}')}\|_w.
	\end{equation}	
	For the estimate of $S_{\ell}^{\ell_1,\ldots,\ell_s}$, with $\ell\in L'$, $\ell_1,\ldots,\ell_s\in L_f$, we proceed as follows.  From  \cite[Eq.(5.39)]{GMR21}, we see that $\mathcal{C}(\boldsymbol{z})$ satisfies: 
	\begin{equation}\label{|C|}
		|\mathcal{C}(\boldsymbol{z})|\le	C_{GH}^s\sum_{\mathcal{T}}\prod_{(z,z')\in\mathcal{T}}M(z-z')
	\end{equation}
	where $C_{GH}$ constant given by the Gram-Hadamard bound  \cite[Lemma D.2]{GMR21} and $M$ as in \cite[Eq.(4.15)]{GMR21}, i.e., such that:
	\begin{equation}\label{M}
		|g_{a,b}^{(0)}(x)|,|\partial_\nu g_{a,b}^{(0)}(x)|,|\partial_{\mu,\nu}g_{a,b}^{(0)}(x)|\le M(x)\equiv C_{\chi^1}e^{-C_{\chi^2}|x/\gamma|^{\sigma}},
	\end{equation}	
	where $C_{\chi^1},C_{\chi^2}$ are constants depending on $\chi$ but independent of $\gamma$ and $\sigma=1/s\in(0,1)$, with $s$ the Gevrey regularity of $\chi$, see the line after \eqref{defchi}. 
	
	Using these estimates, and proceeding as in \cite[Sect. 5.6]{GMR21} and \cite[App.E]{GMR21}, we get, letting $\ell=(n,m,l,\mathbf{p})$ and $\ell_i=(n_i,m_i,l_i,\mathbf{p}_i)$:
	\begin{equation}\label{S}
		\|S_{\ell}^{\ell_1,\ldots,\ell_{s_{v_0}}}(H_{\ell_1}[\tau_1],\ldots,H_{\ell_{s_{v_0}}}[\tau_{s_{v_0}}])\|_w\le
		C_{\gamma}^{{s_{v_0}}-1} C_0^{\sum_{i=1}^{s_{v_0}}l_i-l}N_\ell^{\ell_1,\ldots,\ell_{s_{v_0}}}\prod_{i=1}^{s_{v_0}} \|H_{\ell_i}[\tau_i]\|_{w},
	\end{equation}	
	where $N_{\ell}^{\ell_1,\ldots,\ell_{s_{v_0}}}$ is the number of ways in which $\ell=(n,m,l,\boldsymbol{p})$ can be realized, given $\ell_1,\ldots,\ell_{s_{v_0}}$, via the action of 
	$S_{\ell}^{\ell_1,\ldots,\ell_{s_{v_0}}}$ on $(H_{\ell_1}[\tau_1],\ldots,H_{\ell_{s_{v_0}}}[\tau_{s_{v_0}}])$, which is such that 
	\begin{equation}\label{sumoverp0}\sum_{\boldsymbol{p}}
		N_{(n,m,l,\boldsymbol{p})}^{\ell_1,\ldots,\ell_{s_{v_0}}}\le \binom{\sum_{i=1}^{s_{v_0}} l_i}{l}.\end{equation}
	Moreover,  
	$C_{\gamma}=N^2d^2\|M\|_{w}=\mathrm{Cost}\cdot\gamma^d$ and $C_0$ a constant independent of $\gamma$, see \cite[Remark J.1]{GMR21}. Recall also that $S_\ell^{\ell_1,\ldots,\ell_{s_{v_0}}}$ 
	is non-zero only if $\sum_{i=1}^{s_{v_0}} l_i\ge l+2(s_{v_0}-1)$, $\sum_{i=1}^{s_{v_0}} n_i=n$ and $\sum_{i=1}^{s_{v_0}} m_i=m$. 
	
	Plugging eqs.\eqref{D}--\eqref{Tr} and \eqref{S} into \eqref{T2}, with $R_\ell^{\ell_1,\ldots,\ell_{s_{v_0}}}$ defined as in \eqref{R}, we obtain: 
	\begin{eqnarray}\label{eq:4.30_1}
		\|H_{\ell}[\tau]\|_w&\le& \sum_{(\ell_i)_{i=1}^{s_{v_0}}}^*\bigg\|\frac{D}{1-D}TS^{\ell_1,\ldots,\ell_{s_{v_0}}}	_{\ell}(H_{\ell_1}[\tau_1],\ldots,H_{\ell_{s_{v_0}}}[\tau_{s_{v_0}}])\bigg\|_w	\\
		&\le&(\alpha_\gamma/d_\gamma)^{\mathds 1(\ell=(0,2,0,\emptyset))}\sum_{(\ell_i)_{i=1}^{s_{v_0}}}^*d_\gamma\gamma^{D_{\text{sc}}(\ell)}C_R\gamma^2 C_{\gamma}^{{s_{v_0}}-1} C_0^{\sum_{i=1}^{s_{v_0}}l_i-l}{\mathcal N}_\ell^{\ell_1,\ldots,\ell_{s_{v_0}}}\prod_{i=1}^{s_{v_0}} \|H_{\ell_i}[\tau_i]\|_{w},
		\nonumber\end{eqnarray}
	where ${\mathcal N}_{\ell}^{\ell_1,\ldots,\ell_{s_{v_0}}}$ denotes the number of ways in which $\ell=(n,m,l,\boldsymbol{p})$ can be realized, given $\ell_1,\ldots,\ell_{s_{v_0}}$, via the action of 
	$R_{\ell}^{\ell_1,\ldots,\ell_{s_{v_0}}}$ on $(H_{\ell_1}[\tau_1],\ldots,H_{\ell_{s_{v_0}}}[\tau_{s_{v_0}}])$. Recalling the definition of $R_\ell^{\ell_1,\ldots,\ell_s}$ in \eqref{R}, and in the particular the definition  
	of the components $T_{n,m,l,\boldsymbol{p}}^{n,m,l,\boldsymbol{p}'}$ of the trimming operator given right below \eqref{R}, as well as the bound on $N_{\ell}^{\ell_1,\ldots,\ell_{s_{v_0}}}$
	stated right after \eqref{S}, we find: 
	\begin{equation}\label{sumoverp}\begin{split}\sum_{\boldsymbol{p}}^*\mathcal N_{(n,m,l,\boldsymbol{p})}^{\ell_1,\ldots,\ell_{s_{v_0}}}& \le \binom{\sum_{i=1}^{s_{v_0}} l_i}{l}\cdot\begin{cases} 1 & 
				\text{if}\  (n,m,l)\not\in\mathfrak L\\
				4 & \text{if} \ (n,m,l)\in\mathfrak L  \end{cases}\\
			& \le 4 \binom{\sum_{i=1}^{s_{v_0}} l_i}{l},\end{split} \end{equation}
	where $\sum^*_{\boldsymbol{p}}$ denotes the sum over the $\boldsymbol{p}$'s such that $(n,m,l,\boldsymbol{p})\in L'$, and $4$ is the maximum number of different $\boldsymbol{p}$'s for which, given 
	$(n,m,l)\in\mathfrak{L}$ and $\boldsymbol{p}'\in\{0,1\}^l$, the tuple $(n,m,l,\boldsymbol{p})$ is in $L'$ and the operator 
	$T^{n,m,l,\boldsymbol{p}'}_{n,m,l,\boldsymbol{p}}$ is non zero (such maximum is realized for $(n,m,l)=(0,0,4)$ and $\boldsymbol{p}'=\boldsymbol{0}$; in this case, the different $\boldsymbol{p}$'s with the 
	stated property are: $(1,0,0,0)$, $(0,1,0,0)$, $(0,0,1,0)$, $(0,0,0,1)$). 
	
	By using iteratively the estimates above, and recalling that $|\lambda|,|\nu|\le K\epsilon$ for some $K>0$, we find 
	\begin{equation} \begin{split}
			\|H_\ell[\tau]\|_w\le (\alpha_\gamma/d_\gamma)^{\mathds 1(\ell=(0,2,0,\emptyset))}
			\sum_{\{\ell_v\}} &\Big(\prod_{v\ \text{not e.p.}}
			d_{\gamma}C_{\gamma}^{s_v-1}\gamma^{D_{\text{sc}}(\ell_v)} C_R\gamma^2C_0^{\sum_{i=1}^{s_v} l_{v_i}-l_v}{\mathcal N}_{\ell_v}^{\ell_{v_1},\ldots,\ell_{v_{s_v}}}
			\Big)\\
			\cdot&\Big(\prod_{v\ \text{e.p.}}(K\epsilon)^{\delta_{n_v+m_v,0}}\Big)
	\end{split}\end{equation}		
	where the  sum over $\{\ell_v\}$ in the right hand side runs over $L'_f=L'\cap L_f$ for each $\ell_v$ associated with one the vertices of the tree other than the root $v_0$ and the endpoints; moreover, 
	we denoted $\ell_v=(n_v,m_v,l_v,\mathbf{p}_v)$, and 
	$v_i$ is the $i$-th child of $v$. Let us note that, thanks to the definition of the trimming map, $D_{\text{sc}}(\ell_v)\le -\frac{l_v}{12}<0$ for all $\ell_v\in L_f'$. If we now first sum over 
	the choices of $\boldsymbol{p}_v$, given $(n_v,m_v,l_v)$, for all the vertices $v$ of $\tau$ other than the root $v_0$ and the endpoints, using \eqref{sumoverp} we get: 
	\begin{equation} \begin{split}\label{4.20}
			\|H_\ell[\tau]\|_w\le (\alpha_\gamma/d_\gamma)^{\mathds 1(\ell=(0,2,0,\emptyset))}\gamma^{D_{\text{sc}}(\ell)+l/12}
			\sum_{\{l_v\}} &\Big(\prod_{v\ \text{not e.p.}} K'
			C_{\gamma}^{s_v-1}\gamma^{-l_v/12} C_0^{\sum_{i=1}^{s_v} l_{v_i}-l_v}
			\binom{\sum_{i=1}^{s} l_{v_i}}{l_v}\Big)\\
			\cdot&\Big(\prod_{v\ \text{e.p.}}(K\epsilon)^{\delta_{n_v+m_v,0}}\Big)
	\end{split}\end{equation}		
	where: $\ell\equiv\ell_{v_0}=(n,m,l,\boldsymbol{p})$; the sum over $\{l_v\}$ runs over the positive integers, $l_v\ge 1$, for all the vertices $v$ of $\tau$ other than the root $v_0$ and the endpoints; and $K'=4d_{\gamma}C_R\gamma^2$.  Now, letting $n_{\text{e.p.}}[\tau]$ be the number of endpoints of $\tau$, we have that:
	\begin{enumerate}
		\item the number of vertices of $\tau$ that are not endpoints is $\le 2n_{\text{e.p.}}[\tau]$ (see \cite[eq.(J.10)]{GMR21});
		\item  $\prod_{v\ \text{not e.p.}} C_{\gamma}^{s_v-1}=C_{\gamma}^{n_{\text{e.p.}}[\tau]-1}$; 
		\item $\prod_{v\ \text{not e.p.}} C_0^{\sum_{i=1}^{s_v} l_{v_i}-l_v}=C_0^{-l}\prod_{v \ \text{e.p.}}C_0^{l_v}\le C_0^{4n_{\text{e.p.}}[\tau]-l}$; 
		\item $\sum_{\{l_v\}}\prod_{v\ \text{not e.p.}} 
		\gamma^{-l_v/12}\binom{\sum_{i=1}^{s} l_{v_i}}{l_v}\le (1-\gamma^{-1/12})^{-4n_{\text{e.p.}}[\tau]}$ (see \cite[Appendix A.6.1]{GeM01}). 
	\end{enumerate}
	Therefore, for any $\gamma>1$,
	\begin{equation}\label{TreeBound1}\|H_\ell[\tau]\|_w\le
		(\alpha_\gamma/d_\gamma)^{\mathds 1(\ell=(0,2,0,\emptyset))}\gamma^{D_{\text{sc}}(\ell)}C_\gamma^{-1}(\gamma^\frac1{12}/C_0)^{l}
		(K\epsilon)^{-(n+m)}\left[\left(\frac{C_0}{1-\gamma^{-\frac1{12}}}\right)^4(K')^2C_{\gamma}K\epsilon\right]^{n_{\text{e.p.}}[\tau]}\!\!\!\!\!.
	\end{equation}	 
	In view of \eqref{TreeBound1}, recalling that the number of the trees with $k$ endpoints is less than $4^k$, see e.g. \cite[Lemma A.1]{GeM01}, 
	the sum over trees in the right hand side of 
	\eqref{T1} converges absolutely for any $\gamma>1$ and $\epsilon$ small enough in the weighted $L^1$ norm \eqref{norm}.
	
	\begin{remark}\label{ciao}Note that \eqref{TreeBound1} is increasing in $C_0$, so that, if desired, for any prescribed $\rho\ge 1$,
		we can make the factor $\gamma^{D_{\text{sc}}(n,m,l,\boldsymbol{p})}$ $(\gamma^\frac1{12}/C_0)^{l}$ in the right and side of \eqref{TreeBound1} smaller than 
		$C_{n,m}\rho^{-l}$ for some $C_{n,m}>0$, possibly at the cost of increasing $C_0$, thus getting:
		\begin{equation} \label{eq:4.22}\|H_{n,m,l,\boldsymbol{p}}[\tau]\|_w\le
			(\alpha_\gamma/d_\gamma)^{\mathds 1(\ell=(0,2,0,\emptyset))}C_{n,m}\rho^{-l}(C\rho\epsilon)^{n_{\text{e.p.}}[\tau]}(K\epsilon)^{-n-m},
		\end{equation}	
		for any $\rho\ge 1$ and some $\rho$-independent constants $C_{n,m}, C, K>0$.
	\end{remark}
	
	\subsection{The FPE for the local terms. Analyticity of the scaling exponents.}\label{Caso2}
	
	Consider now the components of the FPE associated with the local parts of the effective potential, \eqref{FPE}.  
	We focus on the last two components of the equation, the first two having been discussed and solved in \cite{GMR21}. 
	Using the definition of $D$ and the second line of \eqref{R}, we see that in those components we can replace $\gamma^{[\psi]-\Delta_1}R^{\ell_1,\ldots,\ell_s}_{(1,0,1,\mathbf{0})}(H_{\ell_1},\ldots,H_{\ell_s})$ by 
	$T_{1,0,1,\mathbf{0}}^{1,0,1,\mathbf{0}}S^{\ell_1,\ldots,\ell_s}_{(1,0,1,\mathbf{0})}(H_{\ell_1},\ldots,H_{\ell_s})$, and similarly for the component with $\ell=(0,1,2,\mathbf{0})$, thus getting that, at the fixed point,
	\begin{equation}\label{TildeH}
		\begin{split}& \zeta_1=\sum_{s\ge1}\sum_{(\ell_i)_{i=1}^s}^* T^{1,0,1,\mathbf{0}}_{1,0,1,\mathbf{0}}S_{\ell}^{\ell_1,\ldots,\ell_s}(V^*_{\ell_1},\ldots,V^*_{\ell_s}),\\
			& \zeta_2=\sum_{s\ge1}\sum_{(\ell_i)_{i=1}^s}^* T^{0,1,2,\mathbf{0}}_{0,1,2,\mathbf{0}}S_{\ell}^{\ell_1,\ldots,\ell_s}(V^*_{\ell_1},\ldots,V^*_{\ell_s}).\end{split}
	\end{equation}	
	To get, out of this, a representation of $\zeta_1,\zeta_2$ in terms of a convergent tree expansion, we insert the rewriting \eqref{T1} in the right sides of \eqref{TildeH}, so that 
	\begin{equation}\label{TreeildeH}
		\zeta_1=\sum_{\tau} \tilde{H}_{1,0,1,\mathbf{0}}[\tau], \qquad  \zeta_2=\sum_{\tau} \tilde{H}_{0,1,2,\mathbf{0}}[\tau],
	\end{equation}	
	where the tree values $\tilde{H}_{\ell}[\tau]$ with $\ell=(1,0,1,\mathbf{0}),(0,1,2,\mathbf{0})$ are defined essentially in the same way as in the previous subsection, with the only difference that the root vertex $v_0$ is associated with the action of an operator 
	$T^\ell_\ell S_\ell^{\ell_1,\ldots,\ell_{s_{v_0}}}$ rather than $(1-D)^{-1}R_\ell^{\ell_1,\ldots,\ell_{s_{v_0}}}$. In other words, for $\ell=(1,0,1,\mathbf{0}),(0,1,2,\mathbf{0})$, we have 
	$\tilde{H}_{\ell}[\tau]=\sum_{(\ell_i)_{i=1}^{s_{v_0}}}T^\ell_\ell S_\ell^{\ell_1,\ldots,\ell_{s_{v_0}}}(H_{\ell_1}[\tau_1],\ldots,H_{\ell_n}[\tau_{s_{v_0}}])$ where, recalling that $\ell_1,\ldots,\ell_{s_{v_0}}\in L'_f$, 
	the values $H_{\ell_i}[\tau_i]$ for $i\in\{1,\ldots,s_{v_0}\}$ have been constructed and bounded in the previous subsection. On the other hand, recalling that the weighted $L^1$ norm of $T^\ell_\ell$ 
	is bounded by $1$, we find that for $\ell=(1,0,1,\mathbf{0}),(0,1,2,\mathbf{0})$ the norms $\|\tilde H_\ell[\tau]\|_w$ are bounded in {a way analogous to} \eqref{TreeBound1}, namely:
	\begin{equation} \|\tilde{H}_{\ell}[\tau]\|_w\le {C'_\gamma (K\epsilon)^{-1}(K''\epsilon)^{n_{\text{e.p.}}[\tau]}, }
	\end{equation}	
	where {we can choose $C'_\gamma=C_\gamma^{-1}(\gamma^\frac1{12}/C_0)^{2}$ and $K''=\big(\frac{C_0}{1-\gamma^{-\frac1{12}}}\big)^4(K')^2C_{\gamma}K$.} Absolute summability over $\tau$ follows by the same considerations after \eqref{TreeBound1}. In conclusion, both $\zeta_1$ and $\zeta_2$ are expressed in terms of  absolutely convergent tree expansions. Recalling that, for $\epsilon_0,\delta_0>0$ small enough:\begin{itemize}
		\item $\lambda=\lambda^*(\epsilon)$ and $\nu=\nu^*(\epsilon)$ are analytic functions of $\epsilon$ of order $\epsilon$, for $|\epsilon|<\epsilon_0$;
		\item the single-scale propagator $g^{(0)}$ depends analytically upon $\epsilon$ for $|\epsilon|<\epsilon_0$; therefore, the connected expectation $\mathcal C(\boldsymbol{z})$ in \eqref{HBxy}  (see \cite[Appendix D]{GMR21} for the 
		explicit representation of ${\mathcal C}(\boldsymbol{z})$ in terms of $g^{(0)}$), which enters the definition of 
		the `integrating out' map $S^{(0)}$ and, as a consequence, of the tree values themselves, is analytic in $\epsilon$ in the same domain, as well;
		\item the dilatation operator $D$ in \eqref{eq:scal}, which enters the definition of the tree values, is analytic in the scaling exponents $\Delta_1$ and $\Delta_2$, for $|\Delta_1-[\psi]|<\delta_0$ and 
		$|\Delta_2-2[\psi]|<\delta_0$;
	\end{itemize}
	all the tree values $H_\ell[\tau]$ are analytic in $(\epsilon, \Delta_1, \Delta_2)$ in the domain $$\mathcal D_0:=\{(\epsilon, \Delta_1, \Delta_2)\in\mathbb C^3 : |\epsilon|<\epsilon_0, |\Delta_1-[\psi]|<\delta_0, |\Delta_2-2[\psi]|<\delta_0\}.$$ Therefore, by absolute convergence of the tree expansion, uniform in $\mathcal D_0$, analyticity of the sums in the 
	right sides of \eqref{TildeH} follows in the same domain, by Weierstrass' theorem 
	on the uniform convergence of sequences of analytic functions. We shall then write:
	\begin{equation}\label{FPEzeta.0}\zeta_1=F_1(\epsilon,\Delta_1,\Delta_2),\qquad \zeta_2=F_2(\epsilon,\Delta_1,\Delta_2),\end{equation}
	with $F_1,F_2$ analytic in $\mathcal D_0$.
	
	\subsubsection{The scaling exponent \texorpdfstring{$\Delta_1$}{Delta1}}\label{subsecDelta_1}
	
	{From the previous considerations and inspection of perturbation theory, it follows that $\zeta_1=0$. 
		In fact, in view of the convergence of the tree expansion $\sum_\tau\tilde H_{1,0,1,\boldsymbol{0}}[\tau]\equiv F_1(\epsilon,\Delta_1,\Delta_2)$,  
		in order to prove that $\zeta_1=0$ it is enough to prove that any tree contributing to $F_1$ has vanishing value. Note that the trees contributing to $F_1$ 
		have one `white square' endpoint (i.e., the first in the second line of Fig.\ref{eq:figtree3.1}, corresponding to the third interaction vertex in Fig.\ref{eq:fig3.2}) 
		and $k\ge 1$ additional endpoints of type $\nu$ or $\lambda$ (i.e., those in the first line of Fig.\ref{eq:figtree3.1}, corresponding to the first two interaction vertices 
		in Fig.\ref{eq:fig3.2}). It is straightforward to check that, for any such $\tau$, by applying the definition of tree value, $\big[\tilde H_{1,0,1,\boldsymbol{0}}[\tau](x_1,z_1)\big]_{a_1,b_1}$ 
		is local, i.e., it is equal to $\delta_{x_1,z_1}F_{a_1,b_1}(\tau)$ for some $F_{a,b}(\tau)$ that is a (in general infinite, absolutely convergent) linear combination of terms of the following form: 
		\begin{equation}\label{z1}
			\sum_{b_2}\int g^{(h)}_{a,b_2}(x_1-z_2)f_{b_2,b}(z_2) dz_2, 
		\end{equation}	
		for appropriate functions $f_{b',b}$ (with the correct $Sp(N)$ invariance properties, such that \eqref{z1} is in fact proportional to $\delta_{a,b}$ and independent of $a$). 
		For example, it is instructive to check that the sum of the values of the trees with one white square endpoint and one additional endpoint, either of type $\nu$ or $\lambda$, is 
		\begin{equation}\label{nu-phi}
			\sum_{b_2}\int \Big[2\nu g_{a,b_2}^{(0)}(x_1-z_2)\Omega_{b_2,b}+4\lambda \sum_{b_3,b_4}q_{b_2b_3b_4b}\sum_{h\ge0}\gamma^{(d+\Delta_1-5[\psi])h}g_{ab_2}^{(h)}(x_1-z_2)g_{b_3b_4}^{(0)}(0)\Big]dz_2, 
		\end{equation}	
		with $q_{abcd}$ the totally antisymmetric tensor defined after \eqref{trimm}. Now, the key remark is that (the summand over $b_2$ in) \eqref{z1} is proportional to $\hat g^{(h)}_{a,b_2}(0)$, which is zero, because the support of the Fourier transform of $g^{(h)}$ does not contain the origin. Therefore, as anticipated above, all the contributions to $F_1$ vanishing, 
		thus implying that $\zeta_1=0$ and, therefore, recalling \eqref{Delta}, $\Delta_1=[\psi]$.}
	
	{\begin{remark} The considerations above, leading to the conclusion that the scaling exponent of the external field $\phi$, coupled linearly to the fluctuation field $\psi$, has scaling 
			dimension $[\phi]=d-[\psi]$, is a general fact, valid for any infrared, critical, even, interacting theory, treatable perturbatively close to a Gaussian fixed point: it implies that, in general, the
			critical exponent describing the asymptotic, large distance, polynomial decay of the interacting two point function $\langle \psi(x)\psi(y)\rangle$ is the same as the one associated with 
			the Gaussian part of the infrared RG fixed point. This is true, in particular, in theories where the fluctuation field $\psi$ has an infrared anomalous scaling dimension (i.e., 
			$[\psi]$ differs from the naive scaling dimension associated with the bare propagator), as it is the case, for example, for models in the `Luttinger liquid' universality class, 
			in their fermionic formulation (i.e., models admitting a large distance effective description in terms of 2D spinless fermions with quartic interaction), see \cite{GeM01}. 
	\end{remark}}
	
	\subsubsection{The scaling exponent \texorpdfstring{$\Delta_2$}{Delta2}}\label{subsecDelta_2}
	
	Contrary to the case of $\zeta_1$, the right hand side of the equation for $\zeta_2$ does not vanish. Recalling that $\Delta_2$ is related to $\zeta_2$ via \eqref{Delta}, we rewrite the second equation 
	of \eqref{FPEzeta.0} as 
	\begin{equation} \label{FPforzetai}\zeta_2=f_2(\epsilon,\zeta_2),\end{equation}
	where $f_2(\epsilon,\zeta_2):=F_2(\epsilon,[\psi],2[\psi]-\log_\gamma(1+\zeta_2))$ is analytic in $(\epsilon,\zeta_2)$ in a small complex neighborhood of the origin. An explicit computation shows that 
	$f_2(\epsilon,\zeta_2)=-2\epsilon\frac{N-2}{N-8}\log\gamma+O(\epsilon^2,\epsilon\zeta_2)$, see Appendix \ref{appA}. Therefore, applying the analytic implicit function theorem, see e.g. \cite[Section 5.11]{Ga83}, it follows that 
	\eqref{FPforzetai} admits a unique analytic solution $\zeta_2(\epsilon)$ in a neighborhood of the origin, satisfying $\zeta_2(\epsilon)=-2\epsilon\frac{N-2}{N-8}\log\gamma+O(\epsilon^2)$. 
	
	\medskip
	
	This concludes the proof of the part of the statement of Theorem \ref{Th:1} concerning the analyticity of $V^*$ and of the scaling exponents, as well as the facts that $\Delta_1=[\psi]$ and $\Delta_2=2[\psi]+\eta_2$, with $\eta_2$ as in \eqref{FO}. 
	
	\section{Pointwise bounds on the response functions}\label{sec5}
	
	In this section, we prove the following result on the pointwise convergence of the limits in \eqref{eq:6}-\eqref{eq:7}. Recall that the scaling exponents satisfy $\Delta_1=[\psi]$ and $\Delta_2=2[\psi]+\eta_2$, with $\eta_2$ as in \eqref{FO}, as proved in the previous section. We make use of the same notations and conventions on trees, tree values, components of the renormalization map, etc., as in the previous two sections. Moreover, whenever possible, we will keep the flavor indices implicit (e.g., in the first line of \eqref{eq:4.2bis} below 
	we drop the $a,b$ indices labelling $\mathcal{G}^*$, which should be then thought of as an $N\times N$ anti-symmetric matrix).
	
	\begin{Proposition}
		There exists $\epsilon_0>0$ such that the limits in \eqref{eq:6}-\eqref{eq:7} exist and are analytic in $|\epsilon|<\epsilon_0$, and, letting $\boldsymbol{x}=(x,0)$ and $\boldsymbol{y}=(y,0)$,
		they can be explicitly written as 
		\begin{equation} \begin{split}
				& \mathcal G^*(x)=2\sum_\tau\sum_{h\in\mathbb Z}\gamma^{2h\Delta_1}\sum_{(\ell_i)_{i=1}^{s_{v_0}}}^*S^{\ell_1,\cdots,\ell_{s_{v_0}}}
				_{2,0,0,\emptyset}(H_{\ell_1}[\tau_1],\ldots,H_{\ell_{s_{v_0}}}[\tau_{s_{v_0}}])(\gamma^{h}\boldsymbol{x}),\\
				& \mathcal F^*(y)=2\sum_\tau\sum_{h\in\mathbb Z}\gamma^{2h\Delta_2}\sum_{(\ell_i)_{i=1}^{s_{v_0}}}^*S^{\ell_1,\cdots,\ell_{s_{v_0}}}
				_{0,2,0,\emptyset}(H_{\ell_1}[\tau_1],\ldots,H_{\ell_{s_{v_0}}}[\tau_{s_{v_0}}])(\gamma^{h}\boldsymbol{y}),\end{split}\label{eq:4.2bis}
		\end{equation}
		with the understanding that the sums in the right hand sides are absolutely summable in $h$ and $\tau$. Moreover, for any $\alpha>0$ small enough, there exists $C_\alpha>0$ such that 
		\begin{equation}\begin{split}
				&	\big| 2\gamma^{2h\Delta_1} V^*_{2,0,0,\emptyset}(\gamma^{h}\boldsymbol{x})-\mathcal G^*(x)\big|\le \frac{C_\alpha}{|x|^{2\Delta_1}}\Big(\min\{1,\gamma^{h}|x|\}\Big)^{2[\psi]-\alpha},\\
				&	\big| 2\gamma^{2h\Delta_2} V^*_{0,2,0,\emptyset}(\gamma^{h}\boldsymbol{y})-\mathcal F^*(y)\big|\le \frac{C_\alpha}{|y|^{2\Delta_2}}\Big(\min\{1,\gamma^{h}|y|\}\Big)^{2[\psi]-\alpha}.
			\end{split}	\label{eq:Tails}
		\end{equation}	 
		\label{Prop:1}
	\end{Proposition}
	As {proved in \eqref{eq:2.44}}, Proposition \ref{Prop:1} implies the scale invariance property \eqref{1.12}. Therefore, in view of the fact that the other statements of Theorem \ref{Th:1}
	have already been proved above, see the comment at the end of Section \ref{Sec:3}, Proposition \ref{Prop:1} implies Theorem \ref{Th:1}.
	
	\begin{proof}[Proof of Proposition \ref{Prop:1}]
		Using the tree representation \eqref{T1}, we write:
		\begin{equation}\label{eq:4.1}\begin{split}
				& 	 \gamma^{2h\Delta_1} V^*_{2,0,0,\emptyset}(\gamma^{h}\boldsymbol{x})=\gamma^{2h\Delta_1} \sum_\tau H_{2,0,0,\emptyset}[\tau](\gamma^{h}\boldsymbol{x}),\\
				& 	 \gamma^{2h\Delta_2} V^*_{0,2,0,\emptyset}(\gamma^{h}\boldsymbol{y})=\gamma^{2h\Delta_2} \sum_\tau H_{0,2,0,\emptyset}[\tau](\gamma^{h}\boldsymbol{y}),\end{split}
		\end{equation}
		with $H_\ell[\tau]$ recursively defined as in \eqref{T2}. 
		We emphasize that the trees $\tau$ contributing to the sum in the first line of \eqref{eq:4.1} have two `white square' endpoints, and those contributing to the second line have two `white rombus' endpoints, see Fig.\ref{eq:figtree3.1}. Using \eqref{R}, the identity $(1-D)^{-1}D=\sum_{k\ge 1}D^k$ with $D$ as in \eqref{eq:scal}, the definition of $\delta_{\text{sc}}$ in 
		\eqref{eq:ScDim} (from which $\delta_{\text{sc}}(2,0,0,\emptyset)=2\Delta_1$ and $\delta_{\text{sc}}(0,2,0,\emptyset)=2\Delta_2$), and the
		fact that for $\ell=(2,0,0,\emptyset),(0,2,0,\emptyset)$
		the trimming map is the identity, we get (after renaming $h+k\equiv h'$): 
		\begin{equation} \begin{split}
				& \gamma^{2h\Delta_1}  H_{2,0,0,\emptyset}[\tau](\gamma^{h}\boldsymbol{x})= \sum_{h'>h}\gamma^{2h'\Delta_1}\sum_{(\ell_i)_{i=1}^{s_{v_0}}}^*S^{{\ell_1,\ldots,\ell_{s_{v_0}}}}
				_{2,0,0,\emptyset}(H_{\ell_1}[\tau_1],\ldots,H_{\ell_{s_{v_0}}}[\tau_{s_{v_0}}])(\gamma^{h'}\boldsymbol{x}),\\
				& \gamma^{2h\Delta_2}  H_{0,2,0,\emptyset}[\tau](\gamma^{h}\boldsymbol{y})= \sum_{h'>h}\gamma^{2h'\Delta_2}\sum_{(\ell_i)_{i=1}^{s_{v_0}}}^*S^{{\ell_1,\ldots,\ell_{s_{v_0}}}}
				_{0,2,0,\emptyset}(H_{\ell_1}[\tau_1],\ldots,H_{\ell_{s_{v_0}}}[\tau_{s_{v_0}}])(\gamma^{h'}\boldsymbol{y}).\end{split}\label{eq:4.2}
		\end{equation}
		If we multiply by $2$ both sides, sum over $\tau$ and take $h\to-\infty$, we obtain the representations \eqref{eq:4.2bis}, provided that the sums in the right hand sides are absolutely summable in $\tau$ and $h$.
		
		The right hand sides of \eqref{eq:4.2} can be bounded via the following lemma that, for later purposes, is formulated in greater generality than required for the moment.  
		In order to state the lemma, we define a mixed $L^1/L^\infty$ norm: using the 
		same notations as in the definition of the weighted $L^1$ norm \eqref{norm}, we let, for $\ell=(n,m,l,\boldsymbol{p})\in L$, 
		\begin{equation}\label{Linfty}
			\tn{H_{\ell}[\tau](\boldsymbol{x},\boldsymbol{y})}:=\max_{\substack{\boldsymbol{a}, \mathbf{B}\, :\\ \boldsymbol{p}(\boldsymbol{B})}}
			\int d\boldsymbol{z}_{\mathbf{B}}|H_{\ell}[\tau](\boldsymbol{a},\boldsymbol{x},\boldsymbol{y},\mathbf{B},\boldsymbol{z}_{\mathbf{B}})|
		\end{equation}
		with the understanding that, if $\ell=(n,m,0,\emptyset)$, then $\tn{H_{n,m,0,\emptyset}[\tau](\boldsymbol{x},\boldsymbol{y})]}$ should be interpreted as being equal to $\max_{\boldsymbol{a}}|H_{n,m,0,\emptyset}[\tau](\boldsymbol{a},\boldsymbol{x},\boldsymbol{y})|$.
		\begin{Lemma}
			Consider a tree $\tau$ contributing to one of the sums in the right sides of \eqref{eq:4.1}, denote by $\tau_1, \ldots, \tau_{s_{v_0}}$ its subtrees rooted in $v_0$, and by 
			$n_{\text{e.p.}}[\tau]$ be the number of its endpoints. Let $\boldsymbol{x}=(x,0)$ and $\boldsymbol{y}=(y,0)$.
			For any $\alpha>0$ sufficiently small, %{and any $\rho\ge 1$,} 
			there exists $C=C(\alpha)>0$ %,\rho)>0$} 
		such that, for any $k\in\mathbb Z$, 
		any $l\ge 0$ even and any $\boldsymbol{p}\in\{0,1\}^l$, 
		\begin{equation}\label{eq:lem:2}\begin{split}
				&\sum_{(\ell_i)_{i=1}^{s_{v_0}}}^*\tn{D^kS^{\ell_1,\ldots,\ell_{s_{v_0}}}_{2,0,l,\boldsymbol{p}}(H_{\ell_1}[\tau_1],\ldots,H_{\ell_{s_{v_0}}}[\tau_{s_{v_0}}])(\boldsymbol{x})} \\
				&	\qquad \le		 C(C\epsilon)^{n_{\text{e.p}}[\tau]-2}%{\rho^{-l}}
				\gamma^{k(2\Delta_1-l[\psi]-\|\boldsymbol{p}\|_1)}e^{-\frac{\bar{C}}2(\gamma^{k-1}|x|)^{\sigma}} \big(\min\{1,\gamma^{k}|x|\}\big)^{-\alpha},\end{split}\end{equation}	
		with $D$ the dilatation operator and $\bar{C}$ the same constant as in \eqref{w}, and 
		\begin{equation}\label{eq:lem:2bis}\begin{split}
				&\sum_{(\ell_i)_{i=1}^{s_{v_0}}}^*\tn{D^kS^{\ell_1,\ldots,\ell_{s_{v_0}}}
					_{0,2,l,\boldsymbol{p}}(H_{\ell_1}[\tau_1],\ldots,H_{\ell_{s_{v_0}}}[\tau_{s_{v_0}}])(\boldsymbol{y})} \\
				&	\qquad \le		 C(C\epsilon)^{n_{\text{e.p}}[\tau]-2}%{\rho^{-l}}
				\gamma^{k(2\Delta_2-l[\psi]-\|\boldsymbol{p}\|_1)}e^{-\frac{\bar{C}}2(\gamma^{k-1}|y|)^{\sigma}} 
				\big(\min\{1,\gamma^{k}|y|\}\big)^{-2\Delta_2+2[\psi]-\alpha}.	\end{split}\end{equation}	
		\label{lem:4.1}	
	\end{Lemma}	
	Assuming the validity of this lemma, the proof of Proposition \ref{Prop:1} goes as follows. 
	Let us focus, e.g., on the component with $\ell=(2,0,0,\emptyset)$, the case $\ell=(0,2,0,\emptyset)$ being analogous. 
	Plugging \eqref{eq:lem:2} with $k=h'$, $l=0$ and $\boldsymbol{p}=\emptyset$ in the right hand side of the first line of \eqref{eq:4.2}, 
	recalling that $D^{h'}S^{\ell_1,\ldots,\ell_{s_{v_0}}}_{2,0,l,\boldsymbol{p}}(H_{\ell_1}[\tau_1],\ldots,H_{\ell_{s_{v_0}}}[\tau_{s_{v_0}}])(\boldsymbol{x})$ is just a rewriting of 
	$\gamma^{2h'\Delta_1}S^{\ell_1,\ldots,\ell_{s_{v_0}}}_{2,0,l,\boldsymbol{p}}(H_{\ell_1}[\tau_1],\ldots$, $H_{\ell_{s_{v_0}}}[\tau_{s_{v_0}}])(\gamma^{h'}\boldsymbol{x})$,
	we find: 
	\begin{equation}\label{wefindthisbound}
		\begin{split} \gamma^{2h\Delta_1} \big| H_{2,0,0,\emptyset}[\tau](\gamma^{h}\boldsymbol{x})\big|&\le 
			C(C\epsilon)^{n_{\text{e.p}}[\tau]-2}|x|^{-2\Delta_1}\\
			&\cdot  \sum_{h'>h}e^{-\frac{\bar{C}}2(\gamma^{h'-1}|x|)^{\sigma}}\max\{(\gamma^{h'}|x|)^{2\Delta_1},(\gamma^{h'}|x|)^{2\Delta_1-\alpha}\}.\end{split}\end{equation}
	Now note that, for any $\beta,\kappa>0$, letting $h_x:=\lfloor \log_\gamma|x|^{-1}\rfloor$, 
	$(\gamma^{h'}|x|)^\beta e^{-\kappa(\gamma^{h'}|x|)^\sigma}$ is bounded from above by $\gamma^{\beta(h'-h_x)}e^{-\kappa\gamma^{\sigma(h'-h_x-1)}}$, which is summable in $h'$ over $\mathbb Z$. Therefore,
	\begin{equation} \sum_{h'>h}(\gamma^{h'}|x|)^\beta e^{-\kappa(\gamma^{h'}|x|)^\sigma}\le 
		C_{\beta,\kappa}:=\sum_{h\in\mathbb Z}\gamma^{\beta h} e^{-\kappa\gamma^{\sigma(h-1)}},\label{largerbound}\end{equation} 
	uniformly in $h$ and $|x|$. For later purpose, let us also observe that
	\begin{equation} \begin{split}\sum_{h'\le h}(\gamma^{h'}|x|)^\beta e^{-\kappa(\gamma^{h'}|x|)^\sigma}&\le\sum_{h'\le h}\gamma^{\beta(h'-h_x)}e^{-\kappa\gamma^{\sigma(h'-h_x-1)}}\\
			&\le 
			\begin{cases} C_{\beta,\kappa} & \text{if}\quad h>h_x\\
				\gamma^{\beta(h-h_x)}/(1-\gamma^{-\beta}) & \text{if}\quad h\le h_x\end{cases}\ \le C_{\beta,\kappa}'(\min\{1,\gamma^h|x|\})^{\beta},
		\end{split}\label{smallerbound}
	\end{equation} 
	for a suitable $C'_{\beta,\kappa}>0$. Using \eqref{largerbound} in \eqref{wefindthisbound}, we find that, for any $\alpha>0$ small enough, 
	\begin{equation} \gamma^{2h\Delta_1} \big| H_{(2,0,0,\emptyset}[\tau](\gamma^{h}\boldsymbol{x})\big|\le 
		C'(C\epsilon)^{n_{\text{e.p}}[\tau]-2}|x|^{-2\Delta_1},\label{eq:5.11}\end{equation}
	for some $C'>0$, uniformly in $h$. Since  the right hand side of this inequality is summable over $\tau$, the sum being
	bounded by (const.)$|x|^{-2\Delta_1}$, this proves the absolute convergence of the sum in the right hand side of the first line of \eqref{eq:4.2bis} and, therefore, as already observed after \eqref{eq:4.2}, 
	it implies the very validity of the first line of \eqref{eq:4.2bis}. 
	
	Concerning the difference $2\gamma^{2h\Delta_1} V^*_{2,0,0,\emptyset}(\gamma^{h}\boldsymbol{x})-\mathcal G^*(x)$, 
	using \eqref{eq:4.2bis}, \eqref{eq:4.1}, \eqref{eq:4.2} and \eqref{eq:lem:2} with $k=h'$, $l=0$ and $\boldsymbol{p}=\emptyset$, we find:
	\begin{equation}\begin{split}
			&\big| 2\gamma^{2h\Delta_1} V^*_{2,0,0,\emptyset}(\gamma^{h}\boldsymbol{x})-\mathcal G^*(x)\big|\\
			&\le 2\sum_\tau
			\sum_{h'\le h}\gamma^{2h'\Delta_1}\sum_{(\ell_i)_{i=1}^{s_{v_0}}}^*\big|S^{\ell_1,\ldots,\ell_{s_{v_0}}}
			_{2,0,0,\emptyset}(H_{\ell_1}[\tau_1],\ldots,H_{\ell_{s_{v_0}}}[\tau_{s_{v_0}}])(\gamma^{h'}\boldsymbol{x})\big|\\
			&\le 2C|x|^{-2\Delta_1}\sum_{\tau}(C\epsilon)^{n_{\text{e.p}}[\tau]-2}
			\sum_{h'\le h}e^{-\frac{\bar{C}}2(\gamma^{h'-1}|x|)^{\sigma}}\max\{(\gamma^{h'}|x|)^{2\Delta_1},(\gamma^{h'}|x|)^{2\Delta_1-\alpha}\}.	
		\end{split}
		\label{eq:4.8.100}
	\end{equation}
	Now, using \eqref{smallerbound}, we find that the sum over $h'\le h$ in the last line is bounded from above by (const.)$(\min\{1,\gamma^h|x|\})^{2\Delta_1-\alpha}$. Moreover, 
	recalling that the number of trees with $k$ endpoints is smaller than $4^k$, see \cite[Lemma A.1]{GeM01}, and noting that the trees contributing to the sum in the last line 
	have at least two endpoints (because $\tau$ has at least two `white square' endpoints), we see that $(C_\alpha\epsilon)^{n_{\text{e.p}}[\tau]-2}$ 
	is summable over $\tau$, and the sum is bounded by a positive constant independent of $\epsilon$. In conclusion, 
	$$\big| 2\gamma^{2h\Delta_1} V^*_{2,0,0,\emptyset}(\gamma^{h}\boldsymbol{x})-\mathcal G^*(x)\big|\le 
	C'|x|^{-2\Delta_1}(\min\{1,\gamma^h|x|\})^{2\Delta_1-\alpha},$$
	which is the desired estimate in the first line of \eqref{eq:Tails}, up  to a redefinition of $C_\alpha$. 
	
	The proof of the second line of \eqref{eq:4.2bis} and of \eqref{eq:Tails} is completely analogous and left to the reader. 
\end{proof}

We are left with proving Lemma \ref{lem:4.1}.

\begin{proof}[Proof of Lemma \ref{lem:4.1}]
	We focus on the proof of \eqref{eq:lem:2}, the one of \eqref{eq:lem:2bis} being analogous (we will make a few comments at the end on the minor differences between the two cases). We recall that $\boldsymbol{x}=(x,0)$. 
	
	In order to obtain a point-wise bound on the left hand side of \eqref{eq:lem:2}, we intend to apply iteratively the definition of tree value, similarly to what we did in Section \ref{Sec:3}.
	We recall that the trees $\tau$ contributing to the sum in the first line of \eqref{eq:4.1} have two `white square' endpoints, which will be 
	denoted $v^*_1$ and $v^*_2$. Note that the coordinates associated with these endpoints are fixed, i.e., not integrated out in the computation of the  tree value: this implies that, 
	for the purpose of recursively deriving bounds on the values of the subtrees of $\tau$, we must be careful in proceeding slightly differently, depending on whether the subtree under consideration 
	contains both $v^*_1$ and $v^*_2$, or just one of them, or none. 
	
	We define $v^*_{12}$ to be the rightmost vertex of $\tau$ that is an ancestor both of $v^*_1$ and of $v^*_2$, and we let $n_\tau$ be the length of the path connecting the root vertex $v_0$ with $v^*_{12}$, see {{Fig.\ref{fig:10}}} and Fig.\ref{fig:9} below. ({{Fig.\ref{fig:10} illustrates the case $n_\tau=0$, where $v^*_{12}\equiv v_0$, while Fig.\ref{fig:9} illustrates the case $n_\tau=n>0$, where $v^*_{12}>v_0$.}})
	
	\bigskip
	
	Let us first discuss the case that $n_\tau=0$, i.e., $v^*_1$ and $v^*_2$ belong to two different subtrees of $\tau$ rooted in two distinct children vertices of $v_0\equiv v_{12}^*$, called $v_1$ and $v_2$, respectively; the two subtrees rooted in $v_1$ and $v_2$ will be denoted by $\tau_1$ and $\tau_2$, respectively, see Figure \ref{fig:10}. 
	
	\begin{figure}[ht]
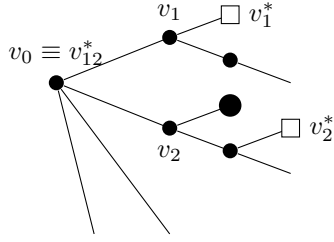

		\begin{center}
			\begin{tabular}{rcl}
				\tikz[baseline=-2pt]{ %orizzontale
					\draw(0,0)node[vertex, label=above:{$v_0\equiv v_{12}^*$}](v0){};
					\draw(v0)--+(0.5,-2)node[]{};
					\draw(v0)--+(1.5,-2)node[]{};
					%% sottoalbero alto vx
					\draw (v0)--+(1.5,0.6)node[vertex,label=above:{$v_1$}](vf1){};
					\draw (vf1)--+(0.8,0.3) node[E,label=right:{$v^*_1$}]{};
					\draw (vf1)--+(0.8,-0.3)node[vertex](V0){};
					\draw(V0)--+(0.8,-0.3);
					%%
					%% sottoalbero basso vx
					\draw (v0)--+(1.5,-0.6)node[vertex,label=below:{$v_2$}](vf1){};
					\draw (vf1)--+(0.8,0.3) node[bigvertex]{};
					\draw (vf1)--+(0.8,-0.3)node[vertex](V0){};
					\draw(V0)--+(0.8,-0.3);
					\draw(V0)--+(0.8,0.3)node[E,label=right:{$v^*_2$}]{};
				}
			\end{tabular}
		\end{center}\caption{Proof of Lemma \ref{lem:4.1}, the case with $n_\tau=0$. The two subtrees rooted in $v_1$ and $v_2$ are denoted by $\tau_1$ and $\tau_2$.}
		\label{fig:10}
	\end{figure}	
	
	In this case, recalling the definition \eqref{Linfty} and proceeding as in the proof of \eqref{S}, we get:
	\begin{equation}\label{4.1}\begin{split}
			&\tn{D^kS^{\ell_1,\ldots,\ell_{s_{v_0}}}
				_{2,0,l,\boldsymbol{p}}(H_{\ell_1}[\tau_1],\ldots,H_{\ell_{s_{v_0}}}[\tau_{s_{v_0}}])(\boldsymbol{x})}e^{\bar{C}(\gamma^{k-1}|x|)^{\sigma}}\\
			&=\gamma^{k\delta_{sc}(2,0,l,\boldsymbol{p})}\gamma^{-kld}\tn{S^{\ell_1,\ldots,\ell_{s_{v_0}}}
				_{2,0,l,\boldsymbol{p}}(H_{\ell_1}[\tau_1],\ldots,H_{\ell_{s_{v_0}}}[\tau_{s_{v_0}}])(\gamma^{k}\boldsymbol{x})}e^{\bar{C}(\gamma^{k-1}|x|)^{\sigma}}\\
			&\hskip2.truecm\le \frac{\gamma^{k(2\Delta_1-l[\psi]-\|\boldsymbol{p}\|_1)}}{s_{v_0}!} C_0^{\sum_{i=1}^{s_{v_0}}l_{i}-l}N_{(2,0,l,\boldsymbol{p})^{\ell_1,\ldots,\ell_{s_{v_0}}}}
			\Big(\sum_{\mathcal T}\prod_{\substack{(z,z')\in \mathcal T \\(z,z')\ne\ell_0}}N^2d^2\|M\|_w\Big)\cdot\\
			&\hskip2.truecm\cdot\sup_{x'}\{N^2d^2M(x')e^{\bar C(|x'|/\gamma)^\sigma}\}\prod_{i=1}^{s_{v_0}}\|H_{\ell_i}[\tau_i]\|_w\,,
		\end{split}	
	\end{equation}	
	where $\bar C$ is the same as in \eqref{w},  $\delta_{sc}(2,0,l,\boldsymbol{p})$ is given by \eqref{eq:ScDim}, and $\gamma^{-kld}$ comes from the rescaling of the $\boldsymbol{z}$ variable in 
	$\tn{\cdot}$. 
	The summation in the right hand side of \eqref{4.1} is over the anchored trees described in \cite[Appendix D.4]{GMR21} (see also \eqref{|C|}), $\ell_0$ is an element of $\mathcal T$, chosen arbitrarily among those in the path from the group of points associated with $\tau_1$ and the one associated with $\tau_2$ (say, the one closest to the group associated with $\tau_1$), and the factor $\sup_{x'}\{M(x')e^{\bar C(|x'|/\gamma)^\sigma}\}$ comes from the factor $M$ associated with $\ell_0$. Now, recalling that $M(x)=C_{\chi_1}e^{-C_{\chi_2}|x/\gamma|^{\sigma}}$ and $\bar C\equiv\frac12 C_{\chi^2}$, bounding the  sum over $\mathcal T$ as in \cite[Appendix D.5]{GMR21}, we find\footnote{The bound \eqref{eq:4.30_2} is similar to \eqref{eq:4.30_1} for the same value of $\ell=(2,0,l,\boldsymbol{p})$ (for which the trimming operator $T$ in the first line of \eqref{eq:4.30_1} acts as the identity and, therefore, ${\mathcal N}_{\ell}^{\ell_1,\ldots,\ell_s}= N_{\ell}^{\ell_1,\ldots,\ell_s}$),
		the most relevant differences being that: (1) the operator $\sum_{h\ge 1}D^h=\frac{D}{1-D}$ in the first line of \eqref{eq:4.30_1} is replaced by the action of $D^k$ in 
		the left hand side of \eqref{eq:4.30_2}; (2) in the left hand side of \eqref{eq:4.30_2} we are not integrating over the coordinate $x$, 
		contrary to the case of \eqref{eq:4.30_1}. {Correspondingly, the proof of \eqref{eq:4.30_2} is similar to that \eqref{eq:4.30_1}, and the details are left to the reader. Let us just mention that the two differences (1) and (2) mentioned above explain}
		the presence of the decay factor $e^{-\bar{C}(\gamma^{k-1}|x|)^{\sigma}}$ in \eqref{eq:4.30_2}, as 
		well as that of 
		the dimensional factor $\gamma^{k[D_{\text{sc}}(2,0,l,\boldsymbol{p})+d]}=\gamma^{k(2\Delta_1-l[\psi]-\|\boldsymbol{p}\|_1)}$ instead of $d_\gamma\gamma^{D_{\text{sc}}(2,0,l,\boldsymbol{p})}$ 
		(roughly speaking, the fact that {in \eqref{eq:4.30_2}} we have `one integration less', in combination with the presence of $D^k$, comes with a dimensional factor $\gamma^{kd}$; {on the other hand, the factor} $d_\gamma\gamma^{D_{\text{sc}}(2,0,l,\boldsymbol{p})}$ {in \eqref{eq:4.30_1}} is a norm bound on $D(1-D)^{-1}$).}
	\begin{equation}\label{eq:4.30_2}\begin{split} 
			&\sum_{(\ell_i)_{i=1}^{s_{v_0}}}^*\tn{D^kS^{\ell_1,\ldots,\ell_{s_{v_0}}}
				_{2,0,l,\boldsymbol{p}}(H_{\ell_1}[\tau_1],\ldots,H_{\ell_{s_{v_0}}}[\tau_{s_{v_0}}])(\boldsymbol{x})}
			\\
			&\le N^2d^2 C_{\chi_1} \gamma^{k(2\Delta_1-l[\psi]-\|\boldsymbol{p}\|_1)} e^{-\bar{C}(\gamma^{k-1}|x|)^{\sigma}} \sum_{(\ell_i)_{i=1}^{s_{v_0}}}^* C_\gamma^{s_{v_0}-2}(4C_0)^{\sum_{i=1}^{s_{v_0}}l_{i}-l}N_{(2,0,l,\boldsymbol{p})}^{\ell_1,\ldots,\ell_{s_{v_0}}}
			\prod_{i=1}^{s_{v_0}} \|H_{\ell_i}[\tau_i]\|_{w}.
	\end{split}\end{equation}
	with $C_\gamma=N^2d^2\|M\|_w$. Now, in the right side of \eqref{eq:4.30_2} we bound $N_{(2,0,l,\boldsymbol{p})}^{\ell_1,\ldots,\ell_{s_{v_0}}}$ from above by $\binom{\sum_{i=1}^{s_{v_0}} l_{i}}{l}$, see 
	\eqref{sumoverp0}, and we bound
	$\|H_{\ell_i}[\tau_i]\|_w\le C(\rho')^{-l_i}(C\epsilon)^{n_{\text{e.p.}}[\tau_i]-n_i}$ via \eqref{eq:4.22}; here $n_i$ is the first component of  $\ell_i=(n_i,m_i,l_i,\boldsymbol{p}_i)$ 
	(by construction $n_1=n_2=1$ and $n_i=0$ for $i>2$) and $C$ is a $\rho'$-dependent constant; therefore, choosing $\rho'\ge 1$ sufficiently large, we can easily sum over $(\ell_i)_{i=1}^{s_{v_0}}$, thus obtaining (noting that $s_{v_0}\le n_{\text{e.p.}}[\tau]$): 
	\begin{equation}\begin{split}
			&\sum_{(\ell_i)_{i=1}^{s_{v_0}}}^*\tn{D^kS^{\ell_1,\ldots,\ell_{s_{v_0}}}
				_{2,0,l,\boldsymbol{p}}(H_{\ell_1}[\tau_1],\ldots,H_{\ell_{s_{v_0}}}[\tau_{s_{v_0}}])(\boldsymbol{x})}\\
			&\hspace{5cm}\le C'(C' \epsilon)^{n_{\text{e.p.}}[\tau]-2}(4C_0)^{-l}\gamma^{k(2\Delta_1-l[\psi]-\|\boldsymbol{p}\|_1)}e^{-\bar{C}(\gamma^{k-1}|x|)^{\sigma}},
		\end{split}
		\label{Step:1}		
	\end{equation}		
	for some $C'>0$, which proves \eqref{eq:lem:2} for $n_\tau=0$ (simply because $1\le (\min\{1,\gamma^{k}|x|\})^{-\alpha}$ for $\alpha>0$), provided we choose $C_0\ge 1/4$ 
	(the freedom to choose $C_0$ as large as desired follows, as already observed in Remark \ref{ciao}, from the monotonicity of the right hand side of \eqref{Step:1} on the choice of $C_0$:
	recall, in fact, that $C'$ is proportional to $C_0^4$, cf. with \eqref{TreeBound1}, and that $4n_{\text{e.p.}}[\tau]\ge l$).
	
	\medskip
	
	Let us discuss next the case $n_\tau\ge 1$, in which $v^*_1$ and $v^*_2$ both belong to a common subtree among those rooted in the vertices immediately following $v_0$ on $\tau$, 
	say  to $\tau_1$. In this case, letting $n\equiv n_\tau$,
	we denote by $v_0$, $v_1$, $\ldots$, $v_n\equiv v^*_{12}$
	the vertices of $\tau$ in the path from $v_0$ to $v^*_{12}$, as shown in Fig.~\ref{fig:9}. 
	\begin{figure}[ht]
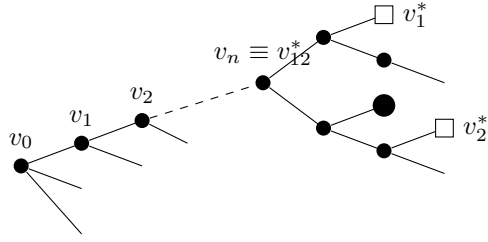

		\begin{center}
			%\begin{equation}
			\begin{tabular}{rcl}
				\tikz[baseline=-2pt]{ %orizzontale
					\draw(0,0)node[vertex, label=above:{$v_0$}](v0){}--+(0.8,0.3)node[vertex,label=above:{$v_1$}](v1){};
					\draw(v1)--+(0.8,0.3)node[vertex,label=above:{$v_2$}](v2){};
					\draw [dashed](v2)--+(1.6,0.5)node [vertex, label=above:{$v_n\equiv v_{12}^*$}](vx){};
					%% sottoalberi v0, v1, v2
					\draw (v0)--(+0.8,-0.3);
					\draw(v0)--(+0.8,-0.9);
					\draw (v1)--(1.6,0);
					\draw (v2)--(2.2,0.3);
					%%
					%% sottoalbero alto vx
					\draw (vx)--+(0.8,0.6)node[vertex](vf1){};
					\draw (vf1)--+(0.8,0.3) node[E,label=right:{$v_1^*$}]{};
					\draw (vf1)--+(0.8,-0.3)node[vertex](V0){};
					\draw(V0)--+(0.8,-0.3);
					%%
					%% sottoalbero basso vx
					\draw (vx)--+(0.8,-0.6)node[vertex](vf1){};
					\draw (vf1)--+(0.8,0.3) node[bigvertex]{};
					\draw (vf1)--+(0.8,-0.3)node[vertex](V0){};
					\draw(V0)--+(0.8,-0.3);
					\draw(V0)--+(0.8,0.3)node[E,label=right:{$v_2^*$}]{};
				}
			\end{tabular}
		\end{center}\caption{Proof of Lemma \ref{lem:4.1}, the case with $n=n_\tau\ge 1$.}
		\label{fig:9}
	\end{figure}	
	For later reference, we introduce (and partly recall) the following notations and definitions. For any vertex $v$ of $\tau$, we denote by $\ell_v=(n_v,m_v,l_v,{\boldsymbol{p}_v})$ the 
	components of the label associated with $v$: if $v=v_0$, then $\ell_{v_0}\equiv(2,0,l,\boldsymbol{p})$; if $v\neq v_0$ is not an endpoint, then $\ell_v$ is summed over $L_f'$, 
	while, if $v$ is an endpoint, $\ell_v$ takes one of the values in $\{(n,m,l,\boldsymbol{0})\}_{(n,m,l)\in\mathfrak{L}}$, depending on the nature of the endpoint, see Fig.\ref{eq:figtree3.1}.
	Moreover, for any vertex $v$ of $\tau$ that is not an endpoint, 
	we denote by  $v_i(v)$ the $i$-th child of $v$ and, if $\tau_v$ is a tree rooted in $v$, $\tau_i(v)$ is the subtree of $\tau_v$ rooted in $v_i(v)$. Correspondingly, 
	we let $\ell_i(v)\equiv \ell_{v_i(v)}$ and denote $\ell_i(v)=(n_i(v),m_i(v),l_i(v),{\boldsymbol{p}}_i(v))$.
	Note that, for $v_0,v_1,\ldots,v_n$ in Figure \ref{fig:9}, we have $\ell_{v_i}=(2,0,l_{v_i},\boldsymbol{p}_{v_i})$, where, for any $1\le i\le n$, $l_{v_i}$ is summed over the even, positive, integers.
	
	We will iteratively use the following basic estimates: 
	\begin{enumerate}
		\item For the vertices $v_j$, $j=0,\ldots,{n-1}$ in Fig.\ref{fig:9}, we use:
		\begin{equation}\begin{split}
				& \tn{S^{\ell_1(v_j),\ldots,\ell_{s_{v_j}}(v_j)}
					_{\ell_{v_j}}(H_{\ell_{1}(v_j)}[\tau_{1}(v_j)],\ldots,H_{\ell_{s_{v_j}}(v_j)}[\tau_{s_{v_j}}(v_j)])(\boldsymbol{x})}\\
				&\qquad\le  \frac{1}{s_{v_j}!}C_0^{\sum_{i=1}^{s_{v_j}}l_i(v_j)-l_{v_j}}
				N_{\ell_{v_j}}^{\ell_1(v_j),\ldots,\ell_{s_{v_j}}(v_j)}
				\Big(\sum_{\mathcal{T}}\prod_{(z,z')\in\mathcal{T}}N^2d^2\|M\|_1\Big)
				\\
				&\qquad \cdot\Big(\prod_{i=2}^{s_{v_j}}\|H_{\ell_i(v_j)}[\tau_i(v_j)]\|_w\Big)\tn{H_{\ell_1(v_j)}[\tau_1(v_j)](\boldsymbol{x})},
			\end{split}\label{item1}
		\end{equation}
		which is proved in the same way as \eqref{S} or \eqref{4.1}. Bounding once again the sum over $\mathcal{T}$ as in \cite[Appendix D.5]{GMR21}, we further obtain
		\begin{equation}\begin{split}
				& \tn{S^{\ell_1(v_j),\ldots,\ell_{s_{v_j}}(v_j)}
					_{\ell_{v_j}}(H_{\ell_{1}(v_j)}[\tau_{1}(v_j)],\ldots,H_{\ell_{s_{v_j}}(v_j)}[\tau_{s_{v_j}}(v_j)])(\boldsymbol{x})}\\
				&\le  C_\gamma^{s_{v_j}-1}(4C_0)^{\sum_{i=1}^{s_{v_j}}l_i(v_j)-l_{v_j}}N_{\ell_{v_j}}^{\ell_1(v_j),\ldots,\ell_{s_{v_j}}(v_j)}
				\Big(\prod_{i=2}^{s_{v_j}}\|H_{\ell_i(v_j)}[\tau_i(v_j)]\|_w\Big)\tn{H_{\ell_1(v_j)}[\tau_1(v_j)](\boldsymbol{x})},
			\end{split}\label{item1bis}
		\end{equation}
		with $C_\gamma=N^2d^2\|M\|_w$. 
		\item Using \eqref{T2}, with $R^{\ell_1,\ldots,\ell_{s_{v_0}}}_\ell=D S^{\ell_1,\ldots,\ell_{s_{v_0}}}_\ell$ for any $\ell$ of the form $\ell=(2,0,l,\boldsymbol{p})$, 
		see \eqref{R}, we rewrite and bound each of the factors $\tn{H_{\ell_{v_i}}[\tau_{v_i}](\boldsymbol{x})}$ 
		in the right hand side of \eqref{item1} (note that, for $j=0,\ldots,n-1$, $\ell_1(v_j)=\ell_{v_i}$ and $\tau_1(v_j)=\tau_{v_i}$, with $i=j+1=1,\ldots,n$) as follows:
		\begin{equation}\label{item3}\begin{split}
				&\tn{H_{\ell_{v_i}}[\tau_{v_i}](\boldsymbol{x})}\\
				&\le\sum_{k_{v_i}\ge1}\sum^*_{(\ell_j(v_i))_{j=1}^{s_{v_i}}}\tn{D^{k_{v_i}}S_{\ell_{v_i}}^{\ell_1(v_i),\dots,\ell_{s_{v_i}}(v_i)}(H_{\ell_1(v_i)}[\tau_1(v_i)],\dots,H_{\ell_{s_{v_i}}(v_i)}[\tau_{s_{v_i}}(v_i)])(\boldsymbol{x})}\\
				&=\sum_{k_{v_i}\ge1}\gamma^{k_{v_i}\delta_{sc}(2,0,l_{v_i},\boldsymbol{p}_{v_i})}\gamma^{-k_{v_i}dl_{v_i}}\cdot\\
				&\hspace{0.9cm}\cdot\sum^*_{(\ell_j(v_i))_{j=1}^{s_{v_i}}}\tn{S_{\ell_{v_i}}^{\ell_1(v_i),\dots,\ell_{s_{v_i}}(v_i)}(H_{\ell_1(v_i)}[\tau_1(v_i)],\dots,H_{\ell_{s_{v_i}}(v_i)}[\tau_{s_{v_i}}(v_i)])(\gamma^{k_{v_i}}\boldsymbol{x})},		 	
		\end{split}\end{equation} 	
		where we used that $D(1-D)^{-1}=\sum_{k_{v_i}\ge1}D^{k_{v_i}}$ together with \eqref{eq:scal}, and $\gamma^{-k_{v_i}dl_{v_i}}$ comes from the rescaling of the $\boldsymbol{z}$ variables
		within the definition of the $\tn{\cdot}$ norm.
	\end{enumerate}
	Let us now explain how to apply these estimates for bounding the left hand side of \eqref{eq:lem:2}: using once again the definition of $D$ in \eqref{eq:scal}, the fact that  $\delta_{sc}(2,0,l,\boldsymbol{p})-ld=2\Delta_1-l[\psi]-\|\boldsymbol{p}\|_1$, and the bound \eqref{item1bis} with $j=0$ (note that $\ell_{v_0}=(2,0,l,\boldsymbol{p})$ and that the labels $\ell_1,\ldots,\ell_{s_{v_0}}$ in \eqref{eq:lem:2} must be identified with $\ell_1(v_0),\ldots,\ell_{s_{v_0}}(v_0)$ in  \eqref{item1bis}), we get 
	\begin{equation}\begin{split}\label{eq:4.1.1} 
			&\sum_{(\ell_i(v_0))_{i=1}^{s_{v_0}}}^*\tn{D^kS^{\ell_1(v_0),\ldots,\ell_{s_{v_0}}(v_0)}
				_{2,0,l,\boldsymbol{p}}(H_{\ell_1(v_0)}[\tau_1(v_0)],\ldots,H_{\ell_{s_{v_0}}(v_0)}[\tau_{s_{v_0}}(v_0)])(\boldsymbol{x})}\\
			& \qquad  \le \gamma^{k(2\Delta_1-l[\psi]-\|\boldsymbol{p}\|_1)}\sum^*_{(\ell_i(v_0))_{i=1}^{s_{v_0}}} C_{\gamma}^{s_{v_0}-1}(4C_0)^{\sum_{i=1}^{s_{v_0}}l_i(v_0)-l_{v_0}}
			N_{\ell_{v_0}}^{\ell_1(v_0),\ldots,\ell_{s_{v_0}}(v_0)}
			\\
			&\qquad\cdot \Big(\prod_{i=2}^{s_{v_0}}\|H_{\ell_i(v_0)}[\tau_i(v_0)]\|_w\Big)\tn{H_{\ell_1(v_0)}{[\tau_1(v_0)](\gamma^{k}\boldsymbol{x})}}.
	\end{split}\end{equation} 
	In order to bound the  last factor in the last line, we apply \eqref{item3} with $i=1$ and $v_1\equiv v_1(v_0)$, thus getting:
	\begin{equation}\begin{split}\label{eq:4.1.100} 
			&\sum_{(\ell_i(v_0))_{i=1}^{s_{v_0}}}^*\tn{D^kS^{\ell_1(v_0),\ldots,\ell_{s_{v_0}}(v_0)}
				_{2,0,l,\boldsymbol{p}}(H_{\ell_1(v_0)}[\tau_1(v_0)],\ldots,H_{\ell_{s_{v_0}}(v_0)}[\tau_{s_{v_0}}(v_0)])(\boldsymbol{x})}\\
			&\qquad  \le \gamma^{k(2\Delta_1-l[\psi]-\|\boldsymbol{p}\|_1)}\sum^*_{(\ell_i(v_0))_{i=1}^{s_{v_0}}} C_{\gamma}^{s_{v_0}-1}(4C_0)^{\sum_{i=1}^{s_{v_0}}l_i(v_0)-l_{v_0}}
			N_{\ell_{v_0}}^{\ell_1(v_0),\ldots,\ell_{s_{v_0}}(v_0)}
			\\
			&\qquad \cdot\Big(\prod_{i=2}^{s_{v_0}}\|H_{\ell_i(v_0)}[\tau_i(v_0)]\|_w\Big)\sum_{k_{v_1}\ge1}\gamma^{k_{v_1}(2\Delta_1-l_{v_1}[\psi]-\|\boldsymbol{p}_{v_1}\|_1)}\\
			&\qquad \cdot\sum^*_{(\ell_i(v_1))_{i=1}^{s_{v_1}}}\tn{S_{\ell_{v_1}}^{\ell_1(v_1),\dots,\ell_{s_{v_1}}(v_1)}(H_{\ell_1(v_1)}[\tau_1(v_1)],\dots,H_{\ell_{s_{v_1}}(v_1)}[\tau_{s_{v_1}}(v_1)])(\gamma^{k+k_{v_1}}\boldsymbol{x})},
	\end{split}\end{equation} 
	where we used once again that $\delta_{sc}(2,0,l_{v_1},\boldsymbol{p}_{v_1})-dl_{v_1}=2\Delta_1-l_{v_1}[\psi]-\|\boldsymbol{p}_{v_1}\|_1$. 
	
	In order to bound the last line of \eqref{eq:4.1.100}, we iteratively apply \eqref{item1bis} and \eqref{item3} $n-1$ more times, write at each step $\delta_{sc}(2,0,l_{v_i},\boldsymbol{p}_{v_i})-dl_{v_i}$ as 
	$2\Delta_1-l_{v_i}[\psi]-\|\boldsymbol{p}_{v_i}\|_1$, 
	thus finding:
	\begin{equation}\label{n}\begin{split}
			&\sum_{(\ell_i(v_0))_{i=1}^{s_{v_0}}}^*\tn{D^kS^{\ell_1(v_0),\ldots,\ell_{s_{v_0}}(v_0)}_{2,0,l,\boldsymbol{p}}(H_{\ell_1(v_0)}[\tau_1(v_0)],\ldots,H_{\ell_{s_{v_0}}(v_0)}[\tau_{s_{v_0}}(v_0)])(\boldsymbol{x})}\\
			&\le \gamma^{k(2\Delta_1-l[\psi]-\|\boldsymbol{p}\|_1)}\sum_{(\ell_i(v_0))_{i=1}^{s_{v_0}}}^*\cdots \sum_{(\ell_i(v_{n-1}))_{i=1}^{s_{v_{n-1}}}}^*\sum_{k_{v_1},\ldots,k_{v_n}\ge 1}\Big(\prod_{i=1}^n\gamma^{k_{v_i}(2\Delta_1-l_{v_i}[\psi]-\|\boldsymbol{p}_{v_i}\|_1)}\Big)\\
			&\cdot\Big(\prod_{i=0}^{n-1}C_\gamma^{s_{v_i}-1}(4C_0)^{\sum_{j=1}^{s_{v_i}}l_j(v_i)-l_{v_i}}
			N_{\ell_{v_i}}^{\ell_1(v_i),\ldots,\ell_{s_{v_i}}(v_i)}
			\prod_{j=2}^{s_{v_i}}\|H_{\ell_{v_j}(v_i)}[\tau_{v_j}(v_i)]\|_w\Big)\\
			&\cdot\sum^*_{(\ell_i(v_n))_{i=1}^{s_{v_n}}}\tn{S_{\ell_{v_n}}^{\ell_1(v_n),\ldots,\ell_{s_{v_n}}(v_n)}(H_{\ell_1(v_n)}[\tau_1(v_n)],\ldots,H_{\ell_{s_{v_n}}(v_n)}[\tau_{s_{v_n}}(v_n)])(
				\gamma^{k+k_{v_1}+\cdots+k_{v_n}}\boldsymbol{x})}.\end{split}
	\end{equation}
	In order to bound the last line, recalling that $v^*_1$ and $v^*_2$ belong to two different subtrees of $\tau_{v_n}$, we use the analogue of Eqs.\eqref{4.1}--\eqref{eq:4.30_2}, which is proved in the same way:
	\begin{equation}\label{analogue}\begin{split}
			&\sum^*_{(\ell_i(v_n))_{i=1}^{s_{v_n}}}\tn{S_{\ell_{v_n}}^{\ell_1(v_n),\ldots,\ell_{s_{v_n}}(v_n)}(H_{\ell_1(v_n)}[\tau_1(v_n)],\ldots,H_{\ell_{s_{v_n}}(v_n)}[\tau_{s_{v_n}}(v_n)])(
				\gamma^{k+k_{v_1}+\cdots+k_{v_n}}\boldsymbol{x})}\\
			&\qquad \le N^2d^2 C_{\chi_1} e^{-\bar{C}(\gamma^{k+k_{v_1}+\cdots+k_{v_n}-1}|x|)^{\sigma}} \\
			&\qquad \cdot\sum_{(\ell_i(v_n))_{i=1}^{s_{v_n}}}^* C_\gamma^{s_{v_n}-2}(4C_0)^{\sum_{i=1}^{s_{v_n}}l_{i}(v_n)-l_{v_n}}
			N_{\ell_{v_n}}^{\ell_1(v_n),\ldots,\ell_{s_{v_n}}(v_n)}
			\prod_{i=1}^{s_{v_n}} \|H_{\ell_i(v_n)}[\tau_i(v_n)]\|_{w}.\end{split}\end{equation}
	Plugging \eqref{analogue} in \eqref{n}, and bounding each factor $\|H_{\ell_v}[\tau_v]\|_w$ as in \eqref{eq:4.22} (with $\rho$ replaced by $\rho'$), we get:
	\begin{equation} \label{n1} \begin{split}
			&\sum_{(\ell_i(v_0))_{i=1}^{s_{v_0}}}^*\tn{D^kS^{\ell_1(v_0),\ldots,\ell_{s_{v_0}}(v_0)}
				_{2,0,l,\boldsymbol{p}}(H_{\ell_1(v_0)}[\tau_1(v_0)],\ldots,H_{\ell_{s_{v_0}}(v_0)}[\tau_{s_{v_0}}(v_0)])(\boldsymbol{x})}\\
			&\le C (C'\rho'\epsilon)^{n_{e.p.}[\tau]-2}(4C_0)^{-l}\sum_{(\ell_i(v_0))_{i=1}^{s_{v_0}}}^*\cdots \sum_{(\ell_i(v_{n}))_{i=1}^{s_{v_{n}}}}^*
			\sum_{k_{v_1},\ldots,k_{v_n}\ge 1}e^{-\bar C(\gamma^{k+\sum_{i=1}^nk_{v_i}-1}|x|)^\sigma}\\
			&\cdot N_{\ell_{v_0}}^{\ell_1(v_0),\ldots, \ell_{s_{v_0}}(v_0)}\gamma^{k(2\Delta_1-l[\psi]-\|\boldsymbol{p}\|_1)}\Big(\prod_{i=1}^n N_{\ell_{v_i}}^{\ell_1(v_i),\ldots, \ell_{s_{v_i}}(v_i)}\gamma^{k_{v_i}(2\Delta_1-l_{v_i}[\psi]-\|\boldsymbol{p}_{v_i}\|_1)}\Big)\\
			&\cdot\Big(\prod_{i=0}^{n-1}\prod_{j=2}^{s_{v_i}}\big(\frac{\rho'}{4 C_0}\big)^{-l_j(v_i)}\Big)\, \Big(\prod_{j=1}^{s_{v_n}}\big(\frac{\rho'}{4 C_0}\big)^{-l_j(v_n)}\Big)
		\end{split}
	\end{equation}
	for some $C,C'>0$. %, and we denoted $k_{v_0}\equiv k$.}}
	We are left with sums over $\ell_{v_1}=(l_{v_1},\boldsymbol{p}_{v_1})$, $\ldots$, $\ell_{v_n}=(l_{v_n},\boldsymbol{p}_{v_n})$, and  over $(\ell_i(v_0))_{i=2}^{s_{v_0}}$, $\ldots$, 
	$(\ell_i(v_{n-1}))_{i=2}^{s_{v_{n-1}}}$, $(\ell_i(v_n))_{i=1}^{s_{v_n}}$.  We first sum over $\boldsymbol{p}_{v_1}, \ldots, \boldsymbol{p}_{v_n}$ and, using \eqref{sumoverp0}, we get: 
	\begin{equation} \label{n1.1} \begin{split}
	&\sum_{(\ell_i(v_0))_{i=1}^{s_{v_0}}}^*\tn{D^kS^{\ell_1(v_0),\ldots,\ell_{s_{v_0}}(v_0)}
		_{2,0,l,\boldsymbol{p}}(H_{\ell_1(v_0)}[\tau_1(v_0)],\ldots,H_{\ell_{s_{v_0}}(v_0)}[\tau_{s_{v_0}}(v_0)])(\boldsymbol{x})}\\
	&\le C (C'\rho'\epsilon)^{n_{e.p.}[\tau]-2}(4C_0)^{-l} \gamma^{k(2\Delta_1-l[\psi]-\|\boldsymbol{p}\|_1)}\sum_{(\ell_i(v_0))_{i=2}^{s_{v_0}}}^*\cdots \sum_{(\ell_i(v_{n-1}))_{i=2}^{s_{v_{n-1}}}}^*\sum_{(\ell_i(v_{n}))_{i=1}^{s_{v_{n}}}}^*\cdot\\
	&\cdot\Biggl(\prod_{i=0}^{n-1}\Big(\prod_{j=2}^{s_{v_i}}\big(\frac{\rho'}{4 C_0}\big)^{-l_j(v_i)}\Big)\Biggr)\, \Big(\prod_{j=1}^{s_{v_n}}\big(\frac{\rho'}{4 C_0}\big)^{-l_j(v_n)}\Big)
	\sum_{k_{v_1},\ldots,k_{v_n}\ge 1}e^{-\bar C(\gamma^{k+\sum_{i=1}^nk_{v_i}-1}|x|)^\sigma}\cdot\\
	&\cdot		 \sum_{l_{v_1},\ldots,l_{v_n}\ge 2}	\binom{\sum_{j=1}^{s_{v_0}}l_{j}(v_0)}{l}
	\Big(\prod_{i=1}^n \binom{\sum_{j=1}^{s_{v_i}}l_{j}(v_i)}{l_{v_i}}\gamma^{k_{v_i}(2\Delta_1-l_{v_i}[\psi])}\Big), 
\end{split}
\end{equation}
{where, if $s_{v_{i_0}}=1$ for some $i_0\in\{0,\ldots,n-1\}$, then $\sum^*_{(\ell_i(v_{i_0}))_{i=2}^{s_{v_{i_0}}}}\Big(\prod_{j=2}^{s_{v_{i_0}}}\big(\frac{\rho'}{4 C_0}\big)^{-l_j(v_{i_0})}\Big)$ should be interpreted as being equal to $1$.}
In the last line, we bound $\binom{\sum_{j=1}^{s_{v_0}}l_{j}(v_0)}{l}$ from above by $2^{\sum_{j=1}^{s_{v_0}}l_{j}(v_0)}$, and, 
recalling that $\Delta_1=[\psi]$ and $k_{v_i}\ge 1$, we rewrite and bound: $\gamma^{k_{v_i}(2\Delta_1-l_{v_i}[\psi])}=\gamma^{-k_{v_i}(l_{v_i}-2)[\psi]} 
\le \gamma^{-(l_{v_i}-2)[\psi]}$, so that, proceeding as in \cite[Appendix A.6.1]{GeM01},
the sum over $l_{v_1},\ldots,l_{v_n}$ in the last line can be bounded as follows: 
\begin{equation}\label{eq:5.25}\begin{split} & 2^{\sum_{j=2}^{s_{v_0}}l_j(v_0)}
	\sum_{l_{v_1},\ldots,l_{v_n}\ge 2} 2^{l_{v_1}}\Big(\prod_{i=1}^n \binom{\sum_{j=1}^{s_{v_i}}l_{j}(v_i)}{l_{v_i}}\gamma^{k_{v_i}(2\Delta_1-l_{v_i}[\psi])}\Big)\\
	& \le 	\gamma^{2[\psi]n}2^{\sum_{j=2}^{s_{v_0}}l_j(v_0)} \sum_{l_{v_1},\ldots,l_{v_n}\ge 0}	2^{l_{v_1}}
	\Big(\prod_{i=1}^n \binom{\sum_{j=1}^{s_{v_i}}l_{j}(v_i)}{l_{v_i}}\gamma^{-l_{v_i}[\psi]}\Big)\\
	&\le \gamma^{2[\psi]n}2^{\sum_{j=2}^{s_{v_0}}l_j(v_0)}(1+2\gamma^{-[\psi]})^{\sum_{j=2}^{s_{v_1}}l_j(v_1)}\cdots \big(\sum_{m=0}^{n-2}\gamma^{-m[\psi]}+2\gamma^{-(n-1)[\psi]}\big)^{\sum_{j=2}^{s_{v_{n-1}}}l_j(v_{n-1})}\cdot\\
	&\qquad \cdot(\sum_{m=0}^{n-1}\gamma^{-m[\psi]}+2\gamma^{-n[\psi]})^{\sum_{j=1}^{s_{v_{n}}}l_j(v_{n})}\\
	& \le\gamma^{2[\psi]n}\big(\frac2{1-\gamma^{-[\psi]}}\big)^{\sum_{i=0}^{n-1}\sum_{j=2}^{s_{v_i}}l_j(v_i)+\sum_{j=1}^{s_{v_{n}}}l_j(v_{n})},
	\end{split}\end{equation}	
	where in the last step we bounded $\max\{2,\max_{n\ge 1}\{\sum_{m=0}^{n-1}\gamma^{-m[\psi]}+2\gamma^{-n[\psi]})\}\}$ by $2/(1-\gamma^{-[\psi]})$. 
	
	Inserting this into \eqref{n1.1} and noting that $1\le n\le 2n_{\text{e.p}}[\tau]$ (the upper bound on 
	$n$ following from \cite[Eq.(J.10)]{GMR21}), so that the factor $\gamma^{2[\psi]n}$ can be re-absorbed in a re-definition of $C,C'$, we find:
	\begin{equation} \label{n1.2} \begin{split}
	&\sum_{(\ell_i(v_0))_{i=1}^{s_{v_0}}}^*\tn{D^kS^{\ell_1(v_0),\ldots,\ell_{s_{v_0}}(v_0)}
		_{2,0,l,\boldsymbol{p}}(H_{\ell_1(v_0)}[\tau_1(v_0)],\ldots,H_{\ell_{s_{v_0}}(v_0)}[\tau_{s_{v_0}}(v_0)])(\boldsymbol{x})}\\
	&\le C (C'\rho'\epsilon)^{n_{e.p.}[\tau]-2}(4C_0)^{-l}\gamma^{k(2\Delta_1-l[\psi]-\|\boldsymbol{p}\|_1)}
	\sum_{(\ell_i(v_0))_{i=2}^{s_{v_0}}}^*\cdots \sum_{(\ell_i(v_{n-1}))_{i=2}^{s_{v_{n-1}}}}^*\sum_{(\ell_i(v_{n}))_{i=1}^{s_{v_{n}}}}^*\cdot\\
	&\cdot\Biggl(\prod_{i=0}^{n-1}\Big(\prod_{j=2}^{s_{v_i}}\big(\frac{\rho'(1-\gamma^{-[\psi]})}{8C_0}\big)^{-l_j(v_i)}\Big)\Biggr)\, \Big(\prod_{j=1}^{s_{v_n}}\big(\frac{\rho'(1-\gamma^{-[\psi]})}{8C_0}\big)^{-l_j(v_n)}\Big)
	\sum_{k_{v_1},\ldots,k_{v_n}\ge 1}e^{-\bar C(\gamma^{k+\sum_{i=1}^nk_{v_i}-1}|x|)^\sigma}.
\end{split}
\end{equation}
%up to a possible re-definition of $C,C'$. 
{As above, if $s_{v_{i_0}}=1$ for some $i_0\in\{0,\ldots,n-1\}$, then $\sum^*_{(\ell_i(v_{i_0}))_{i=2}^{s_{v_{i_0}}}}\Big(\prod_{j=2}^{s_{v_{i_0}}}\big(\frac{\rho'(1-\gamma^{-[\psi]})}{8 C_0}\big)^{-l_j(v_{i_0})}\Big)$ should be interpreted as being equal to $1$.} Choosing $\rho'$ larger than ${8}C_0(1-\gamma^{-[\psi]})^{-1}$, the sums
$$\sum_{(\ell_i(v_0))_{i=2}^{s_{v_0}}}^*\cdots \sum_{(\ell_i(v_{n-1}))_{i=2}^{s_{v_{n-1}}}}^*\sum_{(\ell_i(v_{n}))_{i=1}^{s_{v_{n}}}}^*\Big(\prod_{i=0}^{n-1}\prod_{j=2}^{s_{v_i}}\big(\frac{\rho'(1-\gamma^{-[\psi]})}{8C_0}\big)^{-l_j(v_i)}\Big)\, \Big(\prod_{j=1}^{s_{v_n}}\big(\frac{\rho'(1-\gamma^{-[\psi]})}{8C_0}\big)^{-l_j(v_n)}\Big)$$
are absolutely summable and bounded from above by $(\text{const.})^{n_{\text{e.p.}}[\tau]}$, which can be also re-absorbed into a re-definition of $C,C'$. In conclusion, for some $C''>0$,

\begin{equation}\begin{split}
	&\sum_{(\ell_i(v_0))_{i=1}^{s_{v_0}}}^*\tn{D^kS^{\ell_1(v_0),\cdots,\ell_{s_{v_0}}(v_0)}
		_{2,0,l,\boldsymbol{p}}(H_{\ell_1(v_0)}[\tau_1(v_0)],\ldots,H_{\ell_{s_{v_0}}(v_0)}[\tau_{s_{v_0}}(v_0)])(\boldsymbol{x})}\\
	&\le C''(C''\epsilon)^{n_{\text{e.p}}[\tau]-2}(4C_0)^{-l}\gamma^{k(2\Delta_1-l[\psi]-\|\boldsymbol{p}\|_1)}
	\sum_{k_{v_1},\ldots,k_{v_n}\ge 1}e^{-\bar{C}(\gamma^{k+\sum_{i=i}^n k_{v_i}-1}|x|)^{\sigma}}\\
	&= C''(C''\epsilon)^{n_{\text{e.p}}[\tau]-2}(4C_0)^{-l}\gamma^{k(2\Delta_1-l[\psi]-\|\boldsymbol{p}\|_1)}
	\sum_{\bar k\ge n}\binom{\bar{k}-1}{n-1}e^{-\bar{C}(\gamma^{k+\bar{k}-1}|x|)^{\sigma}}.
\end{split}	
\label{eq:Last:Step}
\end{equation}	
In order to bound the sum over $\bar k$, we use that, for any $\alpha>0$, $$\binom{\bar k-1}{n-1}\le \frac{(\bar k-1)^{n-1}}{(n-1)!}\le (\alpha\log\gamma)^{1-n}\gamma^{(\bar k-1)\alpha},$$
so that 
\begin{equation}\begin{split} \sum_{\bar k\ge n}\binom{\bar{k}-1}{n-1}e^{-\bar{C}(\gamma^{k+\bar{k}-1}|x|)^{\sigma}}&\le  (\alpha\log\gamma)^{1-n} 
	e^{-\frac{\bar{C}}2(\gamma^{k+n-1}|x|)^{\sigma}} 
	\sum_{\bar k\ge n}\gamma^{(\bar k-1)\alpha}e^{-\frac{\bar{C}}2(\gamma^{k+\bar{k}-1}|x|)^{\sigma}}\\
	&\le (\alpha\log\gamma)^{1-n} 
	e^{-\frac{\bar{C}}2(\gamma^{k+n-1}|x|)^{\sigma}} C_{\alpha,\bar{C}/2}(\min\{1,\gamma^k|x|\})^{-\alpha},\end{split} \label{quandofinisce}\end{equation}
	where, in the last inequality, letting $C_{\alpha,\bar{C}/2}$ as in \eqref{largerbound},  we bounded $\sum_{\bar k\ge n}\gamma^{(\bar k-1)\alpha}e^{-\frac{\bar{C}}2(\gamma^{k+\bar{k}-1}|x|)^{\sigma}}$ from 
	above by:
	\begin{equation} \begin{cases} \sum_{\bar k\ge n}\gamma^{(\bar k-1)\alpha}e^{-\frac{\bar{C}}2(\gamma^{\bar{k}-1})^{\sigma}}\le\gamma^{-\alpha} C_{\alpha,\bar{C}/2}, & \text{if}\ \gamma^k|x|\ge 1,\\ 
	(\gamma^k|x|)^{-\alpha}
	\sum_{\bar k\ge n}\gamma^{(k+\bar k-h_x-1)\alpha}e^{-\frac{\bar{C}}2(\gamma^{k+\bar{k}-h_x-2})^{\sigma}}\le
	(\gamma^k|x|)^{-\alpha}C_{\alpha,\bar{C}/2}, & \text{if} \ \gamma^k|x|<1, \end{cases}\end{equation}
	where $h_x$ was defined before \eqref{largerbound}. Plugging \eqref{quandofinisce} in \eqref{eq:Last:Step}, and noting that $1\le n\le 2n_{\text{e.p}}[\tau]$ (the upper bound on 
	$n$ following from \cite[Eq.(J.10)]{GMR21}), we finally obtain the desired bound, \eqref{eq:lem:2}, provided that  $C_0$ is chosen larger than ${1}/4$ (once again, we have the freedom to choose $C_0$ as large as desired, due to the monotonicity of the right hand side of \eqref{eq:Last:Step}
	on $C_0$: recall, in fact, that $C''$ is proportional to $C_0^4$, cf. with \eqref{TreeBound1}, and that $4n_{\text{e.p.}}[\tau]\ge l$). 
	
	The proof of \eqref{eq:lem:2bis} is completely analogous, the only difference being that $x$ must be replaced by $y$ and $\Delta_1$ by $\Delta_2$. Note that, by changing $\Delta_1$ in $\Delta_2$, 
	we get an extra factor $\gamma^{(2\Delta_2-2[\psi])\sum_{i=1}^nk_{v_i}}$ in the analogues of the right hand side of \eqref{eq:5.25} and of the following equations; in particular, it also  reflects into the analogue of 
	\eqref{eq:Last:Step}, which reads: 
	\begin{equation}\begin{split}
	&\sum_{(\ell_i(v_0))_{i=1}^{s_{v_0}}}^*\tn{S^{\ell_1(v_0),\cdots,\ell_{s_{v_0}}(v_0)}
		_{0,2,l,\boldsymbol{p}}(H_{\ell_1(v_0)}[\tau_1(v_0)],\ldots,H_{\ell_{s_{v_0}}(v_0)}[\tau_{s_{v_0}}(v_0)])(\boldsymbol{y})}\\
	&\le C''(C''\epsilon)^{n_{\text{e.p}}[\tau]-2}(4C_0)^{-l}\gamma^{k(2\Delta_2-l[\psi]-\|\boldsymbol{p}\|_1)}
	\sum_{\bar k\ge n}\binom{\bar{k}-1}{n-1}\gamma^{\bar k(2\Delta_2-2[\psi])}e^{-\bar{C}(\gamma^{k+\bar{k}-1}|y|)^{\sigma}}.
\end{split}	
\label{eq:Last:Stepbis}
\end{equation}	
Bounding the sum over $\bar k$ as explained after \eqref{eq:Last:Step} and choosing  $C_0$ larger than ${1}/4$ 
implies the desired estimate, \eqref{eq:lem:2bis}. 
\end{proof}

\section{Stretched exponential decay of the correction terms \texorpdfstring{$\mathcal E_1$}{E1}, \texorpdfstring{$\mathcal E_2$}{E2}}\label{sec6}

In this section we prove Theorem \ref{Th:2}. We focus on the bound on $\mathcal E_1$, {the bound on $\mathcal E_2$ being completely analogous (and, therefore, left to the reader)}. From the definition of $\mathcal{E}_1$ in \eqref{eq:E1E2}, letting $\boldsymbol{x}=(x,0)$, we have: 
\begin{equation}\label{eq:6.1}\big[\mathcal E_1(x)\big]_{a,b}= 2\big[V^*_{2,0,0,\emptyset}(\boldsymbol{x})\big]_{a,b}+\big[\mathfrak{E}_1(x)\big]_{a,b},\end{equation}
where $\big[\mathfrak{E}_1(x)\big]_{a,b}=\big\langle \frac{\delta^2 \mathcal Q^*(\phi,0,\psi)}{\delta\phi_b(0)\delta\phi_a(x)}\big|_{\phi=0}\big\rangle_{H^*}$ with  $\mathcal Q^*(\phi,J,\psi)$ defined in \eqref{rewrv*} and following line, i.e., 
\begin{equation} \label{eqdefQ*}\mathcal Q^*=\{V^*_\ell\}_{\ell\in L_f(2,0)}\end{equation}
(here, letting $L(2,0):=\{(n,m,l,\boldsymbol{p})\in L : (n,m)=(2,0)\}$, we defined $L_f(2,0):=L(2,0)\setminus\{(2,0,0,\emptyset)\}$). 
The first term in the right hand side of \eqref{eq:6.1} has already been analyzed in the previous sections, and proved to be analytic in $\epsilon$ for $\epsilon$ small. Moreover, as we shall see shortly, the tree representation \eqref{eq:4.1} and the bounds derived in the preceding section on the tree values $H_{2,0,0,\emptyset}[\tau](\boldsymbol{x})$ readily imply the stretched exponential decay of $V^*_{2,0,0,\emptyset}(\boldsymbol{x})$. Concerning the second term, we will prove below that 
it can be computed via a convergent expansion in which the kernels $V^*_{2,0,l,\boldsymbol{p}}$ of $\mathcal Q^*$ are `contracted' with those of $H^*$. Not surprisingly, 
$\mathfrak E_1(x)$ can be expressed once again in terms of a tree expansion, slightly different from the one of the preceding sections, in that it includes only a subset of the trees 
contributing to the scale-invariant, fixed-point, potential studied above. As we shall see, the trees contributing to $\mathfrak E_1$ are characterized by a constraint on the `scale indices' (called $h, k_1, k_2,\ldots$ in the following), implying the desired stretched exponential decay. In order to simplify the analysis and to rely on the bounds derived in the preceding sections as much as possible, we shall define the trees contributing to $\mathfrak E_1$ in terms of `dressed endpoints' (rather than of the `bare endpoints' in Figure \ref{eq:figtree3.1}): 
these are the big white and black endpoints of, e.g., Figure \ref{fig:6.1} below, representing the kernels $V^*_{2,0,l,\boldsymbol{p}}$ and $H^*_{l,\boldsymbol{p}}$, respectively. In order to estimate the values of the trees contributing to $\mathfrak E_1$, we shall 
use the bounds on $V^*_{2,0,l,\boldsymbol{p}}$ and $H^*_{l,\boldsymbol{p}}$ derived above and in \cite{GMR21}, without re-expanding from scratch these kernels as sums over trees 
once again.

\medskip

Let us consider the first term in the right hand side of \eqref{eq:6.1}. As mentioned above, we already proved in the previous sections that it is analytic in $\epsilon$ for $\epsilon$ small. Moreover, using the first identity in \eqref{eq:4.1} and the bound 
\eqref{wefindthisbound} with $h=0$, we have (dropping as usual the component labels): 
\begin{equation}\label{eq:6.2}\begin{split}
\big|V^*_{2,0,0,\emptyset}(\boldsymbol{x})\big|
&\le C_\alpha|x|^{-2\Delta_1}\sum_\tau(C_\alpha\epsilon)^{n_{\text{e.p}}[\tau]-2}\sum_{k\ge1} 
e^{-\frac{\bar{C}}2(\gamma^{k-1}|x|)^{\sigma}} \max\{(\gamma^k|x|)^{2\Delta_1},(\gamma^{k}|x|)^{2\Delta_1-\alpha}\}\\
& \le C_\alpha' |x|^{-2\Delta_1}e^{-\frac{\bar C}{4}(|x|/\gamma)^\sigma}\sum_{k\ge1} 
e^{-\frac{\bar{C}}4(\gamma^{k-1}|x|)^{\sigma}} \max\{(\gamma^k|x|)^{2\Delta_1},(\gamma^{k}|x|)^{2\Delta_1-\alpha}\},
\end{split}
\end{equation}	
where, in passing from the first to the second line, we performed the sum over $\tau$ of $(C_\alpha\epsilon)^{n_{\text{e.p}}[\tau]-2}$ (which is summable, and whose sum is bounded by an $\epsilon$-independent constant), and we extracted part of the exponential factor from the sum over $k$. Now, if $|x|\ge 1$, the summand in the sum over $k$ can be bounded from above by 
$e^{-\frac{\bar{C}}4\gamma^{\sigma(k-1)}} \gamma^{2k\Delta_1}|x|^{2\Delta_1}$, so that the right hand side of \eqref{eq:6.2} can be bounded from above by (const.)$e^{-\frac{\bar C}{4}(|x|/\gamma)^\sigma}$, with 
(const.)$=C_\alpha'\sum_{k\ge 1}e^{-\frac{\bar{C}}4\gamma^{\sigma(k-1)}} \gamma^{2k\Delta_1}$. On the other hand, if $|x|\le 1$, the sum over $k$ in the right hand side of \eqref{eq:6.2} can be bounded 
via \eqref{largerbound}, so that the right hand side of \eqref{eq:6.2} can be bounded from above by (const.)$|x|^{-2\Delta_1}e^{-\frac{\bar C}{4}(|x|/\gamma)^\sigma}$. Putting the two cases together, we find
\begin{equation}\label{eq:6.3}
\big|V^*_{2,0,0,\emptyset}(\boldsymbol{x})\big|\le Ce^{-\frac{\bar C}{4}(|x|/\gamma)^\sigma}(\min\{1,|x|\})^{-2\Delta_1}. \end{equation}
Note that the right hand side has the same form as that of the first inequality in \eqref{eq:E}. Therefore, in order to complete the proof of Theorem \ref{Th:2}, we need to prove a comparable bound on $\mathfrak{E}_1(x)$, as well as its analyticity. 

For this purpose, we derive a tree expansion for $\mathfrak{E}_1(x)$. We write (keeping, once again, the flavor indices implicit): 
\begin{equation} \mathfrak{E}_1(x)=\frac{\delta^2 \mathcal Q_{\text{eff}}(\phi,0)}{\delta\phi(0)\delta\phi(x)}\big|_{\phi=0}, 
\end{equation}
where, in analogy with \eqref{eq:1bis} and \eqref{W*phiJ}, 
\begin{equation}\begin{split}\mathcal Q_{\text{eff}}(\phi,J)&=\lim_{h\to-\infty}\log\frac{\int d\mu_{[h,0]}(\psi)e^{H^*(\psi)+\mathcal Q^*(\phi,J,\psi)}}{\int d\mu_{[h,0]}(\psi)e^{H^*(\psi)}}\\
&\equiv\lim_{h\to-\infty}D^h v^{(h)}(\phi,J,0),\end{split}\end{equation}
with
\begin{equation} v^{(h)}(\phi,J,\psi):=R^{|h|}(H^*+\mathcal Q^*)(\phi,J,\psi)-H^*(\psi),\label{definitionvh}	\end{equation}
and $\mathcal Q^*$ as in \eqref{eqdefQ*}.
Denoting by $v^{(h)}_{\ell}$ with $\ell\in L$ the kernels of $v^{(h)}$ (which are by construction non-vanishing only for indices $\ell=(n,m,l,\boldsymbol{p})$ such that $n+m\ge 2$), we can then 
rewrite
\begin{equation}\label{E1v2000}\begin{split} \mathfrak{E}_1(x)&=2\lim_{h\to-\infty}D^hv^{(h)}_{2,0,0,\emptyset}(\boldsymbol{x})\\
&=2\lim_{h\to-\infty}\gamma^{2h\Delta_1}v^{(h)}_{2,0,0,\emptyset}(\gamma^h\boldsymbol{x}).\end{split}
\end{equation}
In view of the definition \eqref{definitionvh} of $v^{(h)}$, of the fact that its kernels $v^{(h)}_\ell$ are non-vanishing only for indices $\ell=(n,m,l,\boldsymbol{p})$ such that $n+m\ge 2$,  
and of the representation \eqref{RGmapincomponents} of the action of the RG map $R$ in components, for any $\ell\in L(2,0)=\{(n,m,l,\boldsymbol{p})\in L : (n,m)=(2,0)\}$, for any $h<0$ we can write: 
\begin{equation} \label{iteravh}
v^{(h)}_\ell=\sum_{s\ge 1}s\sum_{\ell_1\in L_f(2,0)}\sum_{(\ell_i)_{i=2}^s}R^{\ell_1,\ldots,\ell_s}_\ell(v^{(h+1)}_{\ell_1},H^*_{\ell_2},\ldots,H^*_{\ell_s}), \end{equation}
where we recall that $L_f(2,0):=L(2,0)\setminus\{(2,0,0,\emptyset)\}$, and the third sum in the right hand side runs over the $(s-1)$-ples of labels in $L(0,0):=\{(n,m,l,\boldsymbol{p})\in L : (n,m)=(0,0)\}$
(note that for any such label, of the form $\ell=(0,0,l,\boldsymbol{p})$, $V^*_{0,0,l,\boldsymbol{p}}\equiv H^*_{l,\boldsymbol{p}}$, with $H^*$ the FP potential constructed in \cite{GMR21}, see condition (1) in Definition \ref{Def.1}; for this reason, with some abuse of notation, in \eqref{iteravh} we wrote $H^*_\ell$ in place of $V^*_\ell$). From \eqref{R} we see that, for any $\ell\in L(2,0)$, 
$R_\ell^{\ell_1,\ldots,\ell_s}=D S_\ell^{\ell_1,\ldots,\ell_s}$, so that, isolating from the right hand side of \eqref{iteravh} the contribution with $s=1$ and $\ell_1=\ell$, we can rewrite: 
\begin{equation} \label{iteravhbis}
v^{(h)}_\ell=Dv^{(h+1)}_\ell+\sum_{s\ge 1}s\sum_{\ell_1\in L_f(2,0)}^*\sum_{(\ell_i)_{i=2}^s}DS^{\ell_1,\ldots,\ell_s}_\ell(v^{(h+1)}_{\ell_1},H^*_{\ell_2},\ldots,H^*_{\ell_s}), \end{equation}
where the $*$ on the second sum in the right hand side indicates the constraint that, if $s=1$, then $\ell_1\neq\ell$. Note also that, by definition, for $h=-1$, the kernel
$v^{(0)}_{\ell}$ in the right hand side is $v^{(0)}_{\ell}\equiv V^*_\ell$, for all $\ell\in L_f(2,0)$, and zero otherwise. Therefore, for any $\ell\in L(2,0)$, 
\begin{equation} \label{iterav-1bis}
v^{(-1)}_\ell=DV^*_\ell\,\mathds{1}_{\ell\in L_f(2,0)}
+\sum_{s\ge 1}s\sum_{\ell_1\in L_f(2,0)}^*\sum_{(\ell_i)_{i=2}^s}DS^{\ell_1,\ldots,\ell_s}_\ell(V^*_{\ell_1},H^*_{\ell_2},\ldots,H^*_{\ell_s}). \end{equation}
If we graphically represent any kernel $V^*_\ell$ with $\ell\in L(2,0)$ by a big white dot, the first term in the right hand side, whenever it is there (i.e., for $\ell\neq(2,0,0,\emptyset)$), can be represented as  $D\,\tikz[baseline=-3pt]{\draw [fill=none] circle (0.15)  {} ; }$, while the 
second as a sum over trees $\tau$ of length $1$ (the length being the maximal number of branches along a path from the endpoints to the root) as in Fig.\ref{fig:6.1}. 
\begin{figure}[ht]
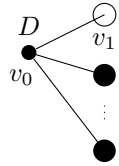

\begin{center}
\begin{tabular}{rcl}
	\tikz[baseline=-3pt]{\draw (0,0) node[vertex, label=below:{$v_0$\;\;}, label=above:{$D$}] (v0) {} -- (1,-0.3) node[bigvertex](v1){}; 
		\draw (v0) -- (1,0.5) circle(0.15) node[label=below:{$v_1$}](v2){}; 
		\draw (v0)-- (1,-1.3) node[bigvertex](v1){};
		\draw[dotted](1,-0.7)--(1,-0.9){};}
\end{tabular}
\end{center}\caption{A tree in $\mathcal{T}_1$, representing one of the contributions $v_\ell[\tau]$ in the right hand side of \eqref{iterav-1bis}, with the action of the dilatation operator $D$ associated with the root 
explicitly indicated. The elements of 
$\mathcal{T}_1$ consists of a root vertex $v_0$, followed by $s_{v_0}\ge 1$ children: the first (i.e., the topmost) one, denoted $v_1$, is a big white dot, representing $V^*_{\ell_{1}}$ (with $\ell_1$ an index to be summed over $L_f(2,0)$, such that $\ell_{1}\neq\ell$), while the others are all big-black-dot endpoints, representing $H^*_{\ell_{i}}$, $i=2,\ldots,s_{v_0}$, where $\ell_{i}$ are indices to be summed over $L(0,0)$.}
\label{fig:6.1}
\end{figure}
We correspondingly write: 
\begin{equation} \label{iterav-1tris}
v^{(-1)}_\ell=DV^*_\ell\,\mathds{1}_{\ell\in L_f(2,0)}+\sum_{\tau\in\mathcal{T}_1} v_\ell[\tau],\end{equation}
where $\mathcal{T}_1$ is the family of trees of length $1$ described in Figure \ref{fig:6.1} and in its caption, and $v_\ell[\tau]$ is its value: 
if $\tau$ has $s_{v_0}$ endpoints, then $v_\ell[\tau]=
s_{v_0}\sum_{\ell_1\in L_f(2,0)}^*\sum_{(\ell_i)_{i=2}^{s_{v_0}}}DS^{\ell_1,\ldots,\ell_{s_{v_0}}}_\ell(V^*_{\ell_1},H^*_{\ell_2},\ldots,H^*_{\ell_{s_{v_0}}})$. 

We can easily iterate this procedure, via \eqref{iteravhbis}, thus ending up with the following tree representation for $v^{(h)}_\ell$: 
\begin{equation} \label{iteravhtris}
v^{(h)}_\ell=D^{|h|}V^*_\ell\,\mathds{1}_{\ell\in L_f(2,0)}+\sum_{\mathfrak{L}=1}^{|h|}\sum_{\substack{k_{\mathfrak{L}}\ge 0, \\ k_0, k_1,\ldots,k_{\mathfrak{L}-1}\ge 1:\\ k_0+\cdots+k_{\mathfrak{L}}=|h|}}
\sum_{\tau\in\mathcal{T}_{\mathfrak{L}}}\, v_\ell[\tau,\boldsymbol{k}],\end{equation}
where $\mathfrak{L}$ represents the length of the tree (i.e., the number of branches crossed by the straight path from the root $v_0$ to the big white end point, see Figure \ref{fig:6.2}), $\mathcal{T}_{\mathfrak{L}}$ is the family of trees of length $\mathfrak{L}$ described in Figure \ref{fig:6.2} and in its caption,
$\boldsymbol{k}:=(k_0,k_1,\ldots,k_{\mathfrak{L}})$ and $v_\ell[\tau,\boldsymbol{k}]$\footnote{Note that for $h=-1$ there is 
only one possible choice of $\boldsymbol{k}=(k_0,k_1)$ compatible with the constraints $k_0\ge1$, $k_1\ge0$, $k_0+k_1=1$ indicated under the second sum in the right hand side of \eqref{iteravhtris}, that is 
$(k_0,k_1)=(1,0)$: therefore, for $h=-1$, the value $v_\ell[\tau]$ in the right hand side of \eqref{iterav-1tris} is the same denoted by $v_\ell[\tau,(1,0)]$ in \eqref{iteravhtris}.} is the value of the tree $\tau$, in the presence of the action of the dilatations $D^{k_0}$, $D^{k_1}$, $\ldots$, $D^{k_{\mathfrak{L}}}$ associated with the vertices $v_0$, $v_1$, $\ldots$, $v_{\mathfrak{L}}$, as in Figure \ref{fig:6.2}. 
\begin{figure}[ht]
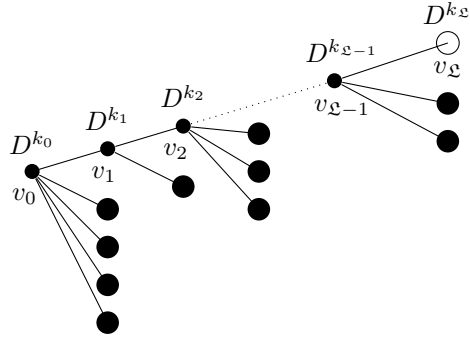

\begin{center}
\begin{tabular}{rcl}
	\tikz[baseline=-3pt]{ %orizzontale
		\draw(0,0)node[vertex, label=below:{$v_0$\;\;}, label=above:{$D^{k_0}$}](v0){}--+(1,0.3)node[vertex,label=below:{$v_1$\;}](v1){};
		\draw (v1)node[label=above:{$D^{k_1}$}]{};
		\draw(v1)--+(1,0.3)node[vertex,label=below:{$v_2$\;\;}](v2){};
		\draw (v2)node[label=above:{$D^{k_2}$}]{};
		\draw [dotted](v2)--+(2,0.6) node[vertex](v3){};
		\draw (4.1,1.2) node[label=above:{$D^{k_{\mathfrak{L}-1}}$}]{};
		\draw (4.1,1.2) node[label=below:{$v_{\mathfrak{L}-1}$}]{};
		\draw (v0)--(+1,-1.5)node[bigvertex]{};
		\draw (v0)--(+1,-2)node[bigvertex]{};
		\draw(v0)--(+1,-1)node[bigvertex]{};
		\draw(v0)--(+1,-0.5)node[bigvertex]{};
		\draw (v1)--+(1,-0.5)node[bigvertex]{};
		\draw (v2)--(3,-0.5)node[bigvertex]{};
		\draw(v2)--(3,+0.5)node[bigvertex]{};
		\draw(v2)--(3,0)node[bigvertex]{};
		\draw(v3)--+(1.5,-0.3)node[bigvertex]{};
		\draw(v3)--+(1.5,0.5)circle(0.15)node[](v4){};
		\draw (v4) node[label=above:{$D^{k_{\mathfrak{L}}}$}]{};
		\draw (v4) node[label=below:{$v_{\mathfrak{L}}$}]{};
		\draw(v3)--+(1.5,-0.8)node[bigvertex ]{};
	}
\end{tabular}
\end{center}\caption{A tree in $\mathcal{T}_{\mathfrak{L}}$ with the action of the dilatations $D^{k_i}$ associated with the vertices $v_i$, $i=0,\ldots, \mathfrak{L}$ explicitly indicated. 
For any $\mathfrak{L}>1$, the elements of $\mathcal{T}_{\mathfrak{L}}$ are recursively characterized by the conditions that the root $v_0$ has $s_{v_0}\ge 1$ children, and that the first (i.e., the topmost) one, denoted $v_1$, is the root of a tree in $\mathcal{T}_{\mathfrak{L}-1}$, while the other children are all big-black-dot endpoints. For $\mathfrak{L}=1$, the set $\mathcal{T}_1$ was described in Fig.\ref{fig:6.1}. 
For any $h$ such that $|h|\ge \mathfrak{L}	$, the integers $k_0,\ldots,k_{\mathfrak{L}}$ satisfy: $k_0,\ldots,k_{\mathfrak{L}-1}\ge 1$, $k_{\mathfrak{L}}\ge 0$, and $k_0+\cdots+k_{\mathfrak{L}}=|h|$; in particular, for $\mathfrak{L}=1$ and $h\le -1$, the vertices $v_0$ and $v_1$ are associated with two dilatations operators $D^{k_0}$ and $D^{k_1}$ with $k_0\ge 1$, $k_1\ge 0$ and $k_0+k_1=|h|$:
the case depicted in Fig.\ref{fig:6.1} (where $v_0$ is associated with $D$, and $v_1$ with the identity) is a special case corresponding  to $h=-1$.}
\label{fig:6.2}
\end{figure}	
Given $\tau\in\mathcal T_{\mathfrak{L}}$ with $\mathfrak{L}>1$,  letting $\tau_{v_1}\in \mathcal T_{\mathfrak{L}-1}$ be the subtree of $\tau$ rooted in $v_1$, $v_\ell[\tau,(k_0,\ldots,k_{\mathfrak{L}})]$, for $\ell\in L(2,0)$, is recursively defined as:
\begin{equation}\label{taurecursive} v_\ell[\tau,(k_0,k_1,\ldots,k_{\mathfrak{L}})]=s_{v_0}\sum^*_{\ell_{1}\in L_f(2,0)}\sum_{(\ell_i)_{i=2}^{s_{v_0}}}D^{k_0}S_{\ell}^{\ell_{1},\ldots,\ell_{s_{v_0}}}(v_{\ell_{1}}[\tau_{v_1},(k_1,\ldots,k_{\mathfrak{L}})],
H^*_{\ell_2},\ldots,H^*_{\ell_{s_{v_0}}}),\end{equation}
where the sums over $\ell_1$ and over $(\ell_i)_{i=1}^{s_{v_0}}$ must be interpreted as in \eqref{iteravhbis}, while, given $\tau\in \mathcal{T}_1$, 
\begin{equation} \label{firststepT1}v_\ell[\tau,(k_0,k_1)]=
s_{v_0}\sum_{\ell_1\in L_f(2,0)}^*\sum_{(\ell_i)_{i=2}^{s_{v_0}}}D^{k_0}S^{\ell_1,\ldots,\ell_{s_{v_0}}}_\ell(D^{k_1}V^*_{\ell_1},H^*_{\ell_2},\ldots,H^*_{\ell_{s_{v_0}}}).\end{equation}
The validity of the tree representation \eqref{iteravhtris} with the tree values defined above 
can be straightforwardly proved by induction, and is left to the reader. 

We now use this tree representation to compute and bound the right hand side of \eqref{E1v2000}. We have: 
\begin{equation}D^hv^{(h)}_{2,0,0,\emptyset}(\boldsymbol{x})=\sum_{\mathfrak{L}=1}^{|h|}\sum_{\substack{k_{\mathfrak{L}}\ge 0, \\ k_0, k_1,\ldots,k_{\mathfrak{L}-1}\ge 1:\\ k_0+\cdots+k_{\mathfrak{L}}=|h|}}
\sum_{\tau\in\mathcal{T}_{\mathfrak{L}}}\, D^h v_{2,0,0,\emptyset}[\tau,\boldsymbol{k}](\boldsymbol{x}).\label{eq:6.15}
\end{equation}
If $\tau\in\mathcal{T}_1$, using \eqref{firststepT1}, the definition of $D$ in \eqref{eq:scal}, and the analogue of \eqref{item1bis}, we get that, for any $k_0\ge1$, $k_1\ge 0$ such that $k_0+k_1=|h|$,  
\begin{equation} \begin{split}
&\big|D^hv_{2,0,0,\emptyset}[\tau,(k_0,k_1)](\boldsymbol{x})\big|\\
&\le s_{v_0}\sum_{\ell_1\in L_f(2,0)}\sum_{(\ell_i)_{i=2}^{s_{v_0}}}
\gamma^{2\Delta_1(h+k_0)}\Big|S^{\ell_1,\ldots,\ell_{s_{v_0}}}_{2,0,0,\emptyset}(D^{k_1}V^*_{\ell_1},H^*_{\ell_2},\ldots,H^*_{\ell_{s_{v_0}}})(\gamma^{h+k_0}\boldsymbol{x})\Big|\\
&\le s_{v_0}\sum_{\ell_1\in L_f(2,0)}\sum_{(\ell_i)_{i=2}^{s_{v_0}}}
\gamma^{2\Delta_1(h+k_0)}C_\gamma^{s_{v_0}-1}(4C_0)^{\sum_{i=1}^{s_{v_0}}l_i}N_{(2,0,0,\emptyset)}^{\ell_1,\ldots,\ell_{s_{v_0}}}
\Big(\prod_{i=2}^{s_{v_0}}\|H^*_{\ell_i}\|_w\Big)\tn{D^{k_1}V^*_{\ell_1}(\gamma^{h+k_0}\boldsymbol{x})}.
\end{split}\label{eq:6.16}
\end{equation}
Now, in the last line, for any $i=2,\ldots,s_{v_0}$, we use that
\begin{equation}\label{eq:6.17}\|H^*_{\ell_i}\|_w\le \sum_{\tau}\|H_{\ell_i}[\tau]\|_w\le C\sum_\tau(C\epsilon)^{n_{\text{e.p.}}[\tau]}\le C'(C'\epsilon)^{\max\{1,\frac{l_i}2-1\}},\end{equation}
where, in the second inequality, we used \eqref{eq:4.22} {with $\rho=1$} and the fact that the trees contributing to $H^*_{\ell_i}$, with $\ell_i=(0,0,l_i,\boldsymbol{p}_i)$, necessarily have 
$n_{\text{e.p.}[\tau]}\ge \max\{1,\frac{l_i}{2}-1\}$. Concerning the factor $\tn{D^{k_1}V^*_{\ell_1}(\gamma^{h+k_0}\boldsymbol{x})}$, we rewrite it and bound it as follows: 
we use once again the tree representation \eqref{T1} with 
$H_\ell[\tau]$ as in \eqref{T2}. Noting that for $\ell=(2,0,l,\boldsymbol{p})$ the trimming operator acts as the identity, and rewriting $\frac{D}{1-D}=\sum_{k\ge 1}D^k$, with $D$ as in \eqref{eq:scal}, we 
find:
\begin{equation}
\tn{D^{k_1}V^*_{\ell_1}(\gamma^{h+k_0}\boldsymbol{x})}
\le\sum_{k\ge1}\sum_\tau\sum_{(\ell_i')_{i=1}^{s_{v_0}}}^*\tn{D^{k+k_1} S^{\ell_1',\cdots,\ell_{s_{v_0}}'}
_{\ell_1}(H_{\ell_1'}[\tau_1],\ldots,H_{\ell_{s_{v_0}}'}[\tau_{s_{v_0}}])(\gamma^{h+k_0}\boldsymbol{x})}\end{equation}
By Lemma \ref{lem:4.1}, we thus obtain:
\begin{equation}	\begin{split}&\gamma^{2\Delta_1(h+k_0)}\tn{D^{k_1}V^*_{\ell_1}(\gamma^{h+k_0}\boldsymbol{x})}
\le C\sum_\tau (C\epsilon)^{n_{\text{e.p}}[\tau]-2}\sum_{k\ge1}
\gamma^{2\Delta_1(h+k_0+k+k_1)}\gamma^{-(k+k_1)(l_1[\psi]+\|\boldsymbol{p}_1\|_1)}\\
&\hskip4.7truecm \cdot e^{-\frac{\bar{C}}2(\gamma^{k+k_1+h+k_0-1}|x|)^{\sigma}} \big(\min\{1,\gamma^{k+k_1+h+k_0}|x|\}\big)^{-\alpha}\\
&\qquad\le C' (C'\epsilon)^{\frac{l_1}2-1}\sum_{k\ge1}
\gamma^{2\Delta_1k}\gamma^{-(k+k_1)(l_1[\psi]+\|\boldsymbol{p}_1\|_1)}e^{-\frac{\bar{C}}2(\gamma^{k-1}|x|)^{\sigma}} \big(\min\{1,\gamma^{k}|x|\}\big)^{-\alpha},
\end{split}\label{eq:6.19}
\end{equation}	
where in the second inequality we performed the sum over $\tau$ (note that, for given $l_1\ge 2$, $n_{\text{e.p.}}[\tau]-2\ge \frac{l_1}2-1$, so that $\sum_\tau (C\epsilon)^{n_{\text{e.p}}[\tau]-2}\le$(const.)$(C'\epsilon)^{\frac{l_1}2-1}$)
and used the fact that $k_0+k_1+h=0$. By the same considerations discussed after \eqref{eq:6.2}, performing the sum over $k\ge1$ we get: 
\begin{equation}\gamma^{2\Delta_1(h+k_0)}\tn{D^{k_1}V^*_{\ell_1}(\gamma^{h+k_0}\boldsymbol{x})}
\le  C' (C'\epsilon)^{\frac{l_1}2-1}\gamma^{-k_1(l_1[\psi]+\|\boldsymbol{p}_1\|_1)}e^{-\frac{\bar{C}}4(|x|/\gamma)^{\sigma}}
(\min\{1,|x|\})^{-2\Delta_1}.\label{eq:6.19bis}
\end{equation}	
Plugging \eqref{eq:6.17} and \eqref{eq:6.19bis} into \eqref{eq:6.16}, and using that $N_{(2,0,0,\emptyset)}^{\ell_1,\ldots,\ell_{s_{v_0}}}\le 2^{\sum_{i=1}^{s_{v_0}}l_i}$, gives: 
\begin{equation} \begin{split}
&\big|D^hv_{2,0,0,\emptyset}[\tau,(k_0,k_1)](\boldsymbol{x})\big|\\
&\le s_{v_0}(C'')^{s_{v_0}}e^{-\frac{\bar{C}}4(|x|/\gamma)^{\sigma}}
(\min\{1,|x|\})^{-2\Delta_1}
\sum_{\ell_1\in L_f(2,0)}(C''\epsilon)^{\frac{l_1}2-1}\gamma^{-k_1(l_1[\psi]+\|\boldsymbol{p}_1\|_1)}\cdot\\
&\cdot\sum_{(\ell_i)_{i=2}^{s_{v_0}}}\prod_{i=2}^{s_{v_0}}(C''\epsilon)^{\max\{1,\frac{l_i}{2}-1\}}\le C'''(C'''\epsilon)^{s_{v_0}-1}e^{-\frac{\bar{C}}4(|x|/\gamma)^{\sigma}}
(\min\{1,|x|\})^{-2\Delta_1}\gamma^{-2k_1[\psi]}.
\end{split}\label{eq:6.20}
\end{equation}

\begin{remark}
From \eqref{eq:6.19}, it is clear that the constraint $k_0+k_1=|h|$ (or, in general, $k_0+k_1+\cdots +k_{\mathfrak L}=|h|$ for trees $\tau\in \mathcal T_{\mathfrak L}$ with $\mathfrak L\ge 1$, to be discussed in the following) is essential in proving the stretched exponential decay in \eqref{eq:6.20}. This constraint, first appearing in \eqref{iteravhtris}, has its origin in 
the recursive equation \eqref{iteravhbis}, which is linear in $\{v^{(h')}_\ell\}^{h'\le 0}_{\ell\in L_f(2,0)}$ (with $v^{(0)}_\ell\equiv V^*_\ell$); in turn, such linearity in $\{v^{(h')}_\ell\}^{h'\le 0}_{\ell\in L_f(2,0)}$ originates from the fact that $\mathcal Q^*$ is quadratic in $\phi$. In order to compare more closely the tree expansion for $D^h v_{2,0,0,\emptyset}^{(h)}$ in \eqref{iteravhtris}
with the one for $D^h V^*_{2,0,0,\emptyset}$ in the first line of \eqref{eq:4.1} we can, if desired, re-expand the kernels $V^*_{\ell'}$ and $H^*_{\ell''}$ associated with the big white and black endpoints of the trees in $\mathcal{T}_{\mathfrak{L}}$ in terms of the tree expansions discussed in Section \ref{sec5} (for $V^*_{\ell'}$) and in \cite[Appendix J]{GMR21} (for $H^*_{\ell''}$): this would lead to a new tree expansion for $D^h v_{2,0,0,\emptyset}^{(h)}$ (more complex, but closer to the tree expansion of the previous section) 
in terms of trees with `bare endpoints' as those in Figure \ref{eq:figtree3.1}. Such trees would be exactly of the same form as those contributing to $D^h V^*_{2,0,0,\emptyset}$, such as those in Figure \ref{fig:9}, with the additional constraint on the scale labels  
$k_0+k_1+\cdots +k_{\mathfrak L}=|h|$, which was absent in the discussion in Section \ref{sec5}; in fact, the analogue of the combination $h+k_0+k_1$ in \eqref{eq:6.19}
and, more generally, of the combination $h+k_0+k_1+\cdots k_{\mathfrak L}$ appearing in the bounds below, 
is what in Section \ref{sec5} was denoted $k+k_{v_1}+\cdots+k_{v_n}$, see \eqref{analogue} and following equations; in those equations, the combination $k+\sum_{i=1}^nk_{v_i}$
could take arbitrarily negative values, and this was (rightly so) at the origin of the polynomial decay of the kernel of $D^h V^*_{2,0,0,\emptyset}$ in the limit $h\to -\infty$, proved in the previous section. 
\end{remark}

We now proceed in a similar fashion, in order to bound the general term in the right hand side of \eqref{eq:6.15}. We take $1<\mathfrak{L}\le |h|$, $\tau\in\mathcal{T}_{\mathfrak{L}}$, 
$\boldsymbol{k}=(k_0,k_1,\ldots,k_{\mathfrak{L}})$ with $k_{\mathfrak{L}}\ge 0$, $k_0,k_1,\ldots,k_{\mathfrak{L}-1}\ge 1$, and $h+k_0+k_1+\cdots+k_{\mathfrak{L}}=0$. Using the same notation introduced before \eqref{item1}, thanks to  \eqref{taurecursive}, we have:
\begin{equation} 
D^h v_{2,0,0,\emptyset}[\tau,\boldsymbol{k}](\boldsymbol{x})=s_{v_0}\sum_{(\ell_i(v_0))_{i=1}^{s_{v_0}}}^*D^{h+k_0}S_{2,0,0,\emptyset}^{\ell_{1}(v_0),\ldots,\ell_{s_{v_0}}(v_0)}(v_{\ell_{v_1}}[\tau_{v_1},\boldsymbol{k}_{v_1}],
H^*_{\ell_2(v_0)},\ldots,H^*_{\ell_{s_{v_0}}(v_0)})(\boldsymbol{x}),\end{equation}
where $\sum_{(\ell_i(v_0))_{i=1}^{s_{v_0}}}^*\equiv \sum^*_{\ell_{1}(v_0)\in L_f(2,0)}\sum_{(\ell_i(v_0))_{i=2}^{s_{v_0}}}$, and
$\boldsymbol{k}_{v_1}:=(k_1,\ldots,k_{\mathfrak{L}})$; moreover, we recall that $\ell_{v_1}\equiv \ell_1(v_0)$. Using (the analogue of) \eqref{eq:4.1.1} we get: 
\begin{equation}\label{...1st}\begin{split} 
&|D^h v_{2,0,0,\emptyset}[\tau,\boldsymbol{k}](\boldsymbol{x})|\le\\
&\qquad \le  s_{v_0}\sum^*_{(\ell_i(v_0))_{i=1}^{s_{v_0}}}\Big|{D^{h+k_0}S_{2,0,0,\emptyset}^{\ell_{1}(v_0),\ldots,\ell_{s_{v_0}}(v_0)}(v_{\ell_{v_1}}[\tau_{v_1},\boldsymbol{k}_{v_1}],
	H^*_{\ell_2(v_0)},\ldots,H^*_{\ell_{s_{v_0}}(v_0)})(\boldsymbol{x})}\Big|\\
&\qquad \le \gamma^{2\Delta_1(h+k_0)} \sum^*_{(\ell_i(v_0))_{i=1}^{s_{v_0}}}
s_{v_0}C_\gamma^{s_{v_0}-1}(4C_0)^{\sum_{i=1}^{s_{v_0}}l_{i}(v_0)}N_{2,0,0,\emptyset}^{\ell_1(v_0),\ldots,\ell_{s_{v_0}}(v_0)}\cdot\\
&\qquad \cdot \Big(\prod_{i=2}^{s_{v_0}} \|H_{\ell_i(v_0)}^*\|_{w}\Big)\tn{v_{\ell_{v_1}}[\tau_{v_1},\boldsymbol{k}_{v_1}](\gamma^{h+k_0}\boldsymbol{x})}.
\end{split}\end{equation}
We now apply again the recursive definition of $v_{\ell_{v_1}}[\tau_{v_1},\boldsymbol{k}_{v_1}]$ and iterate the same procedure until we reach $v_{\mathfrak{L}-1}$, thus getting: 
\begin{equation}\label{...}\begin{split} 
&|D^h v_{2,0,0,\emptyset}[\tau,\boldsymbol{k}](\boldsymbol{x})| \le  \gamma^{2\Delta_1 h}\sum^*_{(\ell_i(v_0))_{i=1}^{s_{v_0}}}\cdots \sum^*_{(\ell_i(v_{\mathfrak{L}-1}))_{i=1}^{s_{v_{\mathfrak{L}-1}}}}\times\\
&\qquad \times\Big(\prod_{j=0}^{\mathfrak{L}-1}s_{v_j}C_\gamma^{s_{v_j}-1}(4C_0)^{\sum_{i=1}^{s_{v_j}}l_i(v_j)-l_{v_j}}N_{\ell_{v_j}}^{\ell_1(v_j),\ldots,\ell_{s_{v_j}}(v_j)}\gamma^{k_j(2\Delta_1-l_{v_j}[\psi]-\|\boldsymbol{p}_{v_j}\|_1)}\cdot\\
&\qquad\ \ \cdot\big(\prod_{i=2}^{s_{v_j}}\|H_{\ell_i(v_j)}^*\|_w\big)\Big)
\tn{D^{k_{\mathfrak{L}}}V^*_{\ell_{v_{\mathfrak{L}}}}(\gamma^{h+k_0+\cdots+k_{\mathfrak{L}-1}}\boldsymbol{x})},
\end{split}\end{equation}
with $\ell_{v_0}=(n_{v_0},m_{v_0},l_{v_0},\boldsymbol{p}_{v_0})\equiv (2,0,0,\emptyset)$. 

{\begin{remark}\label{Rem:6.1} Every collection of labels $\{(\ell_{i}(v_j))_{i=1}^{s_{v_j}}\}_{j=0,\ldots,\mathfrak{L}-1}$ 
contributing to the right hand side of \eqref{eq:6.26}
satisfies the following constraints (recall that we write $\ell_i(v_j)=(n_i(v_j), m_i(v_j), l_i(v_j),\boldsymbol{p}_i(v_j)$, and $l_1(v_j)\equiv l_{v_{j+1}}$): 
for any $j=0,\ldots, \mathfrak{L}-1$, if $v_j$ is trivial, then $l_{v_j}\le l_{v_{j+1}}-2$ (with the understanding that $l_{v_0}\equiv 0$), while, if $v_j$ is non-trivial, then 
$l_{v_j}\le l_{v_{j+1}}+\sum_{i=2}^{s_{v_j}}l_i(v_j)-2(s_{v_j}-1)$. In the following, we use that the sums over $\{(\ell_{i}(v_j))_{i=1}^{s_{v_j}}\}_{j=0,\ldots,\mathfrak{L}-1}$ are performed 
under these constraints. 
\end{remark}}

Now, recalling that $h+k_0+k_1+\cdots+k_{\mathfrak{L}}=0$ and using the analogue of \eqref{eq:6.19bis}, we have
\begin{equation}\begin{split} &\gamma^{2\Delta_1(h+k_0+\cdots+k_{\mathfrak{L}-1})}\tn{D^{k_{\mathfrak{L}}}V^*_{\ell_{v_{\mathfrak{L}}}}(\gamma^{h+k_0+\cdots+k_{\mathfrak{L}-1}}\boldsymbol{x})}\le\\
&\qquad \le  
C' (C'\epsilon)^{\frac{l_{v_{\mathfrak{L}}}}2-1}\gamma^{-k_{\mathfrak{L}}(l_{v_{\mathfrak{L}}}[\psi]+\|\boldsymbol{p}_{v_{\mathfrak{L}}}\|_1)}e^{-\frac{\bar{C}}4(|x|/\gamma)^{\sigma}}
(\min\{1,|x|\})^{-2\Delta_1}.\end{split}\label{eq:6.25}\end{equation}
Inserting \eqref{eq:6.17} and \eqref{eq:6.25} into \eqref{...}, summing over $\boldsymbol{p}_{v_1},\ldots,\boldsymbol{p}_{v_{\mathfrak{L}}}$ and using \eqref{sumoverp0}, implies: 
\begin{equation}\label{eq:6.26}\begin{split} 
&|D^h v_{2,0,0,\emptyset}[\tau,\boldsymbol{k}](\boldsymbol{x})| \le C^{n_{\text{e.p.}}[\tau]}e^{-\frac{\bar{C}}4(|x|/\gamma)^{\sigma}}(\min\{1,|x|\})^{-2\Delta_1}
\sum_{(\ell_i(v_0))_{i=2}^{s_{v_0}}}\cdots \sum_{(\ell_i(v_{\mathfrak{L}-1}))_{i=2}^{s_{v_{\mathfrak{L}-1}}}}\times\\
&\qquad \times \Bigg(\prod_{j=0}^{\mathfrak{L}-1}\Big(\prod_{i=2}^{s_{v_j}}(4C_0)^{l_i(v_j)}(C'\epsilon)^{\max\{1,\frac{l_i(v_j)}{2}-1\}}\Big)\Bigg)\sum_{l_{v_{\mathfrak{L}}}}(C'\epsilon)^{l_{v_{\mathfrak{L}}}/2-1}(8C_0)^{l_{v_{\mathfrak{L}}}}\gamma^{-k_{\mathfrak{L}}l_{v_{\mathfrak{L}}}[\psi]}\cdot\\
&\qquad \cdot 
\sum_{l_{v_1},\ldots,l_{v_{\mathfrak{L}}-1}}\Bigg(\prod_{j=1}^{\mathfrak{L}-1}\binom{\sum_{i=1}^{s_{v_j}}l_i(v_j)}{l_{v_j}}\gamma^{-k_jl_{v_j}[\psi]}\Bigg).
\end{split}\end{equation}
Now, in the last line, recalling that $l_{v_j}\ge 2$ and $k_{j}\ge 1$, we can bound from above each factor $\gamma^{-k_jl_{v_j}[\psi]}$ by $\gamma^{-2k_j[\psi]}\gamma^{-(l_{v_j}-2)[\psi]}$; next, by proceeding 
as in \cite[Appendix A.6.1]{GeM01} and in \eqref{eq:5.25}, we obtain: 
\begin{equation}\label{eq:6.27}\begin{split} & \sum_{l_{v_1},\ldots,l_{v_{\mathfrak{L}}-1}}\Big(\prod_{j=1}^{\mathfrak{L}-1}\binom{\sum_{i=1}^{s_{v_j}}l_i(v_j)}{l_{v_j}}\gamma^{-k_jl_{v_j}[\psi]}\Big)\le \\
&\qquad \le 
\gamma^{2[\psi](\mathfrak{L}-1)}\gamma^{-2[\psi](k_1+\cdots +k_{\mathfrak{L}-1})}(1-\gamma^{-[\psi]})^{-\sum_{j=1}^{\mathfrak{L}-1}\sum_{i=2}^{s_{v_j}}l_i(v_j)}(1-\gamma^{-[\psi]})^{-l_{v_{\mathfrak{L}}}}.
\end{split}\end{equation}
Hence, plugging \eqref{eq:6.27} in \eqref{eq:6.26}, using $\gamma^{-k_{\mathfrak{L}}l_{v_{\mathfrak{L}}}[\psi]}\le \gamma^{-2k_{\mathfrak{L}}[\psi]}$ and $\sum_{j=0}^{\mathfrak{L}-1}(s_{v_j}-1)=n_{\text{e.p.}}[\tau]-1$, we find
\begin{equation}\begin{split}\label{eq:6.28}
&|D^{h}v_{2,0,0,\emptyset}[\tau,\boldsymbol{k}](\boldsymbol{x})|\le (C'')^{n_{\it{e.p.}}[\tau]}e^{-\frac{\bar C}{4}(|x|/\gamma)^{\sigma}}(\min\{1,|x|\})^{-2\Delta_1}\gamma^{2[\psi](\mathfrak{L}-1)}\gamma^{-2[\psi](k_1+\dots+k_{\mathfrak{L}})}\cdot\\
&\qquad\cdot\sum_{(\ell_i(v_0))_{i=2}^{s_{v_0}}}\cdots\sum_{(\ell_i(v_{\mathfrak{L}-1}))_{i=2}^{s_{v_{\mathfrak{L}-1}}}}\Bigg(\prod_{j=0}^{\mathfrak{L}-1}\Bigg(\prod_{i=2}^{s_{v_j}}(C''\epsilon)^{\max\{1,\frac{l_i(v_j)}{2}-1\}}\Big)\Bigg)\sum_{l_{v_{\mathfrak{L}}}}(C''\epsilon)^{l_{v_{\mathfrak{L}}}/2-1}
\end{split}\end{equation}
for a suitable $C''>0$. In order to get the desired bound on $D^h v^{(h)}_{2,0,0,\emptyset}(\boldsymbol{x})$, uniformly as 
$h\to-\infty$ (see \eqref{E1v2000}), in view of \eqref{eq:6.15}, we still need to sum the right hand side of 
\eqref{eq:6.28} over $\mathfrak{L}$, $\boldsymbol{k}$ and $\tau$ and show that the result of the sum is the same as the right hand side of \eqref{eq:6.3} (up, possibly, to a redefinition of the constants).

Now, for the sums over $\boldsymbol{k}$, we simply use
\begin{equation}\sum_{\substack{k_\mathfrak L \ge 0\\k_0,\ldots,k_{\mathfrak L-1}\ge 1\\k_0+\dots+k_{\mathfrak{L}}=|h|}}\gamma^{-2[\psi](k_1+\cdots+k_{\mathfrak L}-\mathfrak{L}+1)}\le 
(1-\gamma^{-2[\psi]})^{\mathfrak{L}},\end{equation}	
so that (letting $\sum_{\boldsymbol{k}}$ be a shorthand notation for $\sum_{k_\mathfrak L \ge 0, \ k_0,\ldots,k_{\mathfrak L-1}\ge 1}^{k_0+\dots+k_{\mathfrak{L}}=|h|}$)
\begin{equation}\begin{split}\label{eq:6.30}
&\sum_{\boldsymbol{k}}|D^{h}v_{2,0,0,\emptyset}[\tau,\boldsymbol{k}](\boldsymbol{x})|\le C^{n_{\it{e.p.}}[\tau]+\mathfrak{L}}e^{-\frac{\bar C}{4}(|x|/\gamma)^{\sigma}}(\min\{1,|x|\})^{-2\Delta_1}\cdot\\
&\qquad\cdot\sum_{(\ell_i(v_0))_{i=2}^{s_{v_0}}}\cdots\sum_{(\ell_i(v_{\mathfrak{L}-1}))_{i=2}^{s_{v_{\mathfrak{L}-1}}}}\Bigg(\prod_{j=0}^{\mathfrak{L}-1}\Big(\prod_{i=2}^{s_{v_j}}(C''\epsilon)^{\max\{1,\frac{l_i(v_j)}{2}-1\}}\Big)\Bigg)\sum_{l_{v_{\mathfrak{L}}}}(C''\epsilon)^{l_{v_{\mathfrak{L}}}/2-1}
\end{split}\end{equation}
up to a redefinition of $C$. If we now performed the sums in the second line using $(\ell_i(v_j))_{i=2}^{s_{v_j}}$, $l_{v_{\mathfrak{L}}}$, by forgetting all the constraints on the indices 
$(\ell_i(v_j))_{i=2}^{s_{v_j}}$, $l_{v_{\mathfrak{L}}}$ but $l_i(v_j)\ge 2$ and $l_{v_{\mathfrak{L}}}\ge 2$, we would find that the second line of \eqref{eq:6.30} would be bounded by $C\prod_{j=0}^{\mathfrak{L}-1}(C\epsilon)^{s_{v_j}-1}$ for some $C>0$. By re-plugging this back into \eqref{eq:6.30}, we would be led to a bound that is summable over  $\tau\in\mathcal{T}_{\mathfrak{L}}$ (i.e., over $s_{v_j}\ge 1$ for $j=0,\ldots,\mathfrak{L}-1$), but not over $\mathfrak{L}$, and we would be in trouble.  
{Therefore, we must take better advantage of the constraints spelled out in Remark \ref{Rem:6.1}, by proceeding as follows.} 

Given $\mathfrak{L}\ge 1$ and $\tau\in \mathcal{T}_{\mathfrak{L}}$, we consider the corresponding set of vertices $\{v_0,\ldots,v_p\}$ as in Fig. \blue{\ref{fig:6.3}}, and rewrite it as the 
union of the set $V_{\text{t}}(\tau)$ of `trivial' vertices (i.e., those with only one child, $s_{v_j}=1$) and of the set $V_{\text{nt}}(\tau)$ of `non-trivial' ones (i.e., those with $s_{v_j}\ge 2$). 
We let $p=p(\tau)$ be the cardinality of $V_{\text{nt}}(\tau)$ and, if $p\ge 1$, we denote $V_{\text{nt}}(\tau)\equiv \{v_{j_1}, \ldots, v_{j_p}\}$, with $j_1<\cdots <j_p$. We shall think the tuple $(v_0,v_1,\ldots,v_{\mathfrak{L}})$ as a concatenation 
of the tuples $(v_0,\ldots,v_{j_1})$, $(v_{j_1+1}, \ldots, v_{j_2})$, $\ldots$, $(v_{j_p+1},\ldots, v_{\mathfrak{L}})$. We let $n_1=j_1+1$, $n_2=j_2-j_1$, $\ldots$, $n_{p+1}=\mathfrak{L}-j_p$ be the lengths of these tuples, and $s_1=s_{v_{j_1}},\ldots, s_p=s_{v_{j_p}}$ 
the numbers of children of $v_{j_1},\ldots, v_{j_p}$. Note that the choice of the integers $p\ge 0$, $n_1,\ldots, n_{p+1}\ge 1$ (with $n_1\ge 2$ if $p=0$) and $s_1,\ldots, s_p\ge 2$ specifies uniquely the choice of $\mathfrak{L}$ and $\tau\in\mathcal{T}_{\mathfrak{L}}$. Therefore, we shall identify the double sum over $\mathfrak{L}\ge 1$ and $\tau\in\mathcal{T}_{\mathfrak{L}}$ with that over $p\ge 0$, $n_1,\ldots, n_{p+1}\ge 1$ (with $n_1\ge 2$ if $p=0$) and $s_1,\ldots, s_p\ge 2$. 

\begin{figure}[ht]
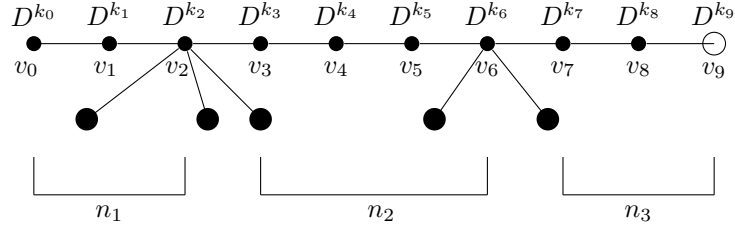

\begin{center}
\begin{tabular}{rcl}
	\tikz[baseline=-3pt]{ %orizzontale
		\draw(0,0)node[vertex, label=below:{$v_0$\;\;}, label=above:{$D^{k_0}$}](v0){}--(1,0)node[vertex,label=below:{$v_1$\;}](v1){};
		\draw (v1)node[label=above:{$D^{k_1}$}]{};
		\draw(v1)--(2,0)node[vertex,label=below:{$v_2$\;\;}](v2){};
		\draw (v2)node[label=above:{$D^{k_2}$}]{};
		\draw(v2)--(3,0)node[vertex,label=below:{$v_3$}](v3){};
		\draw (v3)node[label=above:{$D^{k_3}$}]{};
		\draw (v3)--(4,0) node[vertex,label=below:{$v_4$}](v4){};
		\draw (v4)node[label=above:{$D^{k_4}$}]{};
		\draw(v4)--(5,0)node[vertex,label=below:{$v_5$}](v5){};
		\draw (v5)node[label=above:{$D^{k_5}$}]{};
		\draw(v5)--(6,0)node[vertex,label=below:{$v_6$}](v6){};
		\draw (v6)node[label=above:{$D^{k_6}$}]{};
		\draw (v6)--(7,0) node[vertex,label=below:{$v_7$}](v7){};
		\draw (v7)node[label=above:{$D^{k_7}$}]{};
		\draw(v7)--(8,0)node[vertex,label=below:{$v_8$}](v8){};
		\draw (v8)node[label=above:{$D^{k_8}$}]{};
		\draw(v8)--(9,0)node[label=below:{$v_{9}$}](v9){};
		\draw(v9)circle(0.15);
		\draw (v9)node[label=above:{$D^{k_9}$}]{};
		\draw (v2)--(0.7,-1)node[bigvertex]{};
		\draw(v2)--(2.3,-1)node[bigvertex]{};
		\draw(v2)--(3,-1)node[bigvertex]{};
		\draw (v6)--(5.3,-1)node[bigvertex]{};
		\draw(v6)--(6.8,-1)node[bigvertex]{};
		\draw(0,-1.5)--(0,-2);
		\draw(2,-2)--(2,-1.5);
		\draw(0,-2)--(2,-2);
		\draw(1,-2.6)node[label={$n_1$}]{};
		\draw(3,-1.5)--(3,-2);
		\draw(6,-2)--(6,-1.5);
		\draw(3,-2)--(6,-2);
		\draw(4.6,-2.6)node[label={$n_2$}]{};
		\draw(7,-1.5)--(7,-2);
		\draw(9,-2)--(9,-1.5);
		\draw(7,-2)--(9,-2);
		\draw(8,-2.6)node[label={$n_3$}]{};
	}
\end{tabular}
\end{center}\caption{A tree in $\mathcal{T}_{\mathfrak{L}}$ with $\mathfrak{L}=9$. Its non trivial vertices are $v_2$ and $v_6$, and $s_1\equiv s_{v_2}=4$, $s_2\equiv s_{v_6}=3$. The $10$-ple $(v_0,\ldots,v_9)$ is thought of as a concatenation of 
three tuples of length $n_1=3$, $n_2=4$, $n_3=3$, respectively, namely of $(v_0,v_1,v_2)$, $(v_3,v_4,v_5,v_6)$ and $(v_7,v_8,v_9)$.}
\label{fig:6.3}
\end{figure}
{Let us now use the constraints spelled out in Remark \ref{Rem:6.1}, which we recall here for the reader's convenience:} for $j=0,\ldots, \mathfrak{L}-1$, if $v_j$ is trivial, then $l_{v_j}\le l_{v_{j+1}}-2$  (with $l_{v_0}\equiv 0$), while, if $v_j$ is non-trivial, then 
$l_{v_j}\le l_{v_{j+1}}+\sum_{i=2}^{s_{v_j}}l_i(v_j)-2(s_{v_j}-1)$. Therefore, if $p=0$, then $l_{v_{\mathfrak{L}}}\ge 2(n_1-1)$; while, if $p\ge 1$:
\begin{equation} 	\begin{split}
2(n_1-1)&\le l_{v_{j_1}}\le l_{v_{j_1+1}}+\sum_{i=2}^{s_1}l_i(v_{j_1})+2(s_1-1)\\
l_{v_{j_1+1}}+ 2(n_2-1)&\le l_{v_{j_2}}\le l_{v_{j_2+1}}+\sum_{i=2}^{s_2}l_i(v_{j_2})+2(s_2-1)\\
&\vdots\\
l_{v_{j_{p-1}+1}}+ 2(n_p-1)&\le l_{v_{j_p}}\le l_{v_{j_p+1}}+\sum_{i=2}^{s_p}l_i(v_{j_p})+2(s_p-1)\\
l_{v_{j_p+1}}+ 2(n_{p+1}-1)&\le l_{v_{\mathfrak{L}}},\end{split}\end{equation}
which implies, in particular, that every collection of labels contributing to the right hand side of \eqref{eq:6.30} satisfies
\begin{equation}\sum_{i=2}^{s_1}l_i(v_{j_1})+\cdots +\sum_{i=2}^{s_p}l_i(v_{j_p})+l_{v_{\mathfrak{L}}}\ge 2(n_1-1)+\cdots+2(n_{p+1}-1)+2(s_1-1)+\cdots+2(s_p-1).\end{equation}
Therefore, the right hand side of \eqref{eq:6.30} can be bounded from above by 

\begin{equation}\begin{split}\label{eq:6.33}
&C^{n_{\it{e.p.}}[\tau]+\mathfrak{L}}e^{-\frac{\bar C}{4}(|x|/\gamma)^{\sigma}}(\min\{1,|x|\})^{-2\Delta_1}(\sqrt{\epsilon})^{[n_1+\cdots+n_{p+1}-p-2]_+}
\cdot\\
&\cdot\sum_{(\ell_i(v_{j_1}))_{i=2}^{s_1}}\Big(\prod_{i=2}^{s_1}(C''\sqrt{\epsilon})^{\max\{1,\frac{l_i(v_{j_1})}{2}-1\}}\Big)
\cdots\sum_{(\ell_i(v_{j_p}))_{i=2}^{s_p}}\Big(\prod_{i=2}^{s_p}(C''\sqrt{\epsilon})^{\max\{1,\frac{l_i(v_{j_p})}{2}-1\}}\Big)\sum_{l_{v_{\mathfrak{L}}}}(C''\sqrt{\epsilon})^{\frac{l_{v_{\mathfrak{L}}}}2-1}
\end{split}\end{equation}
where $[\cdot]_+:=\max\{0,\cdot\}$ indicates the positive part. If we now perform 
the sums in the second line by forgetting all the constraints on the indices $(\ell_i(v_{j_k}))_{i=2}^{s_k}$, $l_{v_{\mathfrak{L}}}$, but $l_i(v_{j_k})\ge 2$ and $l_{v_{\mathfrak{L}}}\ge 2$, we find that the second line 
of \eqref{eq:6.33} is bounded by $C\prod_{k=1}^{p}(C\sqrt{\epsilon})^{s_k-1}$ for some $C>0$. In conclusion, plugging this back into \eqref{eq:6.30}, and recalling that $\sum_{k=1}^p(s_k-1)=n_{\text{e.p.}}(\tau)-1$ 
and $\sum_{k=1}^{p+1}n_k=\mathfrak{L}+1$, we find, up to a re-definition of $C$: 
\begin{equation}\begin{split}\label{eq:6.34}
&\sum_{\boldsymbol{k}}|D^{h}v_{2,0,0,\emptyset}[\tau,\boldsymbol{k}](\boldsymbol{x})|\le\\
&\qquad \le C^{n_1+\cdots+n_{p+1}}(\sqrt{\epsilon})^{[n_1+\cdots+n_{p+1}-p-2]_+}(C\sqrt{\epsilon})^{\sum_{k=1}^p(s_k-1)}
e^{-\frac{\bar C}{4}(|x|/\gamma)^{\sigma}}(\min\{1,|x|\})^{-2\Delta_1}, 
\end{split}\end{equation}
with the understanding that, if $p=0$, the factor $(C\sqrt{\epsilon})^{\sum_{k=1}^p(s_k-1)}$ should be interpreted as being equal to $1$. Now, the right hand side of this equation is summable over $p\ge 0$, 
$n_1,\ldots,n_{p+1}\ge 1$ (with $n_1\ge 2$ if $p=0$) and $s_1,\ldots,s_p\ge 2$, the sum being bounded from above by $Ce^{-\frac{\bar C}{4}(|x|/\gamma)^{\sigma}}$ $(\min\{1,|x|\})^{-2\Delta_1}$. 
In conclusion, in view of \eqref{eq:6.15}, we find 
\begin{equation} \big| D^h v_{2,0,0,\emptyset}^{(h)}(\boldsymbol{x})\big|\le Ce^{-\frac{\bar C}{4}(|x|/\gamma)^{\sigma}}(\min\{1,|x|\})^{-2\Delta_1},\end{equation}
as desired. Analyticity of $D^h v_{2,0,0,\emptyset}^{(h)}(\boldsymbol{x})$, uniformly in $h$, is a consequence of the absolute summability of its tree expansion, as well as of the uniform-in-$h$ bounds that we just derived. A slight extension of the discussion above would also allow us to prove the existence of the limit in the right hand side of \eqref{E1v2000}
and to derive explicit estimates on the speed of convergence to the limit. However, in order not to overwhelm an already lengthy discussion, we prefer not to discuss explicitly this point, which is left to the interested reader as a simple exercise. {The bounds on $\mathcal E_2(y)$ and $\mathfrak E_2(y)$ are completely analogous to those discussed above; the minor changes (mostly notational) required for adapting the previous estimates to these functions are left to the reader.} The proof of Theorem \ref{Th:2} is thus concluded.

\appendix	

\section{First order computation of \texorpdfstring{$\zeta_2$}{zeta2}}\label{appA}

In this appendix, we calculate the first order contribution in $\epsilon$ to $\zeta_2=Z_2-1$, which implies \eqref{FO} for $\eta_2$. We recall that $\zeta_2$ is expressed via a convergent 
tree expansion as in \eqref{TreeildeH}; isolating the first order from the higher order contributions implies: 
\begin{equation}\label{labzeta2}\zeta_2=\tilde H_{0,1,2,\boldsymbol{0}}[\tau_0]+\tilde H_{0,1,2,\boldsymbol{0}}[\tau_1]+O(\epsilon^2),\end{equation}
where $\tau_0$ and $\tau_1$ are the two trees in Fig.\ref{figA1} (the analogous trees with an endpoint of type $\nu$ replacing the endpoint of type $\lambda$ give zero contribution). 
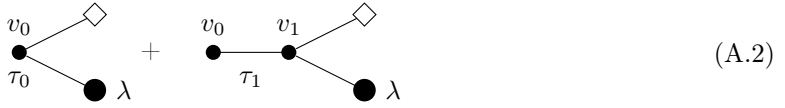
\begin{equation}
\begin{tabular}{rcl}			
\tikz[baseline=-2pt]{ \draw (0,0) node [vertex,label=$v_0$] (v0) {}--(1,0.5) node[ctVertex]{};
	\draw(0,0)node[label=below:{$\tau_0$}]{};
	\draw (v0) -- (1,-0.5) node [bigvertex,label=right:{$\lambda$}]  {};}+\;\;\;
\tikz[baseline=-2pt]{ \draw (0,0) node [vertex,label=$v_0$] (v0) {}--(1,0) node[vertex,label=above:{$v_1$}](v1){};
	\draw(0.5,0)node[label=below:{$\tau_1$}]{};
	\draw(v1)--(2,0.5) node[ctVertex ]{};
	\draw (v1) -- (2,-0.5) node [bigvertex,label=right:{$\lambda$}]  {};}
	\end{tabular}
	\label{figA1}
\end{equation}
In order to compute $\tilde H_{0,1,2,\boldsymbol{0}}[\tau_0]$ and $\tilde H_{0,1,2,\boldsymbol{0}}[\tau_1]$, we recall that the kernel associated with the endpoint of type $\lambda$ has the form in the third line of \eqref{trimm}, with $c_2=\frac{\lambda}3$ and (see \cite[(5.30), (G.7)]{GMR21}): 
\begin{equation}\begin{split}\label{lambda*}
&\lambda=\lambda^*=\frac{-2\epsilon\log\gamma}{I_2}+O(\epsilon^2),	\\
& I_2=-4(N-8)\left[\frac{S_d}{(2\pi)^d}\log\gamma+O(\epsilon(\log\gamma)^2)\right]\end{split}
\end{equation}	
(here $S_d$ is the area of the unit sphere in $\mathbb R^d$). An application of the definition of tree value and a  straightforward computation shows that: 
\begin{equation}\begin{split}
& \tilde H_{0,1,2,\boldsymbol{0}}[\tau_0]=-4(N-2)\lambda^*\int \frac{d^dk}{(2\pi)^d}\frac{f_0(k)^2}{|k|^{d+2\epsilon}}, \\
& \tilde H_{0,1,2,\boldsymbol{0}}[\tau_1]=-8(N-2)\lambda^*\int \frac{d^dk}{(2\pi)^d}\frac{\gamma^{d+\Delta_2-6[\psi] }f_0(k)f_{1}(k)}{|k|^{d+2\epsilon}},
\end{split}
\end{equation}	
where $f_h(k):=\chi(\gamma^{-h}k)-\chi(\gamma^{-h+1} k)$.
Now, plugging these expressions in \eqref{labzeta2}, using \eqref{lambda*} and the fact that $d+\Delta_2-6[\psi]=2\epsilon+\eta_2=O(\epsilon)$, we find:
\begin{equation} \zeta_2= -2\epsilon\frac{N-2}{N-8}\frac{(2\pi)^d}{S_d}\int \frac{d^dk}{(2\pi)^d}\frac{f_0^2(k)+2f_0(k)f_1(k)}{|k|^{d}}+O(\epsilon^2).
\end{equation}
A computation shows that $\int \frac{d^dk}{(2\pi)^d}\frac{f_0^2(k)+2f_0(k)f_1(k)}{|k|^{d}}=\frac{S_d}{(2\pi)^d}\log\gamma$, which gives 
$\zeta_2=-2\epsilon\frac{N-2}{N-8}\log\gamma+O(\epsilon^2)$, as already announced.

\section{Proof of (\ref{G.1}) and (\ref{F.1})}
\label{appB}

Using \eqref{eq:6}-\eqref{eq:7} and \eqref{eq:4.1}, we write: 
\begin{equation}\begin{split}
& \mathcal G^*(x)= 2\lim_{h\to-\infty} \gamma^{2h\Delta_1}\sum_{\tau}H_{2,0,0,\emptyset}[\tau](\gamma^h \boldsymbol{x}),\\
& \mathcal F^*(y)= 2\lim_{h\to-\infty} \gamma^{2h\Delta_2}\sum_{\tau}H_{0,2,0,\emptyset}[\tau](\gamma^h \boldsymbol{y}),
\end{split}\end{equation}
From Proposition \ref{Prop:1} and its proof we know that the limits in the right hand sides exist, and that the contribution to $\mathcal G^*(x)$ (resp. $\mathcal F^*(y)$)
from a given tree $\tau$ is bounded from 
above by $C'(C\epsilon)^{n_{\text{e.p.}}[\tau]-2}|x|^{-2\Delta_1}$ (resp. $C'(C\epsilon)^{n_{\text{e.p.}}[\tau]-2}|y|^{-2\Delta_2}$), for some $C,C'>0$, see \eqref{eq:5.11}. 
Therefore, using the fact that the number of trees with $k$ endpoints is smaller than $4^k$, we find that the sum of the contributions to $\mathcal G^*(x)$ (resp. $\mathcal F^*(y)$)
from the trees with 3 or more endpoints is bounded from above by $C\epsilon |x|^{-2\Delta_1}$ (resp. $C\epsilon |y|^{-2\Delta_2}$). Therefore, in order to prove \eqref{G.1}, it is 
sufficient to prove that the contribution to $\mathcal G^*(x)$ from the tree(s) with 2 endpoints is equal to the right hand side of \eqref{G.1} up, possibly, to an error term of the order $\epsilon |x|^{-2\Delta_1}$, and similarly for \eqref{F.1}. 

\medskip

{\it Dominant contribution to $\mathcal G^*(x)$.} Direct inspection shows that there is only one tree with 2 endpoints contributing to $\mathcal G^*(x)$, which is the following: 
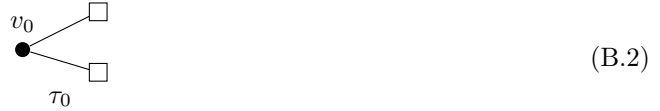
\begin{equation}
\begin{tabular}{rcl}			
\tikz[baseline=-2pt] {
	\draw (0,0) node [vertex,label=$v_0$] (v0) {} ;
	\draw (v0) -- +(1,0.5) node [E]  {};
	\draw (v0) -- +(1,-0.3) node[E,label=below:{$\tau_0\hspace{1cm}$}] {};\,,
}
\end{tabular}
\label{fig:1}
\end{equation}
whose contribution to $\mathcal G^*(x)$ is $\lim_{h\to-\infty}\sum_{h'> h}2^{2h'\Delta_1}g^{(0)}(\gamma^{h'}x)$, which implies \eqref{G.1}. 

\medskip

{\it Dominant contribution to $\mathcal F^*(y)$.} Direct inspection shows that there are two trees with 2 endpoints contributing to $\mathcal F^*(y)$, which are the following:

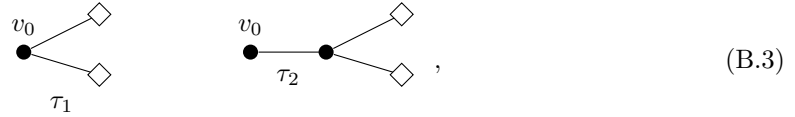
\begin{equation}
\begin{tabular}{rcl}			
\tikz[baseline=-2pt] {
	\draw (0,0) node [vertex,label=$v_0$] (v0) {} ;
	\draw (v0) -- +(1,0.5) node [ctVertex]  {};
	\draw (v0) -- +(1,-0.3) node[ctVertex,label=below:{$\tau_1\hspace{1cm}$}] {};
	%%%%%%%%%
	\draw (3,0)node [vertex,label=$v_0$] (v0) {} ;
	\draw (v0) -- +(1,0) node[vertex,,label=below:{$\tau_2\hspace{1cm}$}](v1){};
	\draw (v1) -- +(1,0.5) node [ctVertex]  {};
	\draw (v1) -- +(1,-0.3) node[ctVertex] {};
}
\end{tabular}\,,
\label{fig:2}
\end{equation}
whose total contribution to $\mathcal F^*(y)$ is (denoting by $\mathfrak{p}_h(x)$ the scalar part of $g^{(h)}(x)$, i.e., $g^{(h)}_{ab}(x)\equiv \Omega_{ab}\mathfrak{p_h}(x)$):
\begin{equation}\label{eq:appF*}\begin{split} &
-2N\lim_{h\to-\infty}\sum_{h'> h}\gamma^{h'(4[\psi]+2\eta_2)}\mathfrak{p}_0(\gamma^{h'}y)\Big[\mathfrak{p}_0(\gamma^{h'}y)+2\sum_{k>0}\gamma^{k(2[\psi]+2\eta_2)}\mathfrak{p}_0(\gamma^{h'+k}y)\Big]=\\
=&-2N\sum_{h'\in\mathbb Z}\mathfrak{p}_{h'}(y)\big[\gamma^{2h'\eta_2}\mathfrak{p}_{h'}(y)+2\sum_{h''>h'}\gamma^{2h''\eta_2}\mathfrak{p}_{h''}(y)\Big].
\end{split}	
\end{equation}	 
We let $f_h(k):=\chi_h(k)-\chi_{h-1}(k)$, with $\chi_h(k):=\chi(\gamma^{-h}k)$, and, after a relabelling of the scale indices, we rewrite \eqref{eq:appF*} as: 
\begin{equation}\begin{split}
&-2N\sum_{h'\in\mathbb{Z}}\gamma^{2h'\eta_2}\left[	\mathfrak{p}_{h'}^2(y)+2\sum_{h<h'}\mathfrak{p}_h(y)\mathfrak{p}_{h'}(y)\right]\\
&=-2N\sum_{h'\in\mathbb{Z}}\gamma^{2h'\eta_2}\int \frac{d^dk}{(2\pi)^d|k|^{d/2+\epsilon}}\int \frac{d^dp}{(2\pi)^d|p|^{d/2+\epsilon}}e^{i(k+p)\cdot y}f_{h'}(p)\left[f_{h'}(k)+2\sum_{h<h'}f_{h}(k)\right]\\
&=-2N\sum_{h'\in\mathbb{Z}}\gamma^{2h'\eta_2}\int \frac{d^dk}{(2\pi)^d|k|^{d/2+\epsilon}}\int \frac{d^dp}{(2\pi)^d|p|^{d/2+\epsilon}}e^{i(k+p)\cdot y}[\chi_{h'}(p)\chi_{h'}(k)-\chi_{h'-1}(p)\chi_{h'-1}(k)]
\label{eq:F.1}
\end{split}\end{equation}	
where we used:
\begin{equation}f_{h'}(k)+2\sum_{h<h'}f_{h}(k)=\chi_{h'}(k)-\chi_{h'-1}(k)+2\chi_{h'-1}(k)=\chi_{h'}(k)+\chi_{h'-1}(k).\end{equation}
Now, noting that 
\begin{equation}
(\gamma^{-h'}|p|)^{-2\eta_2}=1+\int_0^1	ds \frac{d}{ds}(\gamma^{-h'}|p|)^{-2s\eta_2}=1-2\eta_2\log(\gamma^{-h'}|p|)\int_{0}^1ds(\gamma^{-h'}|p|)^{-2s\eta_2},
\end{equation}
multiplying and dividing by $|p|^{-2\eta_2}$ in \eqref{eq:F.1} we get:
\begin{equation}\begin{split}
\eqref{eq:F.1}&=-2N\sum_{h'\in\mathbb Z}\int 	\frac{d^dk}{(2\pi)^d|k|^{d/2+\epsilon}}\int \frac{d^dp}{(2\pi)^{d}|p|^{d/2+\epsilon-2\eta_2}}e^{i(k+p)\cdot y}\left[\chi_{h'}(p)\chi_{h'}(k)-\chi_{h'-1}(p)\chi_{h'-1}(k)\right]\\
&+4\eta_2N\sum_{h'\in\mathbb Z}\int \frac{d^dk}{(2\pi)^d|k|^{d/2+\epsilon}}\int \frac{d^dp}{(2\pi)^d|p|^{d/2+\epsilon-2\eta_2}}e^{i(k+p)\cdot y}
\log(\gamma^{-h'}|p|)\int_{o}^1ds\;(\gamma^{-h'}|p|)^{-2s\eta_2}\cdot\\
&\hspace{5.8cm}\cdot[\chi_{h'}(k)\chi_{h'}(p)-\chi_{h'-1}(k)\chi_{h'-1}(p)].
\label{eq:F.2}
\end{split}
\end{equation}	 
We will prove below that the second addend in the right hand side can be bounded as: 
\begin{equation}\label{eccocidaidai}\begin{split} &4|\eta_2|N\Biggl|\sum_{h'\in\mathbb{Z}}\int \frac{d^dk}{(2\pi)^d|k|^{d/2+\epsilon}}\int \frac{d^dp}{(2\pi)^d|p|^{d/2+\epsilon-2\eta_2}}e^{i(k+p)\cdot y}\log(\gamma^{-h'}|p|)\int_{o}^1ds\;(\gamma^{-h'}|p|)^{-2s\eta_2}\cdot\\
&\hspace{5.8cm}\cdot[\chi_{h'}(k)\chi_{h'}(p)-\chi_{h'-1}(k)\chi_{h'-1}(p)]\Biggr|\le C\epsilon |y|^{-2\Delta_2},\end{split}\end{equation}
that is, this term can be included in the error term $\mathcal F^*_{\text{h.o.}}(y)$ defined and bounded in the lines preceding and following \eqref{F.1}. 
On the other hand, the first term in the right hand side of \eqref{eq:F.2} can be rewritten as follows: recall that the sum over $h'$ in $\mathbb Z$ should be interpreted as the limit as $h\to-\infty$ of the sum over $h'>h$; therefore, using the telescopic structure of the summand and, more specifically, the fact that 
$\lim_{h\to-\infty}\sum_{h'>h}\left[\chi_{h'}(p)\chi_{h'}(k)-\chi_{h'-1}(p)\chi_{h'-1}(k)\right]=\lim_{h\to-\infty}(1-\chi_h(k)\chi_h(p))=1$ (where the identities are in the sense of distributions), 
we see that the first term in the right hand side of \eqref{eq:F.2} is equal to 
\begin{equation}\begin{split}
&\mathcal F_0^*(y)=-2N\int 	\frac{d^dk}{(2\pi)^d|k|^{d/2+\epsilon}}\int \frac{d^dp}{(2\pi)^d|p|^{d/2+\epsilon-2\eta_2}}e^{i(k+p)\cdot y}=\frac{C_0'}{|y|^{d-2\epsilon+2\eta_2}}=\frac{C_0'}{|y|^{2\Delta_2}},
\end{split}
\end{equation}	
as desired. 

\medskip

We are left with proving \eqref{eccocidaidai}. By adding and subtracting $\chi_{h'-1}(k)\chi_{h'}(p)$ to the term in square brackets, we rewrite the second term in the right hand side of  \eqref{eq:F.2} as follows:
\begin{equation}\begin{split}
&4\eta_2N\sum_{h'\in\mathbb{Z}}\int \frac{d^dk}{(2\pi)^d|k|^{d/2+\epsilon}}\int \frac{d^dp}{(2\pi)^d|p|^{d/2+\epsilon-2\eta_2}}e^{i(k+p)\cdot y}\log(\gamma^{-h'}|p|)\cdot\\
&\hspace{2cm}\cdot\int_{0}^1ds\;(\gamma^{-h'}|p|)^{-2s\eta_2}[f_{h'}(k)\chi_{h'}(p)+\chi_{h'-1}(k)f_{h'}(p)]\\	
&=4\eta_2N\sum_{h'\in\mathbb{Z}}\left(\frac{d^dk}{(2\pi)^d|k|^{d/2+\epsilon}}e^{ik\cdot y}f_{h'}(k)\right)\cdot\\
&\hspace{1.8cm}\cdot\left(\int\frac{d^dp}{(2\pi)^d|p|^{d/2+\epsilon-2\eta_2}}e^{ip\cdot y}\log(\gamma^{-h'}|p|)\int_{0}^1ds\;(\gamma^{-h'}|p|)^{-2s\eta_2}\chi_{h'}(p) \right)\\
&+4\eta_2N\sum_{h'\in\mathbb{Z}}\left(\frac{d^dk}{(2\pi)^d|k|^{d/2+\epsilon}}e^{ik\cdot y}\chi_{h'-1}(k)\right)\cdot\\
&\hspace{1.8cm}\cdot\left(\int\frac{d^dp}{(2\pi)^d|p|^{d/2+\epsilon-2\eta_2}}e^{ip\cdot y}\log(\gamma^{-h'}|p|)\int_{0}^1ds\;(\gamma^{-h'}|p|)^{-2s\eta_2}f_{h'}(p) \right)\\
& \equiv 4\eta_2N\sum_{h'\in\mathbb{Z}}\int_0^1ds \Big[\mathfrak{p}_{h'}(y)\sum_{h\le h'} \tilde{\mathfrak{p}}_{h;s,h'}(y)\, +{P}_{\le h'-1}(y)\tilde{\mathfrak{p}}_{h';s,h'}(y)\Big]
\label{eq:reminder}
\end{split}
\end{equation}		
where we recall that $\mathfrak{p}_h(y)$ denotes the scalar part of $g^{(h)}(y)$ (i.e., $g^{(h)}_{ab}(y)\equiv \Omega_{ab}\mathfrak{p_h}(y)$) and, analogously, 
$P_{\le h}(y)$ denotes the scalar part of $G^{(\le h)}(y)$, see \eqref{Glehgeh}; moreover, 
\begin{equation}\label{defptildes}\tilde{\mathfrak{p}}_{h;s,h'}(y):=\int \frac{d^d p}{(2\pi)^d|p|^{d/2+\epsilon-2\eta_2}}e^{ip\cdot y}\log(\gamma^{-h'}|p|)(\gamma^{-h'}|p|)^{-2s\eta_2}f_{h}(p).\end{equation}
Now, recall that, by \eqref{M} and the definition of $\mathfrak{p}_h$, 
\begin{equation}\label{boundg}
|\mathfrak{p}_h(y)|\le C_{\chi^1} \gamma^{h(d/2-\epsilon)}e^{-{C_{\chi^2}}(\gamma^{h-1}|y|)^{\sigma}},\end{equation}
whose proof is given in \cite[Appendix A.2]{GMR21}. In order to get a bound on $\mathfrak{p}_{h;s,h'}(y)$ we proceed as follows: 
in the integral in the right hand side of \eqref{defptildes}, we rescale $p\to \gamma^h p$, thus getting: 
\begin{equation}\begin{split}
\tilde{\mathfrak{p}}_{h;s,h'}(y)&=\gamma^{h(d/2-\epsilon+2\eta_2)}\gamma^{(h'-h)2s\eta_2}\\
&\cdot\int \frac {d^dp}{(2\pi)^d|p|^{d/2+\epsilon+2(s-1)\eta_2}}e^{ip\cdot\gamma^{h}y}\left[\log|p|-(h'-h)\log\gamma\right]f_0(p)
\end{split}
\end{equation}			
The integral in the second line can be bounded by proceeding exactly as in \cite[Appendix A.2]{GMR21} (note, in particular, that $F(p):=\log|p|$ admits an analytic continuation into the polydisk centered at $p\neq 0$ of radius $R=\frac12\max_i|p_i|$ such that the maximum of $\big|F(z)\big|$ in the polydisk is bounded by $C(1+|F(p)|)$: this allows us to proceed as in 
\cite[(A.15)]{GMR21} and following lines); we thus get, for some $C_1,C_2>0$: 
\begin{equation}\label{tildeg}
\left|\tilde{\mathfrak{p}}_{h;s,h'}(y)\right|\le C_1(1+h'-h)\gamma^{h(d/2-\epsilon+2\eta_2)}\gamma^{(h'-h)2s\eta_2}e^{-C_2(\gamma^{h-1}|y|)^{\sigma}}.
\end{equation}
We can now bound \eqref{eq:reminder} via \eqref{boundg} and \eqref{tildeg}, thus getting, for some $C>0$ and $\bar C=\min\{C_{\chi^2},C_2\}$:
\begin{equation}\begin{split}
&\Big|4\eta_2N\sum_{h'\in\mathbb{Z}}\int_0^1ds\Big[\mathfrak{p}_{h'}(y)\sum_{h\le h'} \tilde{\mathfrak{p}}_{h;s,h'}(y)+{P}_{\le h'-1}(y)\tilde{\mathfrak{p}}_{h';s,h'}(y)\Big]\Big|\\
&\le CN|\eta_2|\sum_{h'\in\mathbb{Z}}e^{-\bar{C}(\gamma^{h'-1}|y|)^{\sigma}}\sum_{h\le h'} \Bigg(\gamma^{h'(d/2-\epsilon)}\gamma^{h(d/2-\epsilon+2\eta_2)}(1+h'-h)\int_{0}^1 ds  \gamma^{2s(h'-h)\eta_2}\\
& \hspace{8.5cm}+\gamma^{h(d/2-\epsilon)}\gamma^{h'(d/2-\epsilon+2\eta_2)}\Bigg)\\
&\le C'\epsilon\sum_{h'\in\mathbb{Z}}\gamma^{2h'\Delta_2}e^{-\bar{C}(\gamma^{h'-1}|y|)^{\sigma}}\le C''\epsilon |y|^{-2\Delta_2}		
\end{split}	
\end{equation}	
where we used that $\eta_2= O(\epsilon)$. This conclude the proof of \eqref{F.1}.

\section*{Acknowledgements}
\addcontentsline{toc}{section}{Acknowledgements}

We gratefully acknowledge support from: the European Research Council
(ERC) under the European Union's Horizon 2020 research and innovation programme, ERC CoG UniCoSM, grant agreement n.724939 (A.G.) and ERC StG MaMBoQ, n.802901
(G.S.); MUR, PRIN 2017 project MaQuMA cod. 2017ASFLJR (A.G., V.M., G.S.); MUR, PRIN 2022 project MaIQuFi cod. 20223J85K3; GNFM-INdAM Gruppo Nazionale per la Fisica Matematica. 

\section*{Data availability statement}

The authors declare that the data supporting the findings of this study are available within the paper.

\section*{Conflict of Interest statement}

The first author (A.G.) is on the editorial board of {\it Communications in Mathematical Physics}.

\bibliographystyle{utphys}
\bibliography{biblioRG}
\end{document}